\documentclass[11pt,a4paper]{article}
\usepackage{jheppub}
\usepackage{graphicx}%
\usepackage{subcaption}
\usepackage{dcolumn}% Align table columns on decimal point
\usepackage{bm}% bold math
\usepackage{hyperref}
\usepackage{ragged2e} %fixes a .bbl error though I'm confused why it occurred in the first place, SK 
\usepackage{booktabs}
\usepackage[english]{babel}
\usepackage{amsmath,amssymb,amsbsy,amstext, amsthm, simplewick}
\usepackage{graphicx}
\usepackage{amsfonts}
\usepackage{amssymb}
\usepackage{upgreek}
\usepackage{cancel}
\usepackage{color}
\usepackage{slashed}
\usepackage[dvipsnames]{xcolor}
\usepackage{subcaption}
\usepackage{tikz-cd}
\usepackage{feynmp-auto}
\usepackage{color}
\usepackage{soul}
\usepackage{dsfont}
\usepackage{verbatim}
\usepackage[utf8]{inputenc} 
\usepackage{soul}
\usepackage{epigraph}
\usepackage{todonotes}
\usepackage{csquotes}
\usepackage{tikz}
\usepackage{tikz-3dplot}
\usepackage{braids}
\usetikzlibrary{decorations.markings}
\usepackage[normalem]{ulem}
\renewcommand{\t}[1]{\tilde{#1}}

\def\@fnsymbol#1{\ensuremath{\ifcase#1\or $\Re$\or $\Im$\or  \else\@ctrerr\fi}}

\def\beq{\begin{equation}}
\def\eeq{\end{equation}}
\def\bea{\begin{eqnarray}}
\def\eea{\end{eqnarray}}

\newcommand{\bkA}{\mathcal{A}_2}
\newcommand{\bkB}{\mathcal{B}_2}
\newcommand{\bkC}{\mathcal{C}_2}
\newcommand{\sdoD}{\mathcal{D}}
\newcommand{\nisD}{\mathcal{D}}
\newcommand{\tqftA}{\mathcal{A}}

\begin{document}

\preprint{CERN-TH-2025-204}

\title{Multi-Instantons, Multi-Axions, and Non-Invertible Symmetries in 4d QFT}

\author[a]{Sungwoo Hong,}
\author[a]{Hyungyu Kim,}
\author[a,b]{Sung Mook Lee,}
\author[a]{Dongmin Seo,}

\affiliation[a]{Department of Physics, Korea Advanced Institute of Science and Technology}
\affiliation[b]{Theoretical Physics Department, CERN, CH-1211 Geneve 23, Switzerland}

\emailAdd{sungwooh@kaist.ac.kr}
\emailAdd{hgkim0501@kaist.ac.kr}
\emailAdd{sungmook.lee@cern.ch}
\emailAdd{dm\_seo@kaist.ac.kr}

\abstract{
We study non-invertible global symmetries in 4d quantum field theories, aiming to generalize existing discussions to theories with multiple instantons and axions, and to make the subject more accessible to particle phenomenology.

Building on both the Adler--Bell--Jackiw (ABJ) anomaly construction and the half-space gauging approach, we identify the 3d topological quantum field theories required to describe non-invertible symmetries in the presence of multi-instanton effects. To this end, we introduce a method we call \emph{partial gauging} and show that partial gauging of 3d Chern--Simons theories naturally leads to anomaly inflow actions with general ABJ anomaly matrices. We further generalize the half-space gauging construction to the case of multiple gauge sectors and compute correlation functions of boundary line operators. This enables us to analyze the action of non-invertible operators on various species of 't~Hooft lines and to interpret the results in terms of the Witten effect. For theories with multiple axions, we determine both the non-invertible 0-form and 1-form symmetries, as well as their actions on axion strings and 't~Hooft lines, thereby generalizing the previously known notion of the non-invertible Gauss law.

Our framework unifies several previously disparate constructions, provides a concrete Lagrangian-based formulation of non-invertible symmetries, and naturally extends to theories with multiple instantons and axions, which are of both phenomenological and theoretical interest. These results may also find broader applications in beyond-the-Standard-Model scenarios.
}
\maketitle

\section{Introduction} \label{sec:intro}

Symmetry is a fundamental concept in physics, capturing invariance properties that have long served as a guiding principle in theoretical developments and have profoundly shaped our understanding of the fundamental laws of nature.
The advent of generalized global symmetries (GGSs)~\cite{Gaiotto:2014kfa} has dramatically broadened this program, revealing a richer spectrum of symmetries and suggesting new pathways for uncovering deep structural insights into quantum field theories (QFTs).

In recent years, GGSs have found increasingly important applications in particle physics, ranging from the global structure of the Standard Model~\cite{Tong:2017oea, Anber:2021upc, Choi:2023pdp, Cordova:2023her, Koren:2024xof, Debray:2025iqs, Alonso:2025rkk, Anber:2025gvb, dbsh:2025xx, dbshsk:2025xx, ygshsllw:2025xx}, axion physics~\cite{Hidaka:2020iaz, Hidaka:2020izy, Brennan:2020ehu, Choi:2022fgx, Choi:2023pdp, Cordova:2023her, Anber:2024gis, Craig:2024dnl, Dierigl:2024cxm, Chen:2024tsx, Hidaka:2024kfx, DelZotto:2024ngj, Delgado:2024pcv, Choi:2025vxr, gcshsk:2025xx}, small neutrino mass generation~\cite{Cordova:2022fhg, Kobayashi:2025cwx, Okada:2025kfm}, higher-flavor symmetries~\cite{Cordova:2022qtz}, couplings to topological quantum field theories (TQFTs)~\cite{Brennan:2023kpw}, the monopole--fermion scattering problem~\cite{vanBeest:2023dbu, Brennan:2023tae, vanBeest:2023mbs}, the strong CP problem~\cite{Aloni:2024jpb, Cordova:2024ypu, Liang:2025dkm}, and the electroweak hierarchy problem~\cite{ycshlw:2025xx}, among many other areas (see \cite{Choi:2022rfe,
Putrov:2023jqi,
Das:2023nwl,
Cheung:2024ypq,
Das:2024efs,
Kan:2024fuu,
Garcia-Valdecasas:2024cqn,
Brennan:2024iau,
Hull:2024uwz,
Yang:2024buo,
Wang:2024auy,
Hirono:2025dhz,
Berean-Dutcher:2025ohp,
Hamada:2025cwu,
Koren:2025utp, Gagliano:2025oqv} for further recent developments). For recent reviews on the subject, see~\cite{Gomes:2023ahz, Brennan:2023mmt, Bhardwaj:2023kri, Luo:2023ive, Shao:2023gho, Iqbal:2024pee, Costa:2024wks}, and also the appendices of~\cite{Brennan:2023kpw} for a concise overview.
In this new approach, one type of GGS, known as \emph{non-invertible symmetries}~\cite{Choi:2022jqy, Cordova:2022ieu}, finds broad applications and appears to be particularly powerful.
One reason is that, while higher-form symmetries do not act directly on local operators, non-invertible 0-form symmetries do, providing novel selection rules on admissible operators in the Lagrangian and on scattering amplitudes.
Many of the non-invertible symmetries appearing in 4d QFTs---especially those used in particle physics contexts~\cite{Cordova:2022ieu, Choi:2022jqy, Cordova:2022fhg, Choi:2022fgx, vanBeest:2023dbu, Choi:2023pdp, Cordova:2023her, Cordova:2024ypu, Hidaka:2024kfx, DelZotto:2024ngj, Delgado:2024pcv, gcshsk:2025xx, Kobayashi:2025cwx}---are related to the existence of a $U(1)$ global symmetry, or its discrete subgroup, possessing an Adler--Bell--Jackiw (ABJ) anomaly~\cite{Adler:1969gk, Bell:1969ts}, together with either $U(1)$ instantons or fractional instantons of a non-abelian gauge group.\footnote{See also model-building applications of non-invertible symmetries, including those arising from $\mathbb{Z}_2$ gauging of discrete symmetries~\cite{Okada:2025kfm, Liang:2025dkm, Kobayashi:2024cvp, Kobayashi:2025ldi, Kobayashi:2025lar, Kobayashi:2025thd} and from the Fibonacci~\cite{Suzuki:2025oov}, Ising~\cite{Suzuki:2025oov, Chen:2025awz}, and Tambara--Yamagami~\cite{Nomura:2025sod} fusion rules.}
In the original works~\cite{Choi:2022jqy, Cordova:2022ieu}, the simplest case with a single instanton was considered, while later studies extended the analysis to theories with multiple instantons~\cite{Cordova:2024ypu, Choi:2023pdp, Cordova:2023her, Delgado:2024pcv, gcshsk:2025xx}.
In fact, 4d QFTs with multiple instantons---whether abelian, fractional, or of even more general types~\cite{Anber:2021iip}---are quite ubiquitous, and the Standard Model itself, possibly with a nontrivial global structure, falls into this category~\cite{Tong:2017oea}.

Recent progress has shown that anomalies in theories with $U(1)$ or $PSU(N)$ gauge groups need not imply explicit symmetry breaking by instanton effects. Instead, they can give rise to non-invertible symmetries that impose selection rules---much like ordinary symmetries---and enable a more systematic treatment.

These new types of symmetries are often associated with certain TQFTs supported on defect worldvolumes. We review these aspects in Section~\ref{sec:review_NIS}. However, existing approaches---such as those based on minimal abelian TQFTs---sometimes lack an explicit Lagrangian formulation. It is therefore desirable to develop a more accessible framework, ideally one that provides a clear Lagrangian description of the TQFT and thereby makes the structure and properties of non-invertible symmetries more transparent.

In this paper, we present such a framework, which we call \emph{partial gauging} (see Section~\ref{sec:NIS_from_partial_gauging}). We show that the general 't~Hooft anomaly structure required to cancel 4d ABJ anomalies can be realized using concrete 3d TQFTs---specifically, Chern--Simons (CS) theory and BF theory---with partial gauging of a proper subgroup of the 1-form symmetry via coupling to the 4d theory. These theories admit well-understood Lagrangian descriptions.

One key advantage of this formulation is that it extends naturally to the multi-instanton case, making the computation of fusion rules and the action of symmetries on charged objects more straightforward. Since the transformation of 't~Hooft line operators under non-invertible symmetries is best analyzed through the half-space gauging construction, we generalize the half-space gauging to the case of multiple instantons. In this way, our approach streamlines the analysis of non-invertible symmetries in multi-instanton theories and provides a more direct computational method (see Section~\ref{sec:multi-instanton-case}). We also aim to present our discussion in a way that is accessible and useful not only to formal theorists but also to particle phenomenologists.

Finally, we comment on the case of multiple axions in Section~\ref{sec:multi-axion case}. Many axion-like particle models feature multiple axions (see, for example,~\cite{Kim:2004rp,Higaki:2016jjh,Dunsky:2025sgz,Kondo:2025hdc}) and arise naturally in a variety of contexts, including the string axiverse through extra-dimensional compactification~\cite{Svrcek:2006yi,Arvanitaki:2009fg,Reece:2025thc}. Understanding the structure of non-invertible 0-form and 1-form symmetries in such theories is therefore of both theoretical and phenomenological interest, with important implications for the physics of topological defects and their cosmological consequences~\cite{Benabou:2023npn,Lee:2024toz}.

The rest of the paper is organized as follows. In Section~\ref{sec:review_NIS}, we provide a detailed discussion of non-invertible 0-form and 1-form symmetries.
Readers familiar with the basics of these concepts may skip to later sections, although this section may still be useful for understanding our conventions.
In Section~\ref{sec:NIS_from_partial_gauging}, we introduce the method of partial gauging as a way to obtain the inflow action of a 3d TQFT with a non-minimal anomaly coefficient.
To this end, we take the simple case of axion--Maxwell theory to show that a non-invertible 0-form symmetry defect operator (SDO) with non-minimal ABJ anomalies can be obtained through partial gauging.
In Section~\ref{sec:multi-instanton-case}, we discuss non-invertible 0-form symmetries in the presence of multiple instanton effects.
In Section~\ref{subsec:NIS_from_multiU1}, we study the case with multiple $U(1)$ factors and an arbitrary symmetric ABJ anomaly matrix $K_{ij}$, and we present both ABJ anomaly and half-space gauging constructions.
In Section~\ref{subsec:NIS0_multi-inst_general}, we analyze the fully general case involving multiple $U(1)$ instantons, multiple $SU(N)$ instantons, and multiple fractional instantons of $PSU(N)$ gauge theories.
In Section~\ref{subsec:NIS from CFU}, we further discuss non-invertible symmetries arising from a more general class of instantons, sometimes referred to as color--flavor--$U(1)$ (CFU) instantons.
These results allow us, in Section~\ref{subsec:multi-inst_general_action on t Hooft}, to determine the action of non-invertible SDOs on various species of 't~Hooft line operators, which we further interpret in terms of the Witten effect.
In Section~\ref{sec:multi-axion case}, we discuss theories with multiple axions and the resulting non-invertible 0-form and 1-form symmetries. We present the action of non-invertible 1-form symmetries on Wilson lines, 't~Hooft lines, and axion string operators.
Finally, we conclude in Section~\ref{sec:conclusion}.

Many, often important, details are relegated to several appendices.
In Appendix~\ref{app:spin and anomaly}, we give a brief review of the topological spin of line operators in 3d TQFTs and its relation to the 't~Hooft anomaly.
In Appendix~\ref{app:boundary lines and braiding}, we compute correlation functions of boundary line operators that appear in half-space gauging with multiple instantons and show which 1-form symmetries the 3d boundary TQFT must contain.
In Appendix~\ref{app:fractional_instantons}, we provide a review of fractional instantons.

%%%%%%%%%%%%%%%%%%%%%%%%%%%%%%%%%%%%%%%%%%%%%%%%%%%%%%%%%%%%%%%%%%%%%%%%%%%%%%%%%%
\section{Review of Non-Invertible Symmetries} 
\label{sec:review_NIS}
%%%%%%%%%%%%%%%%%%%%%%%%%%%%%%%%%%%%%%%%%%%%%%%%%%%%%%%%%%%%%%%%%%%%%%%%%%%%%%%%%%

In this section, we review non-invertible 0-form and 1-form symmetries in 4d QFT.
While our discussion below follows closely \cite{Choi:2022jqy, Cordova:2022ieu, Choi:2022fgx, Brennan:2023mmt}, we also aim to make it as accessible as possible and provide some new perspectives while setting our notation.
Our discussion includes non-invertible symmetries arising from fractional instantons as well as from $U(1)$ instantons.
Readers familiar with these basic concepts may skip this section, read only the necessary parts, and proceed directly to Section~\ref{sec:NIS_from_partial_gauging} and later sections for our new results.

%%%%%%%%%%%%%%%%%%%%%%%%%%%%%%%%%%%%%%%%%
\subsection{Non-invertible 0-form symmetries}
\label{subsec:0-form_NIS}
%%%%%%%%%%%%%%%%%%%%%%%%%%%%%%%%%%%%%%%%%

\subsubsection{Non-invertible symmetry from ABJ anomaly}
\label{subsubsec:NIS from ABJ}

To be concrete, let us consider 4d QED with a single Dirac fermion $\psi = \left( \chi, \epsilon \eta^* \right)^t$ charged under a $U(1)_G$ gauge group, where $\chi$ and $\eta$ are two-component Weyl spinors.
The action is given by
\beq
S = \frac{1}{2g^2} \int F_2 \wedge * F_2 + i \int \bar{\psi} \slashed{D} \psi \, ,
\label{eq:4d QED}
\eeq
where $F_2 = dA_1$ is the $U(1)_G$ field strength.
In the limit $g \to 0$, the free fermion theory possesses two $U(1)$ global symmetries, $U(1)_L \times U(1)_R$, or equivalently $U(1)_V \times U(1)_A$ classically.
Under the vector-like $U(1)_{V}$ symmetry, the left-handed and right-handed components of $\psi$ transform in the same way: $\psi \to e^{i \alpha} \psi$.\footnote{
In terms of the two Weyl spinors $\chi$ and $\eta$, the vector-like transformation corresponds to $\chi \to e^{i \alpha} \chi$ and $\eta \to e^{-i \alpha} \eta$.}
This symmetry is anomaly-free and can therefore be gauged.
The resulting theory is precisely the QED shown in Eq.~\eqref{eq:4d QED}.
The other symmetry, $U(1)_A$---often called the axial or chiral $U(1)$---acts on the two Weyl spinors $\chi$ and $\eta$ in the same way, which can be expressed in terms of the Dirac fermion $\psi$ as
\beq
\psi \to e^{i \alpha \gamma_5} \psi\, , \quad J_A^\mu = \bar{\psi} \gamma^\mu \gamma_5 \psi\, ,
\eeq
where $\alpha \sim \alpha + 2\pi$ is the $U(1)_A$ transformation parameter, and $J_A^\mu$ is the Noether current associated with the $U(1)_A$ symmetry.
As is well known, $U(1)_A$ suffers from the ABJ anomaly with $U(1)_G$,\footnote{
For the global symmetry, we use $U(1)_V$, but once it is gauged, we denote it by $U(1)_G$ to emphasize that it is a gauge redundancy rather than a global symmetry.
}
which is captured by the anomalous Ward identity:
\beq
d * J_A = \frac{2}{8 \pi^2} F_2 \wedge F_2 \, .
\label{eq:anom Ward id}
\eeq
Naively, one would conclude that the classical $U(1)_A$ symmetry is broken to $\mathbb{Z}_2$ by the quantum anomaly.
To be clear, this $\mathbb{Z}_2$ acts as $\psi \to -\psi$, which coincides with the $\mathbb{Z}_2$ subgroup of $U(1)_G$. Hence, it is not a genuine part of $U(1)_A$.\footnote{Technically, one says that the global form of the symmetry in the limit $g \to 0$ is
\beq
\frac{U(1)_V \times U(1)_A}{\mathbb{Z}_2}\, .
\eeq
Once $U(1)_V$ is gauged, the remaining global symmetry becomes $U(1)_A / \mathbb{Z}_2$ if there was no anomaly.
}
To take this fact into account, we redefine $\alpha \to \alpha/2$, so that $\alpha/2 \in [0, \pi)$, thereby incorporating the $\mathbb{Z}_2$ quotient.
On the other hand, since $\pi_3(U(1)) = 0$, one might instead expect that there are no genuine instanton effects when the spacetime manifold is $\mathbb{R}^4$ (or $\mathbb{S}^4$), and thus that the entire $U(1)_A$ remains a good quantum symmetry.
As we now explain, the best resolution of this apparent contradiction is that $U(1)_A$ is broken down to---or more precisely, converted into---a set of discrete non-invertible symmetries.
In modern language, the existence of a symmetry is manifested by the existence of a well-defined, and in particular gauge-invariant, topological operator that implements the symmetry transformation.
If we have a conserved 1-form current, $d * J_1 = 0$, the associated SDO $U \left(\alpha, \Sigma_3\right) = \exp \left(\frac{i\alpha}{2} \oint_{\Sigma_3} * J_1 \right) $ is topological in the sense that it does not change under any small and smooth deformation $\Sigma_3 \to \Sigma_3^\prime$, as long as no charged operator is inserted in between.
This follows from the fact that the difference is given by $\exp\left(\frac{i\alpha}{2} \int_{\Sigma_4} d * J_1\right) = 1$ where $\Sigma_4$ is the 4-manifold interpolating between $\Sigma_3$ and $\Sigma_3^\prime$, so that $\partial \Sigma_4 = \Sigma_3 \cup \overline{\Sigma}_3^\prime$ (with $\overline{\Sigma}$ denoting the orientation reversal of $\Sigma$).
In this work, we will refer to the SDO $U \left(\alpha, \Sigma_3\right) = \exp \left( \frac{i\alpha}{2} \oint_{\Sigma_3} * J_1 \right) $ constructed solely from the Noether current as the ``Noether SDO'', in order to distinguish it from the ``non-invertible SDO'' that we will discuss below.
Returning to Eq.~\eqref{eq:anom Ward id}, and using the facts that $d\left(F_2 \wedge F_2\right) = 0$ and $d\left(A_1 \wedge F_2\right) = F_2 \wedge F_2$,
one might be tempted to redefine the current as
\beq
*\hat{J}_A \equiv * J_A - \frac{1}{4\pi^2} A_1 \wedge F_2\, ,
\eeq
since in this case $d * \hat{J}_A = 0$. 
In fact, the corresponding would-be SDO
\beq
\label{naive 0-form SDO}
\hat{U}\left(\alpha, \Sigma_3\right) = \exp \left( \frac{i\alpha}{2} \oint_{\Sigma_3} * \hat{J}_A \right)
\eeq
is indeed topological.
The problem is that this operator is not gauge-invariant for arbitrary values of $\alpha$.
The issue arises because the 3d CS action is gauge-invariant only if its coefficient---the level---is quantized:
\bea 
\label{eq:3d CS action}
S_{\rm CS} = \frac{i N }{4\pi} \int_{\Sigma_3} a_1 \wedge d a_1  \, ,  \quad
N \in \left\lbrace
\begin{array}{ll}
     2 \mathbb{Z} \hspace{0.3cm} {\rm if} \;\; \Sigma_3 =  \text{non-spin manifold}\, , \\
     \mathbb{Z} \hspace{0.5cm} {\rm if} \;\; \Sigma_3 =  \text{spin manifold} \, ,
\end{array} \right.
\eea 
where $a_1$ is a dynamical $U(1)$ 1-form gauge field defined on $\Sigma_3$, and we identify $N = \alpha / 2\pi$.
Here, a manifold being ``spin'' means that fermions (i.e.~a spin structure) can be properly defined and introduced in the theory, and ``non-spin'' otherwise.\footnote{A technical introduction to this concept can be found in the appendix of \cite{Witten:2015aba}.}
If the level $N$ is not quantized as specified above, the theory is not invariant under the large gauge transformation $a_1 \to a_1 + \lambda_1$.
(See \cite{Brennan:2023mmt} for further details.)
To proceed further albeit the caveat discussed above, let us consider a discrete transformation with $\alpha = 2\pi/z$, where $z \in \mathbb{Z}$. The SDO then takes the form
\beq 
\hat{U} \left( \frac{2\pi}{z}, \Sigma_3 \right) = \exp \left[ i \oint_{\Sigma_3} \left( \frac{\pi}{z} * J_A - \frac{1}{4\pi z} A_1 \wedge F_2 \right) \right] \, .
\label{eq:NI_SDO_FHQ}
\eeq 
The coefficient of the CS term still does not satisfy the usual quantization condition.
Interestingly, however, it is by now well known that for this choice of coefficient, there exists a gauge-invariant description, developed in the context of the fractional quantum Hall (FQH) effect in condensed matter physics.
To explain this, let us consider the 3d CS theory given in Eq.~\eqref{eq:3d CS action}.
This theory possesses a 1-form $\mathbb{Z}_N^{(1)}$ symmetry, which acts as $a_1 \to a_1 + \frac{2\pi}{N} \epsilon_1$ with $\oint_{\Sigma_1} \epsilon_1 \in \mathbb{Z}$.
This symmetry has a 't Hooft anomaly, which can be derived as follows.
First, we couple the 3d CS theory to a 2-form background gauge field (BGF) $\mathcal{A}_2$ associated with the $\mathbb{Z}_N^{(1)}$ symmetry.\footnote{For whom may not be familiar with the discrete gauge theory, we refer \cite{Brennan:2023mmt} for the introduction. Even though there is no conserved current in the theory, SDO is still well-defined with the discrete symmetry and this is one of the direct advantages of generalized notion of symmetry.}
The coupled action is given by
\beq 
S_{\rm CS} \left[\bkA\right] = \frac{i N}{4\pi} \int_{\Sigma_3} a_1 \wedge d a_1 + \frac{i N}{2\pi} \int_{\Sigma_3} a_1 \wedge \bkA \, .
\label{eq:3d CS coupled to BGF}
\eeq 
One way to see this is to note that, under the local version of the symmetry transformation $a_1 \to a_1 + \lambda_1$ (where $\lambda_1$ is not a closed form), the 3d CS action shifts by
\beq 
\delta_{\lambda_1} S_{\rm CS} = \frac{i N}{2\pi} \int_{\Sigma_3} a_1 \wedge d \lambda_1 \, . 
\label{eq:Z_N current}
\eeq 
One may recall that the \emph{conserved} Noether current $J^\mu$ for a 0-form symmetry can be obtained by performing the local version of the symmetry transformation, under which the action changes as
\beq
\delta_{\epsilon} S = \int d^dx\, (\partial_\mu \epsilon) J^\mu \, .
\eeq
Motivated by this, one may identify the \emph{closed} 1-form current for the $\mathbb{Z}_N^{(1)}$ symmetry as $j_1 = \frac{N}{2\pi} a_1$, with the corresponding \emph{conserved} 2-form current given by $* j_1$.
Of course, care is required since we are dealing with a discrete symmetry.
In fact, neither $j_1$ nor the associated charge operator $Q(\Sigma_1) = \oint_{\Sigma_1} j_1$ is gauge-invariant, and therefore they should not be regarded as strictly well-defined objects.
Nevertheless, the resulting SDO, $U(\Sigma_1) = \exp \left[\frac{2\pi i}{N} Q (\Sigma_1) \right]$ is gauge-invariant. Keeping these subtleties in mind, we will (ab)use the closed current $j_1$ and write the BGF coupling in the form $\frac{1}{3!} \int d^3 x\, \epsilon^{\mu\nu\rho} \mathcal{A}_{\mu\nu} j_\rho = \int_{\Sigma_3} \mathcal{A}_2 \wedge j_1$.
The result reproduces Eq.~\eqref{eq:3d CS coupled to BGF}.
It is useful to rewrite Eq.~\eqref{eq:3d CS coupled to BGF} on an auxiliary 4d manifold $\Sigma_4$ with $\partial \Sigma_4 = \Sigma_3$.
The 4d action of interest is
\beq  
S = \frac{i N}{4\pi} \int_{\Sigma_4} \left( d a_1 + \bkA \right) \wedge \left( d a_1 + \bkA \right) \, .
\label{eq:3d CS to 4d BGC}
\eeq 
This 4d action almost reduces to Eq.~\eqref{eq:3d CS coupled to BGF} by Stokes' theorem, but not exactly.
In fact, it is straightforward to check that everything except the term proportional to $\bkA \wedge \bkA$ agrees with the 3d action.
Since the term proportional to $\bkA \wedge \bkA$ is intrinsically a 4d local term, it should not be viewed as a local counterterm of the original 3d theory.
If we assign the background gauge transformation $\bkA \to \bkA - d \lambda_1$, then Eq.~\eqref{eq:3d CS to 4d BGC} is invariant under this transformation.
Importantly, this does not imply that the original 3d theory is invariant.
As we just discussed, invariance is achieved only when the term proportional to $\bkA \wedge \bkA$ is added. From this, we find that the anomaly of the 3d CS theory can be canceled by the the term proportional to $\bkA \wedge \bkA$ in 4d---known as the 4d anomaly inflow action~\cite{Callan:1984sa, Witten:2019bou} (see \cite{Harvey:2005it, Hong:2020bvq} for reviews):
\beq 
S_{\rm inflow} = - \frac{i N}{4\pi} \int_{\Sigma_4} \bkA \wedge \bkA \, .
\label{eq:CS_inflow}
\eeq 
We note that $\bkA$ is a 2-form BGF for the $\mathbb{Z}_N^{(1)}$ symmetry, and hence it is normalized as
\beq 
\oint_{\Sigma_2} \frac{\bkA}{2\pi} \in \frac{1}{N} \mathbb{Z} \, .
\label{eq:Z_N BGF holonomy}
\eeq 
Let us take the 2-form BGF to be $\bkA = F_2 / N $. Since $F_2$ is the field strength of the $U(1)$ 1-form gauge field in 4d, the combination $ F_2 / N $ indeed satisfies the desired quantization in Eq.~\eqref{eq:Z_N BGF holonomy}.
The resulting action is
\beq 
S_{\rm CS} \left[ \frac{F_2}{N} \right] = \frac{i N}{4\pi} \int_{\Sigma_3} a_1 \wedge d a_1 + \frac{i}{2\pi} \int_{\Sigma_3} a_1 \wedge F_2 \, .
\label{eq:3d CS coupled to F_2}
\eeq 
We can integrate out $a_1$ by using its equation of motion, $\frac{N}{2\pi} d a_1 = -\frac{F_2}{2\pi}$, which implies $a_1 = - A_1/N$.
Substituting this back into the action yields the effective action for the fractional quantum Hall states:
\beq
S_{ 
\rm FQH} = - \frac{i}{4\pi N} \int_{\Sigma_3} A_1 \wedge F_2 \, .
\eeq 
This is precisely the second term appearing in $\hat{U} \left(\frac{2\pi}{N}, \Sigma_3\right)$ in Eq.~\eqref{eq:NI_SDO_FHQ}, showing that the FQH state indeed provides a gauge-invariant description.\footnote{
It is worth mentioning that the above-described procedure is not completely well-defined, since the LHS of $a_1 = - A_1 / N$ is a $U(1)$ gauge field whereas the RHS is a $\mathbb{Z}_N$ gauge field.
A more rigorous construction of the non-invertible SDO is provided by ``half-space gauging,'' which we discuss in Section~\ref{subsubsec:half space gauging}.}
An alternative viewpoint turns out to be useful.
To this end, let us rewrite the SDO as
\begin{equation}
    \begin{aligned}
    \mathcal{D} \left( \frac{2\pi}{z}, \Sigma_3 \right) &\equiv U \left( \frac{2\pi}{z}, \Sigma_3 \right) \mathcal{A}^{N, 1} \left[ \frac{F_2}{N} \right]  \\  
    &= \int \left[D a_1\right] \exp \left[ i  \oint_{\Sigma_3} \left( \frac{\pi}{z} * J_A + \frac{i N}{4\pi} a_1 \wedge d a_1 + \frac{i}{2\pi} a_1 \wedge F_2 \right) \right] \, .
\end{aligned}
\end{equation}
Here, the term involving the Noether current $J_A$ corresponds to the ``Noether SDO'' $U\left(\frac{2\pi}{z}, \Sigma_3\right)$, while $\tqftA^{N,1}\left[\frac{F_2}{N}\right]$ denotes the partition function of the 3d defect worldvolume TQFT that produces the FQH effective action, nothing but CS action in this case.
The superscript $N$ in $\tqftA^{N,1}$ indicates that this 3d TQFT has a $\mathbb{Z}_N^{(1)}$ symmetry, while the second superscript $1$---later generalized to a symbol $p \in \mathbb{Z}$---labels the size of the 't Hooft anomaly of the $\mathbb{Z}_N^{(1)}$ symmetry.
Finally, the factor $F_2 / N$ in the square brackets specifies that the 2-form BGF of the $\mathbb{Z}_N^{(1)}$ symmetry is taken to be $\mathcal{A}_2 = F_2/N$, where $F_2$ is the field strength of the 4d bulk theory.
We recall that under the action of the ``Noether SDO'' part, the 4d bulk action shifts by the anomaly term under the $U(1)_A$ transformation with $\alpha = 2\pi / z$ as
\beq
S \to S + \frac{2\pi i}{z} \int \frac{F_2 \wedge F_2}{8\pi^2} \, . \quad {(\rm Noether \ SDO)}
\label{eq:S_ABJ}
\eeq 
Turning to the 3d TQFT part, note that while the 4d anomaly inflow action of the 3d defect worldvolume TQFT written in terms of the auxiliary 2-form BGF $\bkA$ vanishes when we set $\bkA = 0$, choosing $\bkA = F_2 / N$ produces a genuine effect.
Namely, since $F_2$ is the field strength of the 4d bulk theory, it is \emph{activated} and cannot be turned off at will: the $\mathbb{Z}_N^{(1)}$ symmetry of the 3d TQFT is gauged by a 4d bulk dynamical gauge field.
This means that under the 3d TQFT part of the SDO, the action further shifts by the 't Hooft anomaly term
\beq
S \to S - \frac{2\pi i}{N} \int \frac{F_2 \wedge F_2}{8\pi^2} \, . \quad {(\rm 3d \ TQFT)}
\label{eq:S_tHooft}
\eeq 
Combining Eq.~\eqref{eq:S_ABJ} with Eq.~\eqref{eq:S_tHooft}, we see that the two anomaly effects cancel for $z = N$. In other words, the theory is invariant under this improved SDO for any rational $U(1)_A$ rotation.
This vividly shows that an anomalous symmetry---previously thought to be lost due to anomaly effects---is converted into a set of discrete symmetries.
These symmetries are \emph{non-invertible}, since there does not exist an inverse SDO for $\sdoD \left( \frac{2\pi}{N}, \Sigma_3 \right)$ for any choice of $N \in \mathbb{Z}$.
Explicitly, one finds
\beq
\begin{aligned}
& \sdoD \left( \frac{2\pi}{N}, \Sigma_3 \right) \times \sdoD^\dagger \left( \frac{2\pi}{N}, \Sigma_3 \right) \\
& \quad = \int \left[D a_1 D b_1\right] \exp \left\{i\oint_{\Sigma_3} \left[\frac{N}{4\pi} \left( a_1 \wedge d a_1 - b_1 \wedge d b_1 \right) + \frac{1}{2\pi} \left(a_1 - b_1\right) \wedge F_2 \right] \right\} \\
& \quad =  \int \left[D c_+ D c_-\right] \exp \left[i\oint_{\Sigma_3} \left(\frac{N}{2\pi} c_+ \wedge d c_-  + \frac{1}{2\pi} c_- \wedge F_2 \right) \right] \neq \mathbf{1} \, ,
\end{aligned}
\label{eq:condensation operator}
\eeq
where we used that the Noether SDO part is unitary and made the change of variables $c_\pm = \frac{1}{2} \left(a_1 \pm b_1\right)$ to obtain the last line.
This shows that the SDO $\sdoD \left(\frac{2\pi}{N}, \Sigma_3 \right)$ is indeed non-invertible and corresponds to a non-group-like symmetry.

As an aside---though it is not used heavily in this paper---it may be useful to comment that the expression in Eq.~\eqref{eq:condensation operator} is an example of a ``condensation operator''~\cite{Choi:2022zal}; see also~\cite{Kong:2014qka,Else:2017yqj,Gaiotto:2019xmp,Kong:2020cie,Johnson-Freyd:2020twl,Roumpedakis:2022aik}.
The basic idea is as follows.
Consider a $d$-dimensional QFT with a $p$-form global symmetry $G^{(p)}$.
In this theory, imagine removing a codimension-$q$ (i.e.~$(d-q)$-dimensional) hypersurface denoted by $\Sigma_{d-q}$. Equivalently, this can be viewed as inserting a codimension-$q$ defect in spacetime.
It is then natural to introduce a ``defect'' QFT living on $\Sigma_{d-q}$.
A particularly interesting case is when the defect QFT gauges the bulk $G^{(p)}$ symmetry on $\Sigma_{d-q}$.
When such gauging is possible, we say that the theory is \emph{$q$-gaugeable}.
Conceptually, $q$-gauging can be thought of as summing over all possible insertions, along $\Sigma_{d-q}$, of $(d - q - p - 1)$-dimensional SDOs (note that this is the dimension of the SDO for a $p$-form symmetry in $(d-q)$ dimensions).
Alternatively, we can perform $q$-gauging by pulling the $G^{(p)}$ symmetry ``current'' to $\Sigma_{d-q}$ and coupling it to a $(p - q + 1)$-form gauge field $a_{p-q+1}$.
In addition---just as we introduce a gauge-kinetic term when gauging a $0$-form symmetry---we may add a nontrivial action for $a_{p-q+1}$.
In the context of non-invertible symmetry, we are interested in a local, topological action of CS type, since we want the resulting defect to be topological and thus suitable for a symmetry operator.
In the literature, such a term is referred to as a symmetry-protected topological (SPT) phase or as discrete torsion.\footnote{If we denote the SPT phase by $\eta\left[a_{p-q+1}\right]$, then formally it is a cohomology class $\eta \in H^{d-q}\left(B^{p}G, U(1)\right)$, meaning that $\eta$ can be regarded as a $(d-q)$-form built from $a_{p-q+1}$, so that integrating it over a $(d-q)$-cycle yields a $U(1)$ phase.
In the context of non-invertible symmetry, one is interested in a ``trivial'' condensation defect, since the theory must be self-dual under higher gauging, i.e.~the theory returns to itself after the higher gauging.
Here, ``trivial'' means that its insertion on any closed manifold can be removed by topological local counterterms on the defect worldvolume; in this way its insertion is effectively trivial and the theory remains unchanged.
A suitable choice of SPT phase is then made to ensure that the condensation defect is trivial.
} With this background in place, let us return to Eq.~\eqref{eq:condensation operator} and examine the structure.
The 4d bulk theory has a $U(1)^{(1)}$ magnetic symmetry with a closed current $F_2/2\pi$.
On $\Sigma_3$, this current is coupled to a 1-form gauge field $c_-$, together with an SPT term proportional to $c_+ \wedge d c_-$, which can be recognized as a 3d $\mathbb{Z}_N$ BF action.
The latter renders $c_-$ a $\mathbb{Z}_N$ field, showing that a subgroup $\mathbb{Z}_N^{(1)} \subset U(1)^{(1)}$ of the bulk 1-form magnetic symmetry is 1-gauged.

Coming back to the main context, we have so far focused on the case where the ABJ anomaly coefficient of the 4d bulk theory is 1 (see Eq.~\eqref{eq:S_ABJ}).
When the anomaly is greater than 1, an essentially identical procedure can be used to construct the associated non-invertible SDO.
The main change is the 3d TQFT defined on the SDO worldvolume, which must compensate the 4d ABJ anomaly by its 't Hooft anomaly.
In the literature, this is done using the 3d ``minimal'' $\mathbb{Z}_N$ TQFT, whose partition function is often denoted by $\tqftA^{N, p}$. 
We now explain the meaning of this expression.
First, the notion of ``minimality'' here means the following.
Any 3d TQFT $\mathcal{T}$ possessing a $\mathbb{Z}_N^{(1)}$ symmetry with a 't Hooft anomaly labeled by an integer $p \sim p+2N$ (or $p \sim p+N$ on spin manifolds), with $\gcd(N,p)=1$ and $pN \in 2\mathbb{Z}$ (or $pN \in \mathbb{Z}$ on spin manifolds), can be factorized as in \cite{Hsin:2018vcg}:
\beq
\mathcal{T} = \tqftA^{N,p} \otimes \mathcal{T}' \, ,
\eeq
where $\tqftA^{N,p}$ consists of $N$ symmetry lines generating the $\mathbb{Z}_N^{(1)}$ symmetry, and $\mathcal{T}'$ denotes a decoupled sector, whose precise meaning we explain below.
More explanation for the periodicity and quantization condition on $p$ will be given below when we discuss a more concrete setup.

The point is that if $\gcd(N,p)=1$ and $pN \in \mathbb{Z}$, then the $N$ symmetry lines by themselves form a consistent 3d TQFT; the condition $\gcd(N,p)=1$ is necessary for the braiding of symmetry lines to be non-degenerate, and hence for the modular $S$-matrix to be unitary \cite{Hsin:2018vcg} (see Appendix~\ref{subapp:spin and anomaly_cs} for a review on braiding of symmetry lines and topological spin).
This means that the entire effect of the $\mathbb{Z}_N^{(1)}$ symmetry is fully captured by the $\tqftA^{N,p}$ factor, while the $\mathcal{T}'$ ``sector'' contains only $\mathbb{Z}_N^{(1)}$-neutral lines; hence $\mathcal{T}'$ corresponds to a decoupled sector, as advertised.
When we couple $\tqftA^{N,p}$ to a 2-form BGF $\bkB$ associated with the $\mathbb{Z}_N^{(1)}$ symmetry, we denote the corresponding partition function by $\tqftA^{N,p}\left[\bkB\right]$.
As we explain in more detail below, the corresponding 4d anomaly inflow action takes the form
\beq
S_{\rm inflow}
= - \frac{ipN}{4\pi} \int_{\Sigma_4} \bkB \wedge \bkB
= - \frac{2\pi ip}{N} \int_{\Sigma_4} \frac{w_2 \wedge w_2}{2} \, ,
\label{eq:CS_inflow_anomaly_p}
\eeq
where $w_2 = N\frac{\bkB}{2\pi}$.
Returning to the discussion of non-invertible symmetry, consider a 4d theory with ABJ anomaly coefficient $K \neq 1$. Under the $U(1)_A$ transformation with $\alpha = 2\pi / N$, the action shifts as
\beq
S \to S + \frac{2\pi i K}{N} \int \frac{F_2 \wedge F_2}{8\pi^2} \, .
\eeq
This can be compensated by the 't~Hooft anomaly discussed above. To avoid confusion with the discrete angle $\alpha$ labeled by $N$, we instead work with a $\mathbb{Z}_M^{(1)}$ symmetry and impose $K/N = p/M$ with $\gcd(M, p) = 1$.
We identify the 2-form BGF associated with the $\mathbb{Z}_M^{(1)}$ symmetry as $\bkB = F_2/M$, which is natural because $\bkB$ is a $\mathbb{Z}_M$ 2-form gauge field, whereas $F_2$ is the field strength of a $U(1)$ 1-form gauge field.
Overall, we see that a non-invertible SDO can be constructed as  
\beq
\sdoD \left( \frac{2\pi}{N}, \Sigma_3 \right) \equiv U \left( \frac{2\pi}{N}, \Sigma_3 \right) \tqftA^{M, p} \left[ \frac{F_2}{M} \right] \, .
\eeq 
Before we end this section, we discuss the connection between the non-invertible symmetries appearing in the QED-type theory above and those in an axion-type theory.
In fact, once the anomalous $U(1)_A$ is spontaneously broken by condensing a scalar, the QED-type theory flows to an axion theory in which CS interaction terms appear that match the ABJ anomalies of $U(1)_A$.
Specifically, all the anomaly data relevant for the study of non-invertible symmetry are captured by, now considering a general case where an axion couples to many $U(1)$ gauge fields
\beq
S \supset \frac{i K_{ij}}{8\pi^2} \int \theta \operatorname{Tr} \left( F_2^i \wedge F_2^j \right) \, ,
\eeq
where $\theta \equiv a/f_a \sim \theta + 2\pi$ is the axion, and $K_{ij}$ is a symmetric integer matrix encoding the ABJ anomaly information.
Here, in defining $a$, we also introduced the axion decay constant $f_a$ to properly normalize its kinetic term.
In the rest of the paper, we often use an axion theory to discuss various aspects of non-invertible symmetries, and it is understood that our discussion is fully general, with the axion-theory description chosen as a convenient parametrization of the ABJ anomaly data.

\subsubsection{Non-invertible symmetry from half-space gauging}
\label{subsubsec:half space gauging}

There is an alternative way to construct the non-invertible SDO: half-space gauging.
This method makes the topological nature of the non-invertible SDO manifest, illustrates the role of the 1-form magnetic symmetry, and is also useful when discussing the action of the non-invertible SDO on 't Hooft line operators.
We reconsider the QED of Section~\ref{subsubsec:NIS from ABJ} and divide spacetime into two regions along a 3-manifold $\Sigma_3$.
Let us take $x\in\mathbb{R}$ to be the coordinate normal to $\Sigma_3$.
We next gauge the subgroup $\mathbb{Z}_M^{(1)} \subset U(1)_m^{(1)}$ on the half-space $x \ge 0$, where $U(1)_m^{(1)}$ denotes the $U(1)$ 1-form magnetic symmetry. This is implemented by first coupling to a 2-form BGF $\mathcal{B}_2$ associated with the $\mathbb{Z}_M^{(1)}$ magnetic symmetry, and then promoting $\mathcal{B}_2$ to a dynamical $\mathbb{Z}_M$ 2-form gauge field $b_2$.
Explicitly, we add the following terms to the QED action:
\beq 
\Delta S = \frac{i p}{2\pi} \int_{x\geq0} b_2 \wedge F_2 + \frac{i M}{2\pi} \int_{x\geq0} b_2 \wedge d c_1  + \frac{ip M}{4\pi} \int_{x\geq0} b_2 \wedge b_2\, , \quad \text{gcd} (M, p) =1 \, .
\label{eq:half space gauging_4d TQFT}
\eeq
Here, the first term describes the coupling of the Noether current of the 1-form magnetic symmetry to the 2-form gauge field, and the second term is the 4d $\mathbb{Z}_M$ BF action, which makes $b_2$ a $\mathbb{Z}_M$ gauge field.
The last term, often called an SPT or discrete torsion term, is an allowed local, gauge-invariant, and topological term.
A detailed discussion of the 4d $\mathbb{Z}_M$ BF theory with an SPT term can be found in \cite{Kapustin:2014gua, Brennan:2023mmt}.
We impose the boundary condition $b_2 \vert_{x=0} = 0$.
This is a good choice because it corresponds to a topological boundary condition: under a smooth deformation of the $x = 0$ surface, one obtains $b_2 |_{x = 0} - b_2 |_{x' = 0} = d b_2 = 0$, 
with the last equality enforced by the 4d $\mathbb{Z}_M$ BF action.

Here, $p$ has to satisfy a quantization condition:
\bea\label{eq:quant_period_p} 
p M \in 2\mathbb{Z} \quad \text{and} \quad  p \sim p + 2 M\, , 
\eea 
as advertised.
These can be seen as follows. First, the addition of an SPT term with coefficient $p$ has the effect of charging $c_1$ under the 1-form gauge symmetry of the $b_2$ field: under the 1-form gauge transformation, $b_2 \to b_2 + d \lambda_1 $ and $c_1 \to c_1 - p  \lambda_1$.
However, from the variation of the action we obtain
\bea 
\delta S = \frac{i M}{2\pi} \int_{x\geq0} d \lambda_1 \wedge d c_1 - 2\pi i \left( \frac{p  M}{2 } \right) \int_{x\geq0} \frac{d \lambda_1}{2\pi} \wedge \frac{d \lambda_1}{2\pi}\, .
\eea 
Using $\oint_{\Sigma_2} \frac{d\lambda_1}{2\pi} \in \mathbb{Z}$ and $\oint_{\Sigma_2} \frac{d c_1}{2\pi} \in \mathbb{Z}$, one finds that this variation is consistent with gauge invariance provided that $p$ satisfies $p M \in 2\mathbb{Z}$.
On the other hand, the periodicity conditions for $p$ can be checked by examining the SPT term with $\oint_{\Sigma_{2}} \frac{b_2}{2\pi} \in \frac{1}{M} \mathbb{Z}$.
Then, it is straightforward to recover the periodicity condition in Eq.~\eqref{eq:quant_period_p}, since the action changes by $2\pi i \mathbb{Z}$ under $p \to p + 2 M$.

We proceed by first integrating out $c_1$.
The result is simply that $b_2$ is restricted to be a $\mathbb{Z}_M$ gauge field.\footnote{This can be seen using a fact from the discrete Fourier transform:
\beq 
\sum_{\left[ dc_1 \right] \in H^2 (\Sigma_4, \mathbb{Z})} \int \left[D b_2 D c_1\right] \exp \left[ 2\pi i \int \left(N \frac{b_2}{2\pi}\right) \wedge \frac{dc_1}{2\pi} \right] = \int \left. \left[D b_2\right] \right\vert_{ \oint_{\Sigma_2} \frac{b_2}{2\pi} \in \frac{1}{N} \mathbb{Z}}\, .
\eeq 
}
Next, it is useful to add the $\theta$-term to see the effect explicitly. Then we can ``complete the square'' to obtain
\beq 
S \supset \frac{i \theta}{8\pi^2} \int_{x<0} F_2 \wedge F_2 + \frac{i}{8\pi^2} \left(  \theta - \frac{2\pi p}{M} \right) \int_{x\geq0} F_2 \wedge F_2 + \frac{i p M}{4\pi} \int_{x\geq0} \left( b_2 + \frac{F_2}{M} \right)^2 \,.
\label{eq:half space gauging_squaring}
\eeq 
Finally, the equation of motion for $b_2$ sets $b_2 = -F_2/M$.
We learn that gauging $\mathbb{Z}_M^{(1)} \subset U(1)_m^{(1)}$ on half of the spacetime, together with the SPT term with parameter $p$, has the effect of shifting the $\theta$ parameter as $\theta \to \theta - 2\pi p/M$.

So far, half-space gauging does not leave the theory invariant.
However, in a theory with an ABJ anomaly, it is possible to restore invariance by compensating for the shift of $\theta$.
On $x \ge 0$, we make an anomalous field redefinition $\psi \to \exp\left[ \frac{i\alpha(x)}{2} \gamma_5 \right] \psi$ with $\alpha (x) = \frac{2\pi p}{M} \Theta (x)$ ($\Theta (x)$ is the Heaviside step function).
The result is
\beq 
S \to S + \frac{2\pi i p}{2M} \oint_{x=0} *J_A + \frac{2\pi i p}{M} \int_{x\geq0} \frac{F_2 \wedge F_2}{8\pi^2}\, .
\eeq 
The combined effect is to leave QED invariant: QED on $x < 0$ and the same QED on $x \ge 0$.
It remains to show that a non-invertible SDO is induced on $x=0$, which we interpret as an SDO mapping the theory to itself.
As noted when discussing the quantization condition in Eq.~\eqref{eq:quant_period_p}, adding the SPT term in the 4d $\mathbb{Z}_M$ BF theory makes $c_1$ charged under the 1-form gauge transformation: $b_2 \to b_2 + d \lambda_1$ and $c_1 \to c_1 - p \lambda_1$. 
This in turn means that, in general, line operators $\exp \left( i \oint_{\Sigma_1} c_1 \right)$ in the bulk are not gauge-invariant.
The gauge-invariant operator in the $x \ge 0$ region of Eq.~\eqref{eq:half space gauging_4d TQFT} is instead given by
\beq
\tilde{W} = \exp \left( i \oint_{\Sigma_1} c_1 \right) \exp \left( i p \int_{\Sigma_2} b_2 \right) \, , \quad \partial \Sigma_2 = \Sigma_1 \,.
\eeq
On the boundary $x = 0$, however, due to the topological boundary condition $b_2 |_{x=0} = 0$, the above operator reduces to the gauge-invariant line operator $W(\Sigma_1) = \exp\left(i \oint_{\Sigma_1} c_1\right)$.
These lines on the 3d boundary satisfy the correlation function (see Appendix~\ref{app:boundary lines and braiding} for a derivation):
\beq
\left\langle W(\Sigma_1) W(\Sigma_1') \right\rangle
= \exp\left[\frac{2\pi i p}{M} \operatorname{Link}(\Sigma_1,\Sigma_1') \right] \, .
\eeq
The coprime condition $\gcd(M, p) = 1$ shows that $M$ is the smallest integer for which $W$ satisfies $W^M = 1$.
These $M$ topological lines $\{ W^s \}_{s = 0, 1, \cdots, M - 1}$ generate a $\mathbb{Z}_M^{(1)}$ symmetry on the 3d boundary.
The coprime condition also ensures that the only line that braids trivially with the rest of the lines is the trivial line 1; that is, the braiding is non-degenerate.
It is known that such $M$ topological lines themselves form a consistent 3d TQFT, known as the 3d minimal $\mathbb{Z}_M$ TQFT $\mathcal{A}^{M,p}$, which we introduced earlier.
We see that the half-space gauging construction clearly shows the reason for the appearance of $\mathcal{A}^{M,p}$ in a natural and rigorous way.
Therefore, the half-space gauging combined with the anomalous change of the fermion variables leads to the generation of a non-invertible SDO localized on the 3d surface $\Sigma_3$ at $x = 0$:
\beq
\mathcal{D} \left( \frac{2\pi p}{M} , \Sigma_3 \right)=\exp\left(\frac{\pi ip}{M} \oint_{x=0} \ast J_A\right) \mathcal{A}^{M,p}\left[ \frac{F_2}{M} \right] \,.
\eeq
Such non-invertible SDOs are manifestly topological since they originate from a 4d TQFT together with a topological boundary condition.
In addition, the half-space gauging construction clarifies that the existence of non-invertible SDOs relies on the presence of an unbroken (and hence gaugeable) 1-form magnetic symmetry in the original theory.

\subsubsection{Action on 't Hooft line operators}

Half-space gauging makes it clear that a non-invertible SDO can be understood as a composition of an axial rotation and the gauging of a $\mathbb{Z}_M^{(1)} \subset U(1)_m^{(1)}$ symmetry.
Given that local operators are charged under the axial symmetry and 't Hooft line operators are charged under the 1-form magnetic symmetry, it is natural to expect that a non-invertible SDO may act on both types.
It acts on local operators invertibly, as is clear from the fact that non-invertibility originates from the 3d TQFT, which in turn arises from gauging a $\mathbb{Z}_M^{(1)}$ magnetic symmetry.
After all, local operators, e.g.~fermions or axions, are not affected by gauging a $\mathbb{Z}_M^{(1)}$ magnetic symmetry.
The action on 't Hooft lines is more interesting and subtle.
If we denote by $T(\gamma, m)$ the charge-$m$ 't~Hooft line supported on a closed curve $\gamma$, then under the $U(1)_m^{(1)}$ symmetry it transforms as
\beq 
T (\gamma, m) \to T (\gamma, m) \exp \left( i m \oint_\gamma \lambda_1 \right) \,,
\eeq 
while the 2-form BGF $\bkB$ transforms as $\bkB \to \bkB + d \lambda_1$.
Once the $\mathbb{Z}_M^{(1)}$ symmetry is gauged on $x \ge 0$, $T(\gamma, m)$ is no longer gauge-invariant.
Instead, when $\gamma$ is a contractible loop, a gauge-invariant operator $\tilde{T}$ can be obtained by attaching $T(\gamma, m)$ to a surface $M_2$:
\beq
\tilde{T} (\gamma,m) =
T (\gamma, m) \exp \left( - i mp \int_{M_2} b_2 \right)\, , \quad \partial M_2 = \gamma\, ,
\eeq 
where $b_2$ is a dynamical $\mathbb{Z}_M$ 2-form gauge field. As discussed below Eq.~\eqref{eq:half space gauging_squaring}, the equation of motion for $b_2$ imposes $b_2 = - F_2 / M$.
Substituting $b_2 = - F_2 / M$ into the above expression, we obtain
\beq
T (\gamma, m) \quad \text{at} \quad x<0 \quad \to \quad  T (\gamma, m) \exp \left(\frac{2\pi imp}{M} \int_{M_2} \frac{F_2}{2\pi} \right) \quad \text{at} \quad x\geq0\, .
\eeq 
In other words, under the action of the non-invertible SDO, $T(\gamma, m)$ is mapped to $T(\gamma, m)$ attached to the topological surface operator $\exp\left(\frac{2\pi im p}{M} \int_{M_2} \frac{F_2}{2\pi} \right)$.

This topological surface operator may be thought of as an improperly quantized Wilson surface with a fractional charge $m p/M$.
This allows us to reinterpret it in terms of the Witten effect as follows.
In Section~\ref{subsubsec:half space gauging}, we showed that gauging the $\mathbb{Z}_M^{(1)}$ symmetry on $x \ge 0$ has the effect of shifting $\theta \to \theta - 2\pi p/M$ there.
The Witten effect states that a charge-$m$ magnetic monopole in the background of a nontrivial $\theta$-term picks up an (possibly improperly quantized) electric charge $\frac{\theta}{2\pi} m$.
In our case, while a charge-$m$ monopole remains a monopole on $x < 0$ (i.e., before the action of the non-invertible SDO), it becomes a dyon with electric charge $m p / M$ on $x \ge 0$ (i.e., after sweeping through the SDO).

\subsubsection{Non-invertible symmetry from fractional instantons} 
\label{subsubsec:NI0_frac_Inst}

To discuss non-invertible symmetry in a theory with fractional instantons, we consider a 4d $PSU(N) = SU(N)/\mathbb{Z}_{N}$ gauge theory with an anomalous $U(1)_A$ symmetry. See Appendix~\ref{app:fractional_instantons} for a review of fractional instantons.

More concretely, let us take an axion--$SU(N)$ theory with action
\begin{equation}\label{eq:axion-PSUN}
S = \frac{f_a^2}{2} \int d\theta \wedge * d\theta 
+ \frac{1}{g^{2}} \int \operatorname{Tr}\left(F_2 \wedge * F_2\right) 
+ \frac{iK}{8 \pi^{2}} \int \theta \operatorname{Tr}\left(F_2 \wedge F_2\right) \, ,
\end{equation}
where $\theta \equiv a/f_a \sim \theta + 2\pi$ is the axion, and $F_2 = dA_1 + A_1 \wedge A_1$ is the $SU(N)$ field strength. To obtain the axion--$PSU(N)$ theory from the axion--$SU(N)$ theory, we gauge the $\mathbb{Z}_N^{(1)}$ center by first introducing a 2-form BGF $\mathcal{B}_2$ associated with the $\mathbb{Z}_N^{(1)}$ electric symmetry, and then promoting $\mathcal{B}_2$ to a dynamical $\mathbb{Z}_N$ 2-form gauge field $B_2$. As described in Appendix~\ref{app:fractional_instantons}, there exist $1/N$-valued fractional instantons in the $PSU(N)$ theory, and $K$ must be quantized as $K \in N\mathbb{Z}$ to ensure invariance under $\theta \to \theta + 2\pi$.\footnote{This condition can also be understood by obtaining $PSU(N)$ from $SU(N)$ via gauging the $\mathbb{Z}_N^{(1)}$ center. In the axion--$SU(N)$ theory, a nontrivial higher-group symmetry arises between the $\mathbb{Z}_N^{(1)}$ (child) electric symmetry and the $U(1)^{(2)}$ (parent) winding symmetry when $K \notin N\mathbb{Z}$. In that case, the child symmetry cannot be gauged without simultaneously gauging the parent symmetry. However, the higher-group symmetry becomes trivial (i.e., splits) when $K \in N\mathbb{Z}$; in this case one can gauge the $\mathbb{Z}_N^{(1)}$ center to obtain $PSU(N)$ without affecting the $U(1)^{(2)}$ symmetry.
}
Under a shift transformation $\theta \to \theta + 2\pi/z$ with $z \in \mathbb{Z}$, the action shifts as
\bea
S \to S + \frac{2\pi i K}{z} \left(n + \frac{N-1}{N} \int \frac{w_2 \wedge w_2}{2}\right) \, ,
\label{eq:NIS_rac_Inst_S_shift}
\eea
where $w_2 \equiv N \frac{B_2}{2\pi}$. Here, $n \in \mathbb{Z}$ denotes the integer-valued $SU(N)$ instanton number, and $\int \frac{w_2 \wedge w_2}{2} \in \mathbb{Z}$, as explained in Appendix~\ref{app:fractional_instantons}.
One then sees that the $SU(N)$ instantons break the shift symmetry $U(1)^{(0)} \to \mathbb{Z}_K^{(0)}$ (without leaving any non-invertible symmetry), while the effects of fractional instantons require some care.
Naively, fractional instantons would appear to further break the shift symmetry $\mathbb{Z}_K^{(0)} \to \mathbb{Z}_{K/N}^{(0)}$ (note that $K \in N\mathbb{Z}$).
There is, however, a key difference between the $SU(N)$ instantons and the fractional instantons.
While the former are defined in terms of the non-abelian gauge field $A_1$, the latter are defined in terms of the abelian field $w_2$.
Because of this, the effect of fractional instantons is not to break the shift symmetry $\mathbb{Z}_K^{(0)} \to \mathbb{Z}_{K/N}^{(0)}$; instead, they convert the elements of $\mathbb{Z}_K^{(0)} \setminus \mathbb{Z}_{K/N}^{(0)}$ (those in $\mathbb{Z}_K^{(0)}$ but not in $\mathbb{Z}_{K/N}^{(0)}$) into non-invertible symmetries. Note that, for $U(1)$ instantons, a countable yet infinite set of rational shifts survives as non-invertible symmetries, while in the fractional instanton case only finitely many non-invertible symmetries persist.
To see how the non-invertible symmetries arise in this case, note that the equation of motion for $\theta$ is given by
\begin{equation}
d * J_1 = \frac{K}{8 \pi^{2}} \operatorname{Tr}\left(F_2 \wedge F_2\right) + \frac{K(N-1)}{2N} w_2 \wedge w_2 \, ,
\end{equation}
where $J_1 = - i f_a^{2} d\theta$ is the (non-conserved) Noether current associated with the $U(1)^{(0)}$ shift symmetry.
Let us first consider the SDO associated with the \emph{invertible} $\mathbb{Z}_{K/N}^{(0)}$ symmetry, given by
\begin{align}
\hat{U}\left(\frac{2\pi \ell}{K/N}, \Sigma_3\right) &= \exp \left\{ i \oint_{\Sigma_3}  \left[ \frac{2\pi \ell}{K/N} *\hat{J}_1 - \frac{(N-1)\ell}{4\pi} B_1 \wedge dB_1\right]\right\}\, ,
\end{align}
where $*\hat{J}_1 \equiv *J_1 - \frac{K}{8\pi^2} {\rm Tr} \left(A_1 \wedge dA_1 + \frac{2}{3} A_1 \wedge A_1 \wedge A_1\right)$ and $\ell \in \{0, 1, \dots, K/N-1\}$.
Here we have used the relation $B_2 = dB_1/N$, as explained in Appendix~\ref{app:fractional_instantons}.
In this case, the 3d CS terms, both the non-abelian $A_1$ term in $*\hat{J}_1$ and the abelian $B_1$ term, come with properly quantized levels.
Importantly, this means that the CS terms correspond to \emph{invertible} SPT phases (in contrast to the TQFT partition function appearing in the non-invertible case below).
This explains why the $\mathbb{Z}_{K/N}^{(0)}$ symmetry is invertible even though its charge operator is not constructed solely from the Noether current.
Next, we discuss the non-invertible symmetry by considering the SDO associated with the shift $\theta \to \theta + 2\pi \ell/K$ for $\ell \in \{0,1,\dots,K-1\}$. In this case, while the non-abelian $A_1$ term is properly quantized, the abelian $B_1$ term can be improperly quantized. Explicitly, consider
\begin{align}
\hat{U}\left(\frac{2\pi\ell}{K}, \Sigma_3\right)
&= \exp\left\{ i \oint_{\Sigma_3} \left[ \frac{2\pi \ell}{K} *\hat{J}_1
- \frac{(N-1)\ell}{4\pi N} B_1 \wedge dB_1 \right] \right\} \,.
\end{align}
Indeed, for $\ell \in \{0,1,\dots,K-1\} \setminus N\mathbb{Z}$, the $B_1$ term is improperly quantized, signaling that the corresponding $\mathbb{Z}_K^{(0)} \setminus \mathbb{Z}_{K/N}^{(0)}$ (as a set, not a group) symmetry is non-invertible.
At this point, we can follow exactly the same procedure as in the $U(1)$ instanton case (see \cite{Cordova:2022ieu} for an early discussion).
The only nontrivial point is that we now couple a 3d TQFT to the 4d bulk via $\tqftA^{M,p}\left[\bkB = \frac{2\pi w_2}{M}\right]$ and impose $\frac{(N-1)\ell}{N} = \frac{p}{M}$ with $\gcd(M,p) = 1$, or, equivalently,
\beq
M = \frac{N}{\gcd(N,\ell)}\,, \quad p = \frac{(N-1)\ell}{\gcd(N,\ell)}\,.
\label{eq:fractional_instanton_M_and_p}
\eeq
Note that the $\gcd(N,\ell)$ factor is introduced to ensure $\gcd(M,p)=1$.
We gauge the $\mathbb{Z}_M^{(1)}$ symmetry of the 3d TQFT by the 4d $\mathbb{Z}_M$ field $w_2$.
The shift of the action induced by fractional instantons with $\alpha = 2\pi \ell / K$ can then be canceled by the 't~Hooft anomaly of the 3d TQFT.
In Section~\ref{subsection:Fractional instantons and partial gauging}, we will also revisit the construction of the SDO in terms of partial gauging.
One of the main questions we address in this paper is how to think about and construct non-invertible symmetries appearing in 4d QFTs with \emph{multiple instantons}.
See \cite{Cordova:2022fhg, Choi:2023pdp, Cordova:2023her, Cordova:2024ypu, Delgado:2024pcv, gcshsk:2025xx} for recent discussions in the particle physics context.
In fact, many realistic particle physics models (including the Standard Model itself, possibly with a nontrivial global structure \cite{Tong:2017oea}) and a variety of 4d QFTs of formal interest exhibit multiple instantons, including any of $U(1)$, non-abelian, and fractional instantons.
If a $U(1)$ global symmetry is broken by multiple instanton effects, how can we determine whether the remaining symmetries are invertible or non-invertible?
Given a general form of ABJ anomalies
\beq
S \to S + \frac{2\pi i \alpha K_{ij}}{8\pi^2} \int \operatorname{Tr} \left( F_2^i \wedge F_2^j \right) \, ,
\eeq 
which types of ABJ anomalies can (or cannot) be compensated by a suitable 3d TQFT?
For amenable ABJ anomalies, what is the structure---i.e., the symmetries and 't Hooft anomalies---of such a 3d TQFT? How does such a non-invertible symmetry act on local and extended operators of multiple types?
We will answer these questions in detail in the rest of the paper.

\subsection{Non-invertible 1-form symmetries}
\label{subsec:1-form_NIS}

In this section, we review the basic notions of non-invertible 1-form symmetry, or the non-invertible Gauss law, following \cite{Choi:2022fgx}.
Let us again consider our QED example with a scalar that spontaneously breaks the anomalous $U(1)_A$ symmetry.
As discussed above, this theory flows to an axion theory in the IR.
For simplicity, let us focus on the case with a single $U(1)$ gauge group.
The action is given by
\beq 
S = \frac{f_a^2}{2} \int d\theta \wedge * d\theta 
+ \frac{1}{2g^{2}} \int F_2 \wedge * F_2 
+ \frac{iK}{8 \pi^{2}} \int \theta F_2 \wedge F_2 \, ,
\eeq 
where $\theta \equiv a/f_a \sim \theta + 2\pi$ is the axion, $F_2 = dA_1$ is the $U(1)$ field strength, and $K \in \mathbb{Z}$.
The equation of motion for $A_1$ is given by
\beq 
d * J_2 = \frac{K}{4\pi^2} d \theta \wedge F_2\, ,
\label{eq:1-form anomaly}
\eeq
where $J_2 = \frac{i}{g^2} F_2$ is the (non-conserved) Noether current associated with the 1-form electric symmetry $U(1)^{(1)}_e$.
This equation may be viewed as an anomalous conservation law: $U(1)^{(1)}_e$ appears to be broken down to $\mathbb{Z}_K^{(1)}$.
This means that a naive ``Noether SDO''
\beq 
U (\alpha, \Sigma_2) = \exp \left( i \alpha \oint_{\Sigma_2} * J_2 \right)
\eeq 
fails to be topological for $\alpha \notin \frac{2\pi}{K} \mathbb{Z}$. Following~\cite{Choi:2022fgx}, the non-invertible SDO for $K=1$ is constructed as
\beq 
\sdoD^{(1)} \left( \frac{2\pi p}{N}, \Sigma_2 \right) = \int \left[D \phi D c_1\right] \exp \left[ i \oint_{\Sigma_2} \left(\frac{2\pi p}{N} * J_2 + \frac{N}{2\pi} \phi d c_1 + \frac{p}{2\pi}c_1 \wedge d \theta + \frac{1}{2\pi} \phi F_2 \right)\right]\,,
\label{eq:NI_1_SDO}
\eeq 
where $\phi \sim \phi + 2\pi$ is a scalar defined on $\Sigma_2$, $c_1$ is a dynamical $U(1)$ 1-form gauge field defined on $\Sigma_2$, and $\gcd (N,p) = 1$. The first term in the exponent is just the naive ``Noether SDO'' with $\alpha = 2\pi p/N$.
The rest can be understood as a 2d $\mathbb{Z}_N$ BF theory living on the 2-manifold $\Sigma_2$, coupled to the 4d bulk in a specific manner.
The 2d $\mathbb{Z}_N$ BF theory has a $\mathbb{Z}_N^{(0)} \times \mathbb{Z}_N^{(1)}$ symmetry acting on $\phi$ and $c_1$, respectively.
We can couple this theory to 1-form and 2-form BGFs $\mathcal{A}_1$ and $\mathcal{B}_2$ to obtain
\beq 
S = \frac{i N}{2\pi} \int_{\Sigma_2} \phi d c_1 + \frac{i N}{2\pi} \int_{\Sigma_2} c_1 \wedge \mathcal{A}_1 + \frac{i N}{2\pi} \int_{\Sigma_2} \phi \bkB\, .
\eeq 
There exists a mixed anomaly with inflow action
\beq 
S_{\rm inflow} = - \frac{iN}{2\pi} \int_{\Sigma_3} \mathcal{A}_1 \wedge \bkB\, .
\eeq 
The non-invertible SDO corresponds to taking $\mathcal{A}_1 = pd\theta/N$ and $\bkB = F_2/N$, which leads to the anomaly
\beq 
S_{\rm inflow} = - \frac{ip}{2\pi N} \int_{\Sigma_3} d \theta \wedge F_2\, .
\label{eq:inflow for 2d TQFT}
\eeq 
One recognizes that this is exactly what is needed to compensate the anomaly in Eq.~\eqref{eq:1-form anomaly} (take $K = 1$ for now).
More explicitly, under a smooth deformation $\Sigma_2 \to \Sigma_2'$, the ``Noether SDO'' part gives rise to an ABJ anomaly $\frac{i p}{2\pi N} \int_{\Sigma_3} d\theta \wedge F_2$.  Simultaneously, the 2d TQFT generates a 't~Hooft anomaly $- \frac{i p}{2\pi N} \int_{\Sigma_3} d\theta \wedge F_2$, exactly canceling the former.

The non-invertible SDO $\sdoD^{(1)} \left( \frac{2\pi p}{N}, \Sigma_2 \right)$ is gauge-invariant, yet it is non-invertible precisely because of the 2d TQFT stacked to make the operator topological.
It can be checked by computing 
\begin{equation}
    \begin{aligned}
        & \sdoD^{(1)} \left( \frac{2\pi p}{N}, \Sigma_2 \right)  \times \sdoD^{(1) \dagger} \left( \frac{2\pi p}{N}, \Sigma_2 \right) \\
        & = \int \left[D \phi D \bar{\phi} D c_1 D \bar{c}_1\right] \exp \left\{ i \oint_{\Sigma_2} \left[\frac{N}{2\pi} \phi d c_1 - \frac{N}{2\pi} \bar{\phi} d \bar{c}_1 + \frac{p}{2\pi} \left(c_1-\bar{c}_1\right) \wedge d \theta + \frac{1}{2\pi} \left(\phi - \bar{\phi}\right) F_2 \right]\right\} \\
        &\neq \mathbf{1} \,, 
    \end{aligned}
\end{equation}
where we have used that the ``Noether SDO'' part is unitary.
This is an example of a condensation defect.
This may be seen most clearly by making the change of variables $\bar{\phi} \to \phi' = \phi - \bar{\phi}$ and $c_1 \to c_1' = c_1 - \bar{c}_1$. In terms of the new variables, the right-hand side becomes
\begin{equation} \label{eq:NI-1_fusion}
\begin{aligned}
&\int \left[D \phi D c'_1\right] \exp \left[ i \oint_{\Sigma_2} \left(\frac{N}{2\pi} \phi d c'_1 + \frac{p}{2\pi} c'_1 d \theta \right) \right] \\
&\times \int \left[D\phi' D\bar{c}_1\right] \exp \left[ i \oint_{\Sigma_2} \left(\frac{N}{2\pi} \phi' d \bar{c}_1 + \frac{1}{2\pi} \phi' F_2 \right) \right] \,.
\end{aligned}
\end{equation}
The first factor describes the condensation defect from 2-gauging the 2-form winding symmetry $\mathbb{Z}_N^{(2)} \subset U(1)^{(2)}$: the bulk closed current $d\theta/2\pi$ is pulled back to $\Sigma_2$ and gauged by $c'_1$, which is made a $\mathbb{Z}_N$ gauge field by the 2d $\mathbb{Z}_N$ BF term.
The second factor corresponds to a condensation defect obtained by 2-gauging the 1-form magnetic symmetry $\mathbb{Z}_N^{(1)} \subset U(1)_m^{(1)}$: the bulk closed current $F_2/2\pi$ is pulled back to $\Sigma_2$ and gauged by $\phi'$.
Overall, we see that the Gauss law becomes non-invertible in the axion phase.
In the case of non-invertible 0-form symmetry, we observed symmetry matching between the QED-like (UV) and axion-like (IR) phases.
One may wonder whether the same holds for the non-invertible 1-form symmetry.
The answer is no.
In the QED-like phase (e.g., a KSVZ-like UV completion of the axion theory), in the presence of charge-1 fermions, the 1-form electric symmetry is completely broken.
This breaking is not due to a topological term analogous to Eq.~\eqref{eq:1-form anomaly}; the 1-form electric symmetry is genuinely lost.
Once these fermions become massive---for instance, via Yukawa interactions with a scalar whose vacuum expectation value breaks $U(1)_A$---one might expect the 1-form electric symmetry to re-emerge below the mass scale.
If this were the whole story, why would the 1-form electric symmetry appear as a non-invertible symmetry rather than as an invertible symmetry?
The key point is that the axion phase does include an ``electric current'': it is a quantum number carried by a soliton, often referred to as the ``Goldstone--Wilczek current''~\cite{Goldstone:1981kk}.
Strikingly, in the axion theory, the electric charge carried by the axion is fixed by a topological datum, namely the ABJ anomaly.
Viewing the CS interaction between the axion and the gauge field as $A_1 \wedge *J_1$, one derives the electric current (up to sign conventions) as
\beq 
*J_1 = \frac{1}{i} \frac{\delta S}{\delta A_1} =  \frac{K}{4\pi^2} d \theta \wedge F_2\, .
\eeq 
This is precisely the anomaly appearing in Eq.~\eqref{eq:1-form anomaly}. In this way, we obtain an alternative interpretation: in the axion phase, the 1-form electric symmetry is broken by the charge-$K$ Goldstone--Wilczek current.
However, since this solitonic current is purely topological, the net effect is to transmute the symmetry into a non-invertible one.

\subsubsection{Action on line and axion string operators}

In this section, we summarize how the SDO of the non-invertible 1-form symmetry $\mathcal{D}^{(1)}\left( \frac{2\pi p}{N}, \Sigma_{2} \right)$
constructed above acts on Wilson and 't~Hooft line operators as well as on axion string worldsheets, noting that we have gauged the 1-form magnetic symmetry and the 2-form winding symmetry (see \cite{Choi:2022fgx} for more details).

\begin{figure}
  \centering
  \begin{subfigure}
  {0.45\textwidth}
  \includegraphics[width=\linewidth]{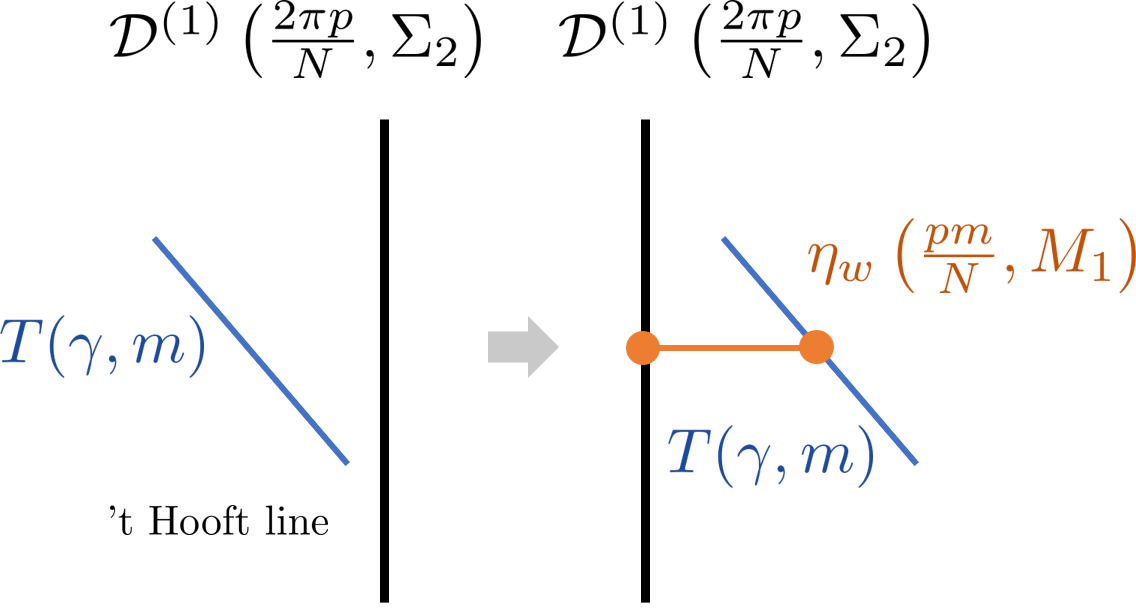}
    \caption{'t Hooft line} \label{fig:SDO_action_thooft_line}
  \end{subfigure}
  \hspace*{0.5cm}
  \begin{subfigure}{0.4\textwidth}
\includegraphics[width=\linewidth]{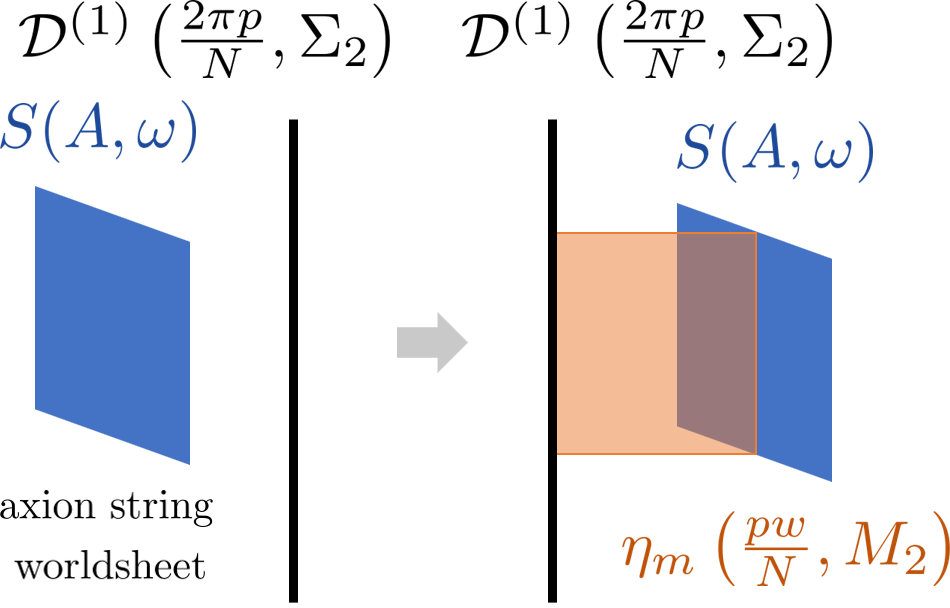}
    \caption{axion string worldsheet} \label{fig:SDO_action_axion_string_worldsheet}
  \end{subfigure}%
\caption{The action of the non-invertible SDO on (a) the 't~Hooft line and (b) the axion string worldsheet. 
One spatial dimension is suppressed so that the 2-manifold $\Sigma_{2}$ supporting the non-invertible SDO is represented as a 1-dimensional object in the figure.
} \label{fig:SDO_action}
\end{figure}

\begin{itemize}
    \item[(i)] For the Wilson line $W(\gamma,q)$ of electric charge $q$ supported on a curve $\gamma$, the non-invertible SDO acts invertibly:
    \begin{align}
     \mathcal{D}^{(1)}\left( \frac{2\pi p}{N}, \Sigma_{2} \right) \quad:\quad W(\gamma,q) \to W(\gamma,q) \exp \left( \frac{2 \pi i qp}{N} \right) \,.
    \end{align}
    This invertible action may be understood as follows. As suggested by the discussion around Eq.~\eqref{eq:NI-1_fusion} (see \cite{Choi:2022fgx} for details), the SDO can be constructed via half 1-gauging of $\mathbb{Z}_N^{(1)} \times \mathbb{Z}_N^{(2)}$, where $\mathbb{Z}_N^{(1)}$ is a subgroup of the 1-form magnetic symmetry $U(1)_m^{(1)}$ and $\mathbb{Z}_N^{(2)}$ is a subgroup of the 2-form winding symmetry $U(1)^{(2)}$. None of these gauging operations affect the 1-form electric symmetry; hence the resulting symmetry acts invertibly on Wilson lines.

    \item[(ii)] For the 't~Hooft line $T(\gamma, m)$ of magnetic charge $m$ supported on a curve $\gamma$, the non-invertible SDO acts non-invertibly:
    \begin{align}
     \mathcal{D}^{(1)}\left( \frac{2\pi p}{N}, \Sigma_{2} \right) \quad:\quad T( \gamma, m) \to T( \gamma,m) \exp \left( \frac{2\pi i mp}{N}  \int_{M_{1}} \frac{d\theta}{2\pi} \right) \,. 
    \end{align}
    Here, $\eta_w\left(\alpha, M_1\right) = \exp\left( i \alpha \int_{M_{1}} d\theta \right)$ denotes the SDO associated with the 2-form winding symmetry, with the winding number as its charge, and $M_{1}$ is a path connecting a point on the 't~Hooft line to the intersection point of the 't~Hooft line and the non-invertible SDO. See Fig.~\ref{fig:SDO_action_thooft_line} for an illustration.

    \item[(iii)] For the axion string worldsheet $S(A,w)$ of winding number $w$ supported on a surface $A$, the non-invertible SDO acts non-invertibly:
    \begin{align}
    \mathcal{D}^{(1)}\left( \frac{2\pi p}{N}, \Sigma_{2} \right) \quad:\quad S(A,w) \to  S (A,w) \exp \left( \frac{2\pi i wp}{ N }  \int_{M_{2}} \frac{F_2}{2\pi} \right) \,. 
    \end{align}
    Here, $\eta_m\left(\alpha, M_2\right) = \exp\left( i \alpha \int_{M_{2}} F_2 \right)$ denotes the SDO associated with the 1-form magnetic symmetry, with the magnetic flux as its charge, and $M_{2}$ is a surface connecting a line on the axion string worldsheet to the intersection line between the axion string worldsheet and the non-invertible SDO. See Fig.~\ref{fig:SDO_action_axion_string_worldsheet} for an illustration.

\end{itemize}

\section{Non-Invertible Symmetries from Partial Gauging}
\label{sec:NIS_from_partial_gauging}

In the previous section, we discussed that when a non-invertible symmetry in a 4d QFT arises from an ABJ anomaly, the 3d TQFT used in the literature to compensate the ABJ anomaly through its 't~Hooft anomaly is a 3d minimal $\mathbb{Z}_N$ TQFT $\tqftA^{N,p}$ with $\gcd(N, p) = 1$.
At first glance, this appears to be (and indeed is) a sensible choice because when $p = 1$, we have $\tqftA^{N,1} \leftrightarrow U(1)_N$, and the 3d TQFT has a 't~Hooft anomaly with coefficient $1$.

One may wonder how this should be generalized when the 4d bulk theory involves multiple ABJ anomalies. 
To model the situation in a concrete setup, let us consider an axion theory with the following action:
\beq
S = \frac{f_a^2}{2} \int d \theta \wedge * d \theta 
+ \sum_{i=1}^n \frac{1}{2g_i^2} \int F_2^i \wedge * F_2^i 
+ \sum_{i,j=1}^n \frac{i K_{ij}}{8\pi^2} \int \theta F_2^i \wedge F_2^j\,,
\label{eq:axion with general ABJ}
\eeq
where $\theta = a / f_a \sim \theta + 2\pi$ is the axion, $F_2^i = d A_1^i$ are the $U(1)$ field strengths, and $K_{ij}$ is a symmetric ABJ anomaly matrix with integer entries.
The entries of $K_{ij}$ must be integers to be consistent with the gauge identification $\theta \sim \theta + 2\pi$, once the gauge fields are normalized so that $\oint_{\Sigma_2} \frac{F_2^i}{2\pi} \in \mathbb{Z}$.
To be general, we allow not only unmixed (diagonal) anomalies of the form $\propto K_{ii} F_2^i \wedge F_2^i$, but also mixed (off-diagonal) anomalies $\propto K_{i \ne j} F_2^i \wedge F_2^j$.
Our discussion will be fully general in the sense that it can be applied not only to $U(1)$ instantons but also to fractional and CFU instantons as well (see Section~\ref{sec:multi-instanton-case}).
In addition, while we took an axion theory to demonstrate our idea, the entire discussion can be repeated for any 4d QFT with non-invertible symmetries induced by ABJ anomalies.

Can we always find a 3d TQFT with a sufficiently rich 't~Hooft anomaly structure to compensate for arbitrary ABJ anomalies characterized by a symmetric matrix $K_{ij}$? 
Certainly, a naive stack of 3d minimal $\mathbb{Z}_{N_i}$ TQFTs of the form $\prod_{i=1}^n \tqftA^{N_i, p_i}$ will not work, since such a 3d TQFT contains only diagonal anomalies.
The requirement for the 3d TQFT can be stated as follows.  
If we consider a $U(1)$ global symmetry transformation with parameter $\alpha = 2\pi / z$, $z \in \mathbb{Z}$, then, given the input anomaly data specified by a symmetric matrix $K_{ij}$, we need a 3d TQFT with symmetry $\prod_{i=1}^n \mathbb{Z}_{N_i}^{(1)}$ and a 't~Hooft anomaly characterized by a matrix $p_{ij}$, such that these 't~Hooft anomalies cancel the ABJ anomalies of the general form
\beq
S \to S + \sum_{i,j=1}^n\frac{i K_{ij}}{2\pi z} \int F_2^i \wedge F_2^j\, .
\eeq
Constructing such a 3d TQFT does not appear to be a completely trivial task, though it is certainly not impossible.
One of the goals of this work is to provide a systematic answer to this problem.
In this section, considering simple examples of axion--Maxwell theory---again mainly as a way to model ABJ anomaly structures---we introduce the concept of \emph{partial gauging} as a method to obtain 't~Hooft anomalies of 3d TQFTs with \emph{general} anomaly coefficients from simple 3d TQFTs such as 3d CS theory or 3d BF theory.
A similar idea was used in the discussion of non-invertible symmetries in theories with a single instanton and/or a single axion~\cite{Choi:2022jqy,Choi:2022fgx,DelZotto:2024ngj}.
This method is then used to construct non-invertible SDOs, first focusing on simple cases with a single diagonal or a single mixed anomaly.
The case with a single diagonal anomaly corresponds to the scenario discussed in detail in Section~\ref{sec:review_NIS}, but we revisit the same problem here using the method of partial gauging to illustrate the key idea.
Along the way, we demonstrate the relation between our construction and the existing one based on the 3d minimal $\mathbb{Z}_N$ TQFT $\tqftA^{N,p}$ by comparing the symmetries and anomalies on both sides.
While these relations are well known to experts, we nevertheless find it useful to discuss them in some detail.

\subsection{\texorpdfstring{$U(1)$}{U(1)} anomaly and 3d Chern--Simons theory}
\label{subsec:CS_partial gauging}

Let us begin by revisiting an axion theory with a single anomaly term,
\beq
S = \frac{f_a^2}{2} \int d \theta \wedge * d \theta + \frac{1}{2g^2} \int F_2 \wedge * F_2 + \frac{i K}{8\pi^2} \int \theta F_2 \wedge F_2 \, ,
\eeq
where $\theta \equiv a/f_a \sim \theta + 2\pi$ is the axion, $F_2 = dA_1$ is the $U(1)$ field strength, and $K \in \mathbb{Z}$. The topological interaction with integer coupling $K$ breaks the $U(1)^{(0)}$ shift symmetry down to the \emph{invertible} subgroup $\mathbb{Z}_K^{(0)}$, since the action shifts under $\theta \to \theta + \alpha$ as
\beq 
S \to S + \frac{i K \alpha}{8\pi^2} \int  F_2 \wedge F_2\, .
\eeq 
The remainder of $\left(\mathbb{Q}/\mathbb{Z}\right)^{(0)}$ (as a set, not a group) is converted into a set of non-invertible symmetries.
In essence, this was previously achieved by dressing the Noether SDO with the 3d minimal $\mathbb{Z}_N$ TQFT $\tqftA^{N,p}$, as described in detail in Section~\ref{sec:review_NIS}.

We now show that a 3d abelian CS theory $U(1)_{pN}$ with $\gcd(N,p)=1$ can achieve the same goal.
As we will describe shortly, it is known that $U(1)_{pN}$ is related to the 3d minimal $\mathbb{Z}_N$ TQFT $\tqftA^{N,p}$ by \cite{Hsin:2018vcg}
\beq 
U(1)_{pN} \quad \leftrightarrow \quad \tqftA^{pN,1} = \tqftA^{N,p} \otimes \tqftA^{p,N} \quad \text{if} \quad \gcd(N,p) = 1\, .
\label{eq:UpN_ANp_ApN_map}
\eeq  
Let us first discuss the symmetry of the $U(1)_{pN}$ theory.
The action is 
\beq  
S = \frac{i p N}{4\pi} \int_{\Sigma_3} a_1 \wedge d a_1\, ,
\label{eq:U1_pN}
\eeq  
for some 3-manifold $\Sigma_{3}$.
This theory has a $\mathbb{Z}_{pN}^{(1)}$ symmetry, which factorizes as $\mathbb{Z}_{pN}^{(1)} = \mathbb{Z}_N^{(1)} \times \mathbb{Z}_p^{(1)}$ since $\gcd(N,p) =1$.
The inflow action for the 't~Hooft anomaly of $\mathbb{Z}_{pN}^{(1)}$ is given by (see Eq.~\eqref{eq:CS_inflow})
\beq 
S_{\rm inflow} = - \frac{i pN}{4\pi} \int_{\Sigma_4} \bkA \wedge \bkA\, ,
\eeq 
where $\bkA$ is the 2-form BGF for the $\mathbb{Z}_{pN}^{(1)}$ symmetry and $\partial \Sigma_{4} = \Sigma_{3}$.
The idea of \emph{partial gauging} is the following.
Instead of coupling the 3d TQFT to the 4d bulk theory via its full symmetry, here $\mathbb{Z}_{pN}^{(1)}$, we couple only to a subgroup.
This, in fact, amounts to \emph{partially} gauging the 1-form symmetry of the 3d TQFT by the 4d bulk field strength.
Let us demonstrate how this works. If we denote the 2-form BGF for $\mathbb{Z}_{pN}^{(1)}$ by $\bkA$ and the 2-form BGF for $\mathbb{Z}_N^{(1)} \subset \mathbb{Z}_{pN}^{(1)}$ by $\bkB$, we can couple Eq.~\eqref{eq:U1_pN} to $\bkB$ via
\beq  
S = \frac{i p N}{4\pi} \int_{\Sigma_3} a_1 \wedge d a_1 + \frac{ipN}{2\pi} \int_{\Sigma_3} a_1 \wedge \bkB \, , \quad \oint_{\Sigma_2} \frac{\bkB}{2\pi} \in \frac{1}{N} \mathbb{Z}\, ,
\label{eq:U1_pN_partial_gauging}
\eeq  
where we have explicitly written the allowed holonomy of $\bkB$ to emphasize that it is a $\mathbb{Z}_N$ gauge field rather than a $\mathbb{Z}_{pN}$ gauge field. We now identify $\bkB = F_2 / N$, which effectively couples the 4d bulk theory to the 3d defect worldvolume TQFT only through its subgroup $\mathbb{Z}_N^{(1)} \subset \mathbb{Z}_{pN}^{(1)}$. Plugging this into the inflow action, we find
\beq 
S_{\rm inflow} = - \frac{i pN}{4\pi} \int_{\Sigma_4} \bkB \wedge \bkB = - \frac{i p}{4\pi N} \int_{\Sigma_4} F_2 \wedge F_2 \, .
\label{eq:U1_pN_inflow_partial_gauging_1}
\eeq 
This can also be written in terms of a (generalized) Stiefel--Whitney class $w_2 \in H^2 \left( \Sigma_4, \mathbb{Z}_N \right)$. Defining $w_2 \equiv N \frac{\bkB}{2\pi}$, the inflow action becomes
\beq 
S_{\rm inflow} = - \frac{2 \pi i p}{N} \int_{\Sigma_4} \frac{w_2 \wedge w_2}{2} \, .
\label{eq:U1_pN_inflow_partial_gauging_2}
\eeq 
From Eq.~\eqref{eq:U1_pN_inflow_partial_gauging_1} or Eq.~\eqref{eq:U1_pN_inflow_partial_gauging_2}, one sees that this is exactly the 't~Hooft anomaly of $\tqftA^{N,p}$ (see Appendix~\ref{subapp:spin and anomaly_cs}).
It is worth emphasizing what has been achieved.
We started with a 4d bulk theory with a non-minimal ABJ anomaly coefficient $K > 1$.
For the 3d TQFT, we took the $U(1)_{pN}$ theory, whose 't~Hooft anomaly for the full symmetry $\mathbb{Z}_{pN}^{(1)}$ has coefficient $1$.
However, focusing on a subgroup yields the non-minimal 't~Hooft anomaly in Eq.~\eqref{eq:U1_pN_inflow_partial_gauging_2}.
Therefore, the non-invertible SDO can be constructed in terms of a $U(1)_{pN}$ theory, provided it is coupled to the 4d bulk via the partial gauging in Eq.~\eqref{eq:U1_pN_partial_gauging}:
\begin{equation}
    \begin{aligned}
        \sdoD \left( \frac{2\pi}{z}, \Sigma_3 \right) & \equiv U \left( \frac{2\pi}{z}, \Sigma_3 \right) \tqftA^{pN, 1} \left[\frac{F_2}{N} \right] \\  
& =  \int \left[Da_{1}\right] \exp \left[ i  \oint_{\Sigma_3} \left( \frac{2\pi}{z} * J_1 + \frac{p N}{4\pi} a_1 \wedge d a_1 + \frac{p}{2\pi} a_1 \wedge F_2 \right) \right]\, ,
    \end{aligned}
\end{equation}
where $J_1 = -i f_a^2 d\theta$ is the (non-conserved) Noether current associated with the $U(1)^{(0)}$ shift symmetry.
It is instructive to understand this result in terms of the factorization of $U(1)_{pN}$ into two 3d minimal abelian TQFTs in Eq.~\eqref{eq:UpN_ANp_ApN_map}.
On the $U(1)_{pN}$ side, let the generator of $\mathbb{Z}_{pN}^{(1)}$ be $W(\gamma) = \exp\left( i \oint_{\gamma} a_1 \right)$.
Since $W(\gamma)$ generates $\mathbb{Z}_{pN}^{(1)}$, it satisfies $W^{pN}(\gamma) = 1$.
The two subgroups $\mathbb{Z}_N^{(1)}$ and $\mathbb{Z}_p^{(1)}$ are then generated by
\begin{equation}
\begin{array}{ccccc}
\mathbb{Z}_N^{(1)} & \quad:\quad & \hat{n} = W^p & \quad\to\quad & \hat{n}^N = 1 \,, \\[0.5em]
\mathbb{Z}_p^{(1)} & \quad:\quad & \ \hat{p} = W^N & \quad\to\quad & \ \hat{p}^p = 1 \, .
\end{array}
\end{equation}
For $\gcd(N,p)=1$, the smallest integers $\alpha,\beta \in \mathbb{Z}$ satisfying $\hat{n}^\alpha = \hat{p}^\beta$ are $\alpha = N$ and $\beta = p$, showing again that the 1-form symmetry decomposes as $\mathbb{Z}_{pN}^{(1)} = \mathbb{Z}_N^{(1)} \times \mathbb{Z}_p^{(1)}$.\footnote{To see this, use that for any two integers $a$ and $b$, $ab = \gcd(a,b) \times \mathrm{lcm}(a,b)$. For coprime integers, $ab = \mathrm{lcm}(a,b)$.}
The global symmetry of $\tqftA^{N,p} \otimes \tqftA^{p,N}$ is clearly $\mathbb{Z}_N^{(1)} \times \mathbb{Z}_p^{(1)}$.
Furthermore, we have already shown that the 't~Hooft anomaly of $\mathbb{Z}_N^{(1)}$ matches on both the $U(1)_{pN}$ and $\tqftA^{N,p}$ sides (see Eq.~\eqref{eq:U1_pN_inflow_partial_gauging_2}).
It remains to prove that the 't~Hooft anomaly of $\mathbb{Z}_p^{(1)}$ also agrees on both sides of the map.
This can be done in exactly the same way as before: couple $U(1)_{pN}$ only to the 2-form BGF for $\mathbb{Z}_p^{(1)}$, call it $\bkC$, and write the inflow action in terms of the associated Stiefel--Whitney class $\tilde{w}_2 = p \frac{\bkC}{2\pi} \in H^2(\Sigma_4,\mathbb{Z}_p)$ as
\beq 
S_\text{inflow} = - \frac{i pN}{4\pi} \int_{\Sigma_4} \bkC \wedge \bkC =  - \frac{2 \pi i N}{p} \int_{\Sigma_4} \frac{\tilde{w}_2 \wedge \tilde{w}_2}{2}\, .
\eeq 
As expected, the 't~Hooft anomaly for the $\mathbb{Z}_p^{(1)}$ symmetry matches on both sides.
These facts strongly suggest---although they do not constitute a complete proof---that Eq.~\eqref{eq:UpN_ANp_ApN_map} holds.
A more rigorous proof can be established by checking the fusion rules, spins, and central charge; see Section~2.3 of \cite{Hsin:2018vcg}.
The fact $U(1)_{pN} \leftrightarrow \tqftA^{pN,1} = \tqftA^{N,p} \otimes \tqftA^{p,N}$ (if $\gcd(N,p)=1$) in the context of non-invertible symmetry can be interpreted as follows.
Given an anomalous $U(1)$ (with anomaly coefficient $K$), we need a 3d TQFT with a $\mathbb{Z}_N^{(1)}$ symmetry and a 't~Hooft anomaly labeled by an integer $p$ such that $K/z = p/N$, where $z \in \mathbb{Z}$ is related to the transformation parameter $\alpha = 2\pi/z$.
It is a general fact that any 3d TQFT $\mathcal{T}$ with this property can be decomposed as $\mathcal{T} = \tqftA^{N,p} \otimes \mathcal{T}'$ when $\gcd(N,p)=1$.
On the other hand, the second sector $\mathcal{T}'$ is invariant under the $\mathbb{Z}_N^{(1)}$ symmetry in the sense that it contains only neutral lines; it is a \emph{decoupled} sector. 
While non-invertible SDOs can be constructed minimally using $\tqftA^{N,p}$ (as has been extensively done in the literature), what we have demonstrated above is that, alternatively---and perhaps less esoterically---we can add an appropriate decoupled sector $\mathcal{T}' = \tqftA^{p,N}$ and use the $U(1)_{pN}$ theory in place of $\tqftA^{N,p}$.
In fact, the assignment of 2-form BGFs $\bkB = F_2/N$ and $\bkC = 0$ corresponds to
\beq
\tqftA^{pN,1} \left[ \bkB = \frac{F_2}{N}, \bkC = 0 \right] = \tqftA^{N,p} \left[ \bkB = \frac{F_2}{N} \right] \otimes \tqftA^{p,N} \left[ \bkC = 0 \right]\, .
\eeq 
This shows clearly that \emph{partial gauging}---i.e., $\bkB = F_2/N$ and $\bkC = 0$---projects out the $\tqftA^{p,N}$ factor; hence, as far as anomaly cancellation and the construction of the non-invertible SDO are concerned, $U(1)_{pN}$ and $\tqftA^{N,p}$ have identical effects.

At first glance, this construction may appear unnecessarily redundant. 
So, what is the benefit of doing this?
First, the $U(1)_{pN}$ theory admits an explicit Lagrangian description, whereas $\tqftA^{N,p}$ may not.
At the very least, symmetry analysis and the computation of 't~Hooft anomalies are often more straightforward when a Lagrangian formulation exists.
Although we do not elaborate on this point here, such a formulation may also simplify the computation of fusion rules.
Second, this relates to the question we posed at the beginning.
Namely, when dealing with a theory with multiple instantons, the 3d TQFT required for constructing the non-invertible SDO must have an appropriate symmetry and 't~Hooft anomaly structure.
Since a naive ``superposition'' of 3d minimal $\mathbb{Z}_{N_i}$ TQFTs of the form $\prod_{i=1}^n \tqftA^{N_i, p_i}$ does not work, searching for the desired 3d TQFT is much better approached through a Lagrangian theory (see Section~\ref{sec:multi-instanton-case}).

\subsection{Mixed anomaly and 3d BF theory}
\label{subsec:BF_partial gauging}

In this section, we discuss the non-invertible symmetry associated with a \emph{mixed} ABJ anomaly.
We again use an axion theory with action
\beq
S = \frac{f_a^2}{2} \int d \theta \wedge * d \theta + \frac{1}{2e^2} \int F_2 \wedge * F_2 + \frac{1}{2g^2} \int G_2 \wedge * G_2 + \frac{i K}{4\pi^2} \int \theta F_2 \wedge G_2\, ,
\eeq
where $\theta \equiv a/f_a \sim \theta + 2\pi$ is the axion, $F_2 = dA_1$ and $G_2 = dB_1$ are the $U(1)$ field strengths, and $K \in \mathbb{Z}$.
It is easy to see that the mixed anomaly implies the invertible shift symmetry is $\mathbb{Z}_K^{(0)}$.
Our discussion also applies if the second gauge-theory sector is replaced by a 4d $\mathbb{Z}_N$ TQFT.
Such a setup and its physical implications were studied in \cite{Brennan:2023kpw} as a possible 4d $\mathbb{Z}_N$ TQFT coupling realized in particle physics.

We next show that the theory has a $\left(\mathbb{Q}/\mathbb{Z}\right)^{(0)}$ (as a set, not a group) worth of non-invertible symmetries, while the $\mathbb{Z}_{K}^{(0)}$ subgroup enjoys a group-like invertible symmetry. The corresponding SDO is a stack of a ``Noether'' part and the partition function of a 3d $\mathbb{Z}_N$ BF theory, rather than a CS theory.
It will be useful to first review basic properties of the 3d $\mathbb{Z}_N$ BF theory, which we do next.

\subsubsection{3d BF theory}

Consider the 3d $\mathbb{Z}_N$ BF theory with action
\beq
S = \frac{i N }{2\pi} \int_{\Sigma_3} a_1 \wedge d b_1\, ,
\label{eq:BF action}
\eeq 
where $a_1$ and $b_1$ are dynamical $U(1)$ 1-form gauge fields. This theory has a $\mathbb{Z}_N^{(1)}(a) \times \mathbb{Z}_N^{(1)}(b)$ symmetry, which acts as
\begin{equation}
\begin{array}{ccccc}
\mathbb{Z}_N^{(1)} (a) & \quad:\quad & a_1 \to a_1 + \lambda_a\, , & \quad & \displaystyle \oint_{\gamma} \frac{\lambda_a}{2\pi} \in \frac{1}{N} \mathbb{Z} \,, \\[1em]
\mathbb{Z}_N^{(1)} (b) & \quad:\quad & b_1 \to b_1 + \lambda_b\, , & \quad & \displaystyle \oint_{\gamma} \frac{\lambda_b}{2\pi} \in \frac{1}{N} \mathbb{Z} \,.
\end{array}
\end{equation}
Here, $\lambda_a$ and $\lambda_b$ are the 1-form transformation parameters.
The charged objects are Wilson line operators, which are classified by $\mathbb{Z}_N$:
\begin{equation}
\begin{array}{ccc}
\displaystyle W_a (\gamma, m) = \exp\left( i m \oint_\gamma a_1 \right) \, , & \quad & m = 0, 1, \cdots, N-1  \,, \\[1em]
\displaystyle W_b (\gamma, n) = \exp \left( i n \oint_\gamma b_1 \right) \, , & \quad & \ n = 0, 1, \cdots, N-1 \, .
\end{array}
\end{equation}
The $\mathbb{Z}_N^{(1)}(a) \times \mathbb{Z}_N^{(1)}(b)$ symmetry has a mixed 't~Hooft anomaly.
To see this, we couple the theory to 2-form BGFs of the global symmetries.
Denoting the BGFs for $\mathbb{Z}_N^{(1)}(a)$ and $\mathbb{Z}_N^{(1)}(b)$ by $\bkA$ and $\bkB$, respectively, the action with BGF couplings is
\beq
S = \frac{i N }{2\pi} \int_{\Sigma_3} a_1 \wedge d b_1 + \frac{i N}{2\pi} \int_{\Sigma_3} b_1 \wedge \bkA + \frac{i N}{2\pi} \int_{\Sigma_3} a_1 \wedge \bkB\, .
\label{eq:3d BF with BGF}
\eeq 
It is useful to rewrite the above action on an auxiliary 4-manifold in the form
\beq
S = \frac{i N}{2\pi} \int_{\Sigma_4} \left( d a_1 + \bkA \right) \wedge \left( d b_1 + \bkB \right)  - \frac{i N}{2\pi} \int_{\Sigma_4} \bkA \wedge \bkB\, ,
\label{eq:3d BF with BGF in 4d}
\eeq 
where $\partial \Sigma_4 = \Sigma_3$.
Under background gauge transformations, the BGFs transform as
\beq 
\bkA \to \bkA - d \lambda_a \, , \quad \bkB \to \bkB - d \lambda_b \,.
\eeq 
From this, one sees that the first term in Eq.~\eqref{eq:3d BF with BGF in 4d} is manifestly invariant under the background gauge transformations, whereas the second term is not. 
Therefore, the second term captures the 't~Hooft anomaly, as it encodes the obstruction to gauging. 
We conclude that the anomaly inflow action for the mixed 't~Hooft anomaly between $\mathbb{Z}_N^{(1)}(a)$ and $\mathbb{Z}_N^{(1)}(b)$ is
\beq
S_\text{inflow} = - \frac{i N}{2\pi} \int_{\Sigma_4} \bkA \wedge \bkB\, .
\label{eq:3d Bf inflow}
\eeq 
It is worth noting that although Eq.~\eqref{eq:3d Bf inflow} is a local action built solely from the BGFs, it is not a local counterterm, since it is defined only on the auxiliary 4-manifold rather than on $\Sigma_3$, where the theory lives.

\subsubsection{Non-invertible symmetry for \texorpdfstring{$K=1$}{K=1}}

We now return to the discussion of non-invertible symmetries, focusing first on the $K = 1$ case.
The presence of a mixed anomaly implies that under a shift $\theta \to \theta + 2\pi / N$, $N \in \mathbb{Z}$, the action shifts as 
\beq  
S \to S + \frac{2\pi i}{N} \int \frac{F_2 \wedge G_2}{4\pi^2}\, .
\eeq 
Equivalently, the same shift follows if one uses an operator constructed solely from the (non-conserved) Noether current $J_1 = - i f_a^2 d\theta$:
\beq 
U \left( \frac{2\pi}{N}, \Sigma_3 \right) = \exp \left( \frac{2\pi i}{N} \oint_{\Sigma_3} *J_1 \right)\, .
\eeq  
Instead, the non-invertible symmetry is generated by 
\beq  
\nisD \left( \frac{2\pi}{N}, \Sigma_3 \right) = \int \left[D a_{1} D b_{1}\right] \exp \left[ i \oint_{\Sigma_3} \left(\frac{2\pi}{N} *J_1 + \frac{N }{2\pi} a_1 \wedge d b_1 + \frac{1}{2\pi} b_1 \wedge F_2 + \frac{1}{2\pi} a_1 \wedge G_2 \right)\right]\, ,
\label{eq:NISDO_mixed_K1}
\eeq 
where the additional terms correspond to the 3d $\mathbb{Z}_N$ BF theory with BGFs chosen to be the 4d bulk field strengths: $\bkA = F_2/N$ and $\bkB = G_2/N$.
Since $F_2$ and $G_2$ are $U(1)$ field strengths, $F_2/N$ and $G_2/N$ define $\mathbb{Z}_N$ 2-form fields suitable as BGFs for $\mathbb{Z}_N$ 1-form symmetries.
With these choices, the anomaly inflow for the mixed anomaly \eqref{eq:3d Bf inflow} becomes 
\beq
S_\text{inflow} = - \frac{2\pi i}{N} \int_{\Sigma_4} \frac{F_2 \wedge G_2}{4\pi^2}\, ,
\eeq 
which exactly cancels the ABJ anomaly, thereby leaving the action invariant.

\subsubsection{Non-invertible symmetry for general \texorpdfstring{$K$}{K} with partial gauging}

Let us next consider the general $K \in \mathbb{Z}$.
The key question is which 3d TQFT living on the SDO worldvolume, and for what form of coupling to the 4d bulk theory, we can obtain an inflow action with a general coefficient
\beq
S_{\rm inflow} = - \frac{2\pi i p}{M} \int_{\Sigma_4} \frac{F_2 \wedge G_2}{4\pi^2} \,, \quad M, p \in \mathbb{Z} \quad \text{and} \quad \gcd(M, p) = 1 \,.
\eeq 
We now show that this can be achieved with the same type of 3d BF theory described above, when we couple this 3d TQFT to the 4d bulk theory via partial gauging.
Consider the following 3d TQFT:
\beq
S = \frac{i pM}{2\pi} \int_{\Sigma_3} a_1 \wedge d b_1 + \frac{i pM}{2\pi} \int_{\Sigma_3} b_1 \wedge \bkA + \frac{i pM}{2\pi} \int_{\Sigma_3} a_1 \wedge \bkB\, .
\label{eq:3d BF partial gauging 1}
\eeq 
So far, this is identical to Eq.~\eqref{eq:3d BF with BGF} with a replacement $N \to pM$.
In this form, $\bkA$ and $\bkB$ are $\mathbb{Z}_{pM}$ BGFs, satisfying
\beq
\oint_{\Sigma_2} \frac{\bkA}{2\pi} \in \frac{1}{pM} \mathbb{Z} \, , \quad \oint_{\Sigma_2} \frac{\bkB}{2\pi} \in \frac{1}{pM} \mathbb{Z} \, .
\eeq 
Instead of coupling the 3d $\mathbb{Z}_{pM}$ BF theory to the most refined 2-form BGFs with minimal holonomy, let us couple it to 2-form BGFs associated with a $\mathbb{Z}_M^{(1)}$ subgroup.
If we denote the latter as $\hat{\mathcal{A}}_2$ and $\hat{\mathcal{B}}_2$, they satisfy 
\beq
\oint \frac{\hat{\mathcal{A}}_2}{2\pi} \in \frac{1}{M} \mathbb{Z} \, , \quad \oint \frac{\hat{\mathcal{B}}_2}{2\pi} \in \frac{1}{M} \mathbb{Z} \, ,
\eeq 
and for our purpose we can simply choose $\hat{\mathcal{A}}_2 = F_2 / M$ and $\hat{\mathcal{B}}_2 = G_2 / M$. 
This means that we couple the 3d $\mathbb{Z}_{pM}$ BF theory with the $\mathbb{Z}_{pM}^{(1)} (a) \times \mathbb{Z}_{pM}^{(1)} (b)$ symmetry to the 4d bulk theory via 2-form BGF coupling, but only for the subgroup $\mathbb{Z}_{M}^{(1)} (a) \times \mathbb{Z}_{M}^{(1)} (b)$.
Crucially, the resulting anomaly inflow becomes
\beq
S_\text{inflow} = - \frac{i pM}{2\pi} \int_{\Sigma_4} \hat{\mathcal{A}}_2 \wedge \hat{\mathcal{B}}_2 = - \frac{2 \pi i p}{M} \int_{\Sigma_4} \frac{F_2 \wedge G_2}{4\pi^2} \,.
\label{eq:inflow_3d BF_partial gauging}
\eeq 
One quickly recognizes that this is exactly the anomaly needed to compensate the ABJ anomaly.
We conclude that for any given $K \in \mathbb{Z}$, the theory enjoys $\left(\mathbb{Q} / \mathbb{Z}\right)^{(0)}$ (as a set, not a group) worth of non-invertible symmetry (up to the residual invertible $\mathbb{Z}_{K}^{(0)}$ symmetry), and the SDO is given by
\beq  
\mathcal{D} \left( \frac{2\pi}{N}, \Sigma_3 \right) = \int \left[D a_{1} D b_{1}\right] \exp \left[ i \oint_{\Sigma_3} \left(\frac{2\pi}{N} *J_1 + \frac{pM}{2\pi} a_1 \wedge d b_1 + \frac{p}{2\pi} b_1 \wedge F_2 + \frac{p}{2\pi} a_1 \wedge G_2 \right)\right]\, ,
\label{eq:NISDO_mixed_Kgeneral}
\eeq 
with $K/N = p / M$. 

\subsection{Fractional instantons and partial gauging}
\label{subsection:Fractional instantons and partial gauging}

Before moving to the multi-instanton case, let us explain the non-invertible SDO originating from fractional instanton effects and its construction via partial gauging.
In fact, the steps are identical to the $U(1)$ instanton case discussed in Section~\ref{subsec:CS_partial gauging}, so our discussion will be brief.
Basically, we seek a 3d TQFT whose 't~Hooft anomaly inflow cancels the anomalous transformation due to fractional instantons, given in Eq.~\eqref{eq:NIS_rac_Inst_S_shift}.
Among these, we focus on the non-invertible part $\mathbb{Z}_{K}^{(0)} \setminus \mathbb{Z}_{K/N}^{(0)}$ (as a set, not a group).
Let us consider the following 3d TQFT action defined on a 3-manifold $\Sigma_3$:
\begin{align}
S = \frac{ipM}{4\pi} \int_{\Sigma_3} a_1 \wedge da_1 + \frac{i pM}{2\pi} \int_{\Sigma_3} a_1 \wedge \mathcal{B}_2\, , \quad \gcd(M,p) = 1\,,
\end{align}
where $\mathcal{B}_2$ is a 2-form BGF associated with the $\mathbb{Z}_{pM}^{(1)}$ symmetry, satisfying $\oint_{\Sigma_{2}} \frac{\mathcal{B}_2}{2\pi} \in \frac{1}{pM} \mathbb{Z}$.
Instead of coupling the 3d CS theory to the most refined BGF with minimal holonomy, let us couple it to a BGF associated with a $\mathbb{Z}_{M}^{(1)}$ subgroup.
Denoting the latter by $\hat{\mathcal{B}}_2$, it satisfies $\oint_{\Sigma_{2}} \frac{\hat{\mathcal{B}}_2}{2\pi} \in \frac{1}{M} \mathbb{Z}$.
In terms of the 4d bulk field strength, we may take $\hat{\mathcal{B}}_2 = 2\pi w_2 / M$.
In this way, we couple the 3d CS theory to the 4d bulk theory via partial gauging, i.e. gauging only the $\mathbb{Z}_{M}^{(1)}$ subgroup.
The resulting anomaly inflow is
\begin{equation}\label{eq:inflow_fract_inst}
S_\text{inflow} = -\frac{i pM}{4\pi} \int_{\Sigma_4} \hat{\mathcal{B}}_2 \wedge \hat{\mathcal{B}}_2 = -\frac{2 \pi i p}{M} \int_{\Sigma_4} \frac{w_2 \wedge w_2}{2}\, ,
\end{equation} 
which is precisely what we need to cancel the ABJ anomaly in Eq.~\eqref{eq:NIS_rac_Inst_S_shift}. More explicitly, for an axion shift $\theta \to \theta + 2\pi \ell/K$, it is sufficient to impose Eq.~\eqref{eq:fractional_instanton_M_and_p}.
For completeness, we write the explicit expression for the non-invertible SDO (see Section~\ref{subsubsec:NI0_frac_Inst}):
\begin{align}
\mathcal{D}\left(\frac{2\pi\ell}{K}, \Sigma_3\right) = \int [D a_1] \exp \left[i \oint_{\Sigma_3} \left(\frac{2\pi \ell}{K}*\hat{J}_1 + \frac{pM}{4\pi} a_1 \wedge da_1 + p a_1 \wedge w_2\right)\right].
\end{align}
Half-space gauging for the $PSU(N)$ case proceeds as follows. First add the following action to the original one:
\beq\label{eq:half-space-gauging_fract_inst_1}
\begin{aligned}
\Delta S = ip \int_{x\geq0} b_2 \wedge w_2 + \frac{i M}{2\pi} \int_{x\geq0} b_2 \wedge d c_1  + \frac{ip M}{4\pi} \int_{x\geq0} b_2 \wedge b_2\, .
\end{aligned}
\eeq
Integrating out $c_1$ constrains $b_2$ to be a $\mathbb{Z}_M$ gauge field.
After that, we can ``complete the square'' to obtain
\beq \label{eq:half-space-gauging_fract_inst_2}
S \supset \frac{iK(N-1)}{2N}\int_{x<0} \theta w_2 \wedge w_2 + \frac{i}{2} \int_{x\geq0} \left[ \frac{K(N-1)}{N} \theta - \frac{2\pi p}{M} \right] w_2 \wedge w_2 + \frac{i p M}{4\pi} \int_{x\geq0} \left( b_2 + \frac{2\pi w_2}{M} \right)^2 \,.
\eeq 
The second term shows that gauging the magnetic symmetry $\mathbb{Z}_M^{(1)} \subset \mathbb{Z}_N^{(1)}$ shifts
\beq
\frac{K(N-1)}{N} \theta \to \frac{K(N-1)}{N} \theta - \frac{2\pi p}{M}\, .
\eeq
This can be compensated by a corresponding anomalous axion field redefinition, which is precisely the non-invertible shift transformation.
We can then integrate out $b_2$, which gives $b_2 = -2\pi w_2/M$.
The remainder of the procedure is the same as in Section~\ref{subsubsec:half space gauging}.

\section{Multi-Instantons and Non-Invertible Symmetries}
\label{sec:multi-instanton-case}

\subsection{Non-invertible symmetry from multiple \texorpdfstring{$U(1)$}{U(1)} instantons}
\label{subsec:NIS_from_multiU1}

The previous discussion shows that studying non-invertible symmetries in a 4d QFT induced by ABJ anomalies with multiple instanton effects reduces to finding a 3d TQFT with a $\prod_{i=1}^{n} \mathbb{Z}_{M_i}^{(1)}$ symmetry and a 't~Hooft anomaly characterized by a matrix $p_{ij}$ (see the discussion at the beginning of Section~\ref{sec:NIS_from_partial_gauging}). 
Our analysis in Section~\ref{subsec:CS_partial gauging} showed that unmixed (diagonal) anomalies of the form $\propto K_{ii} F_2^i \wedge F_2^i$ can be handled by a 3d CS theory with partial gauging by a 4d bulk field strength.
On the other hand, Section~\ref{subsec:BF_partial gauging} demonstrated that mixed (off-diagonal) anomalies of the form $\propto K_{i \ne j} F_2^i \wedge F_2^j$ can be canceled by a 3d BF theory with partial gauging.
These observations suggest that the 3d TQFT required for a general symmetric ABJ anomaly matrix $K_{ij}$ in Eq.~\eqref{eq:axion with general ABJ} should be an appropriate ``admixture'' of 3d CS and BF theories.
In this section, we first construct 3d TQFTs with the desired symmetry structure and analyze their 't~Hooft anomalies.
We then use these to study non-invertible symmetries with multiple instantons, including a generalization of half-space gauging to multi-gauge-sector setups and the action on multi-species 't~Hooft lines.

\subsubsection{ABJ anomaly construction}
\label{subsubsec:NIS_from_ABJ_multiU1}

Let us now consider the case where an axion couples to multiple $U(1)$ gauge bosons:
\begin{equation}
S = \frac{f_a^2}{2} \int d \theta \wedge * d \theta + \sum_{i=1}^n \frac{1}{2g_i^2} \int F_2^i \wedge * F_2^i + \sum_{i,j=1}^n \frac{i K_{ij}}{8\pi^2} \int \theta F_2^i \wedge F_2^j \, ,
\label{eq:multi-instanton_axion theory}
\end{equation}
where $\theta = a / f_a \sim \theta+2 \pi$ is the axion, $F_2^i = d A_1^i$ are the $U(1)$ field strengths, and $K_{ij}$ is the symmetric ABJ anomaly matrix with integer entries.
Under the shift $\theta \to \theta + 2\pi/N$, $N \in \mathbb{Z}$, the action shifts anomalously as
\beq 
S \to S + \sum_{i,j=1}^n \frac{2\pi i K_{ij}}{N} \int  \frac{F_2^i \wedge F_2^j}{8\pi^2}\, .
\label{eq:multiU1_ABJ_anom}
\eeq 
For the discussion of non-invertible symmetry, let us consider the following 3d TQFT:
\begin{align}
S = \sum_{i,j=1}^n \frac{p_{ij} M_{i} M_{j}}{4\pi M_{ij}} \int a_{1}^{i} \wedge da_{1}^{j}\, ,
\end{align}
where $M_i \in \mathbb{Z}$ and $M_{ij} \equiv \gcd(M_i, M_j)$. Here, $p_{ij}$ is a symmetric integer matrix subject to certain quantization and periodicity conditions, which we will discuss below.  
One notices that this theory is a matrix generalization of a combination of 3d CS and BF theories. It possesses a large set of 1-form symmetries, but for our purposes it suffices to focus on the subgroup under which each $a_1^i$ undergoes a $\mathbb{Z}_{M_i}^{(1)}$ transformation.  
Denoting the BGFs for each $\mathbb{Z}_{M_i}^{(1)}$ symmetry by $\mathcal{A}_{2}^{i}$, we can write the action on an auxiliary 4-manifold $\Sigma_4$ with coupling to these BGFs as
\begin{equation}
\label{eq:4d_auxiliary_action_multiU1}
S = \sum_{i,j=1}^n \frac{ip_{ij} M_{i} M_{j}}{4\pi M_{ij}} \int_{\Sigma_4} \left(da_1^i + \mathcal{A}_2^i\right) \wedge \left(da_1^j + \mathcal{A}_2^j\right) - \sum_{i,j=1}^n \frac{ip_{ij} M_i M_j}{4\pi M_{ij}} \int_{\Sigma_4} \mathcal{A}_2^i  \wedge \mathcal{A}_2^j \, .
\end{equation}
Each 2-form BGF $\mathcal{A}_2^i$ is quantized according to $\mathbb{Z}_{M_i}^{(1)}$:
\begin{equation}
\oint_{\Sigma_2} \frac{\mathcal{A}_2^i}{2\pi} \in \frac{1}{M_i} \mathbb{Z}\, .
\end{equation}
For our purposes, we simply choose $\mathcal{A}_2^i = F_2^i / M_i$. This means we couple a 3d TQFT whose 1-form symmetry $G^{(1)}$ contains $\prod_{i=1}^n \mathbb{Z}_{M_i}^{(1)}$ to the 4d bulk theory via a 2-form BGF coupling, but only through the subgroup $\prod_{i=1}^n \mathbb{Z}_{M_i}^{(1)} \subset G^{(1)}$.
It is then straightforward to show that such a partial gauging leads to the anomaly inflow action
\begin{equation}
S_{\rm inflow} = - \sum_{i,j=1}^n \frac{ip_{ij} M_i M_j}{4\pi M_{ij}} \int_{\Sigma_4} \mathcal{A}_2^i  \wedge \mathcal{A}_2^j = - \sum_{i,j=1}^n \frac{ip_{ij}}{4\pi M_{ij}} \int_{\Sigma_4} F_2^i \wedge F_2^j\, .
\end{equation} 
We recognize that this precisely matches the form of the anomaly required to cancel the ABJ anomaly in Eq.~\eqref{eq:multiU1_ABJ_anom}, provided that $p_{ij}$ and $M_i$ satisfy $p_{ij}/M_{ij} = K_{ij}/N$ for the given $K_{ij}$ and $N$, which can always be ensured.
We conclude that the theory admits a set of non-invertible symmetries labeled by $\left(\mathbb{Q}/\mathbb{Z}\right)^{(0)}$ (as a set, not a group), with the understanding that $\mathbb{Z}_K^{(0)}$, where $K \equiv \gcd(K_{ij})$, corresponds to the invertible symmetry.
In other words, the entire effect of multiple $U(1)$ instantons can be compensated by a 3d TQFT, at the price of turning the symmetry into a non-invertible one.
The SDO associated with the non-invertible symmmetry is
\begin{equation}
\label{eq:SDO_definition_multiU1}
\begin{aligned}
\mathcal{D}\left(\frac{2\pi}{N}, \Sigma_3\right) &= \int \left[\prod_{i=1}^n D a_1^i\right] \exp \left\{i\oint_{\Sigma_3} \left[\frac{2\pi}{N}*J_1 +\sum_{i,j=1}^n \left(\frac{p_{ij}M_iM_j}{4\pi M_{ij}} a_1^i \wedge da_1^j + \frac{p_{ij} M_i}{2\pi M_{ij}} a_1^i \wedge F_2^j \right)\right]\right\}\, ,
\end{aligned}
\end{equation}
where $J_1 = - i f_a^{2} d\theta$ is the (non-conserved) Noether current associated with the $U(1)^{(0)}$ symmetry. For each fixed $j$, the last term represents the coupling of the closed current
$\sum_{i=1}^{n} \frac{p_{ij} M_i M_j}{2\pi M_{ij}} a_1^{i}$ of the 1-form symmetry associated with $a_1^{j}$, to the 4d bulk field strength $F_2^{j}/M_j$.

\subsubsection{Half-space gauging construction}
\label{subsubsec:half_space_gauging_multiU1}

To begin our discussion, we again consider the theory introduced in Section~\ref{subsubsec:NIS_from_ABJ_multiU1} and imagine dividing the spacetime manifold into two regions by a 3-manifold $\Sigma_3$.
Let $x \in \mathbb{R}$ be the coordinate normal to $\Sigma_3$.
We then gauge $\prod_{i=1}^n \mathbb{Z}_{M_i}^{(1)} \subset \prod_{i=1}^n U(1)_{m,i}^{(1)}$ on $x \ge 0$, where $U(1)_{m,i}^{(1)}$ is the 1-form magnetic symmetry associated with the $i$-th photon $A_1^i$.
This is achieved by first coupling the theory to 2-form BGFs $\mathcal{B}_2^i$ associated with the $\mathbb{Z}_{M_i}^{(1)}$ magnetic symmetries, and then promoting $\mathcal{B}_2^i$ to dynamical $\mathbb{Z}_{M_i}$ 2-form gauge fields $b_2^i$.
Explicitly, we add the following action to the theory on $x \ge 0$, for some symmetric integer matrix $p_{ij}$:
\bea
\begin{aligned}
\Delta S &= \sum_{i, j=1}^n \frac{i p_{ij} M_i}{2\pi M_{ij}} \int_{x \geq 0} b_2 ^i \wedge F_2^j + S_{\rm MBF} \label{eq:half-space-gauging_multiple} \, ,\\
S_{\rm MBF} &= \sum_{i=1}^n \frac{i M_i}{2\pi} \int_{x \geq 0} b_2^i \wedge d c_1^i + \sum_{i,j=1}^n \frac{ip_{ij} M_i M_j}{4\pi M_{ij}} \int_{x \geq 0} b_2^i \wedge b_2^j \, ,
\label{eq:M-BF action}
\end{aligned}
\eea 
where $M_{ij} \equiv \gcd(M_i, M_j)$ as before, and $c_1^i$ are 1-form Lagrange multiplier fields.
Here, $b_2^i$ are restricted to be $\mathbb{Z}_{M_i}$ gauge fields by the 4d 
$\mathbb{Z}_{M_i}$ BF term proportional to $b_2^i \wedge d c_1^i$.
The first term can be viewed as $i \sum_{j=1}^n \tilde{b}_2^{j} \wedge * J_2^{m,j}$, where $\tilde{b}_2^{j} \equiv \sum_{i=1}^n \frac{p_{ij} M_i}{M_{ij}} b_2^{i}$ is the linear combination of the $b_2^i$ fields, and $J_2^{m,j} = * \frac{F_2^{j}}{2\pi}$ is the Noether current associated with the $U(1)_{m,j}^{(1)}$ symmetry.
In our construction, we perform the half-space gauging using a matrix generalization of the 4d BF theory, equipped with SPT terms parametrized by $p_{ij}$.

As in the single-instanton case, $p_{ij}$ must satisfy quantization conditions analogous to Eq.~\eqref{eq:quant_period_p}:
\bea\label{eq:quant_period_pij_v2}
\left\lbrace 
\begin{array}{cccc}
\text{(i)} \quad & i = j & \quad:\quad & p_{ii} M_i \in 2\mathbb{Z} \quad \text{and} \quad p_{ii} \sim p_{ii} + 2 M_i \, , \\
\text{(ii)} \quad & i \neq j & \quad:\quad & p_{ij} \sim p_{ij} + M_{ij} \, .
\end{array}
\right.
\eea 
The derivation parallels that of Eq.~\eqref{eq:quant_period_p}, and we sketch it here while emphasizing the differences arising from the multiplicity of sectors. See also Appendix~\ref{subapp:spin and anomaly_general} for an alternative derivation from the 3d TQFT perspective.
First, the addition of SPT terms parametrized by $p_{ij}$ has the effect of charging $c_1^i$ under the 1-form gauge symmetry of the $b_2^j$ fields: $b_2^j \to b_2^j + d \lambda_1^j$ and $c_1^i \to c_1^i - \sum_{j=1}^n \frac{p_{ij} M_j}{M_{ij}} \lambda_1^j$.
The invariance of the action imposes quantization conditions on $p_{ij}$.
Under the 1-form gauge transformations, the MBF sector varies (after canceling a few terms) as
\bea 
\delta S_{\rm MBF} = \sum_{i=1}^n \frac{i M_i}{2\pi} \int_{x \ge 0} d \lambda_1^i \wedge d c_1^i - 2\pi i \sum_{i,j=1}^n \frac{p_{ij} M_i M_j}{2 M_{ij}} \int_{x \ge 0} \frac{d \lambda_1^i}{2\pi} \wedge \frac{d \lambda_1^j}{2\pi} \, .
\eea 
Using $\oint_{\Sigma_2} \frac{d\lambda_1^{i}}{2\pi} \in \mathbb{Z}$ and $\oint_{\Sigma_2} \frac{d c_1^{i}}{2\pi} \in \mathbb{Z}$, we see that gauge invariance requires $p_{ii} M_i \in 2\mathbb{Z}$. For $i \neq j$, the variation $\delta S_{\rm MBF} \in 2\pi i\mathbb{Z}$, so no additional quantization condition on $p_{ij}$ is needed.

The periodicity conditions for $p_{ij}$ can be checked by examining the SPT terms, which can be conveniently rewritten as
\beq 
\sum_{i,j=1}^n \frac{2\pi i p_{ij}}{2 M_{ij}} \int_{x \ge 0} \left( M_i \frac{b_2^i}{2\pi} \right) \wedge \left( M_j \frac{b_2^j}{2\pi} \right)\, .
\eeq  
Using $\oint_{\Sigma_2} \frac{b_2^i}{2\pi} \in \frac{1}{M_i} \mathbb{Z}$, it is quick to realize the periodicity conditions in Eq.~\eqref{eq:quant_period_pij_v2}, because the action shifts by $2\pi i \mathbb{Z}$.

Going back to the discussion of half-space gauging, as discussed in Section~\ref{subsubsec:half space gauging}, we impose the boundary condition $b_2^i \vert_{x=0} = 0$ for all $i$.
These are valid topological boundary conditions since, under a smooth deformation of the $x=0$ surface, we obtain 
$b_2^i \vert_{x=0} - b_2^i \vert_{x'=0} = d b_2^i = 0$, the last equality being enforced by the bulk equations of motion.

Integrating out $c_1^i$ constrains $b_2^i$ to be a $\mathbb{Z}_{M_i}$ gauge field.
Using this, Eq.~\eqref{eq:half-space-gauging_multiple} can be rewritten as 
\begin{equation}
\label{eq:half_space_gauging_squaring_multiU1}
\begin{aligned}
S \supset \sum_{i,j=1}^n \left[\frac{i K_{ij}}{8\pi^2} \int_{x<0} \theta F_2^i \wedge F_2^j + \frac{i}{8\pi^2} \int_{x\geq0} \left( K_{ij} \theta - \frac{2\pi p_{ij}}{M_{ij}} \right) F_2^i \wedge F_2^j \right. \\
+ \left.  \frac{ip_{ij}M_iM_j}{4\pi M_{ij}} \int_{x\geq0} \left( b_2^i + \frac{F_2^i}{M_i}\right) \wedge \left( b_2^j + \frac{F_2^j}{M_j}\right)\right] \, .
\end{aligned}
\end{equation}
We have combined the terms in Eq.~\eqref{eq:half-space-gauging_multiple} into the expression shown in the second line.
Crucially, this must be accompanied by the shift
\beq 
K_{ij} \theta \to K_{ij} \theta - \frac{2\pi p_{ij}}{M_{ij}}\, .
\label{eq:multi instanton_theta shift}
\eeq 
For an axion theory as in Eq.~\eqref{eq:multi-instanton_axion theory}, this simply corresponds to a change in the axion--gauge--boson couplings.
In the QED-like theory discussed in Section~\ref{subsec:0-form_NIS}, this should be understood as a shift of the matrix--generalized $\theta$-term.

Next, the equation of motion for $b_2^i$ sets
\beq
\sum_{j=1}^n \frac{p_{ij}M_j}{M_{ij}} \left( b_2^j + \frac{F_2^j}{M_j}\right) = 0\, .
\label{eq:multi_instanton_b2-eom}
\eeq
Thus, gauging $\prod_{i=1}^n \mathbb{Z}_{M_i}^{(1)} \subset \prod_{i=1}^n U(1)_{m,i}^{(1)}$ on half of the spacetime has the net effect shown in
Eq.~\eqref{eq:multi instanton_theta shift}.
This change can be compensated by an anomalous axion field redefinition.
Specifically, if we perform the change of variables
$\theta \to \theta + 2\pi/N$ on $x \ge 0$, the action shifts as
\begin{equation}
S \to S + \frac{2\pi i}{N} \oint_{x=0} *J_1 + \sum_{i,j=1}^n \frac{2\pi i K_{ij}}{N} \int_{x\geq0} \frac{F_2^i \wedge F_2^j}{8\pi^2}\, .
\end{equation}
Therefore, if we choose $p_{ij}/M_{ij} = K_{ij}/N$, the combined effects leave the theory invariant. The induced topological defect at $x = 0$ is identified as the non-invertible SDO mapping the theory on $x < 0$ onto itself on $x \ge 0$.

To show that a non-invertible SDO is indeed induced on $\Sigma_3$ at $x = 0$, we first recall that the addition of the SPT term makes $c_1^i$ charged under 1-form gauge transformations: $b_2^j \to b_2^j + d\lambda_1^j$ and $c_1^i \to c_1^i - \sum_{j=1}^n \frac{p_{ij} M_j}{M_{ij}} \lambda_1^j$.
As a result, bulk line operators $\exp\left(i \oint_{\Sigma_1} c_1^i\right)$ are, in general, not gauge-invariant.
The gauge-invariant operator in $x \ge 0$ is instead
\begin{equation}
\tilde{W}_i = \exp \left( i \oint_{\Sigma_1} c_1^i \right) \exp \left(\sum_{j=1}^n \frac{ip_{ij}M_j}{M_{ij}} \int_{\Sigma_2} b_2^j \right)\, , \quad \partial \Sigma_2 = \Sigma_1 \,.
\end{equation}
However, on the boundary at $x = 0$, due to the boundary condition $b_2^j \vert_{x=0} = 0$, this reduces to a gauge-invariant line operator $W_i(\Sigma_1) = \exp \left( i \oint_{\Sigma_1} c_1^i \right)$.
As we prove in Appendix~\ref{app:boundary lines and braiding}, these boundary line operators satisfy the correlation function
\beq 
\left\langle W_i (\Sigma_1) W_j (\Sigma_1^\prime)  \right\rangle = \exp \left[ \frac{2\pi i p_{ij}}{M_{ij}} \text{Link} (\Sigma_1, \Sigma_1^\prime) \right]\, .
\label{eq:braiding of boundary lines}
\eeq 
This shows that the 3d TQFT must include line operators with braiding characterized by $p_{ij}$.
As described in Appendix~\ref{subapp:spin and anomaly_general}, this in turn implies that the 3d TQFT must possess 1-form global symmetries, at least $\prod_{i=1}^n \mathbb{Z}_{M_i}^{(1)}$, with 't~Hooft anomaly captured by $p_{ij}$.
One such theory is what we introduced in Section~\ref{subsubsec:NIS_from_ABJ_multiU1}.
There, we chose a 3d TQFT whose 1-form symmetry is larger than $\prod_{i=1}^n \mathbb{Z}_{M_i}^{(1)}$, and a suitable addition of sectors decoupled from the 4d bulk theory allowed a simple Lagrangian formulation and an easy determination of the 't~Hooft anomaly.

\subsection{\texorpdfstring{$U(1)$}{U(1)} and fractional instantons, all together}
\label{subsec:NIS0_multi-inst_general}

We finally study the general case in which the axion couples to multiple $U(1)$ and non-abelian gauge-theory sectors.
The ABJ anomaly is encoded in the axion--gauge couplings
\begin{equation}
S \supset  \sum_{i,j=1}^n \frac{i K_{ij}}{8\pi^2} \int \theta \operatorname{Tr} \left( F_2^i \wedge F_2^j \right) \, ,
\label{eq:multi-instanton_axion theory_abj}
\end{equation}
where, to simplify the notation, we use ``$\mathrm{Tr}$'' even when $F_2^i$ corresponds to a $U(1)$ field strength.
It is worth noting that, while ``mixed anomalies'' for two different $U(1)$ factors are natural to consider, it is not a priori clear whether mixed terms between two $w_2$ fields of different non-abelian sectors, or between a $w_2$ and a $U(1)$ field strength, are possible.
Even if such terms are mathematically acceptable, a further question is whether they can arise from 4d QFT and/or extra-dimensional UV completions of axion theories, and whether they might provide any distinction between 4d and extra-dimensional axion theories.
In this work, while we consider mixed terms among $U(1)$ factors (as in the previous sections), we analyze only diagonal terms for fractional instantons, leaving a more comprehensive study for future investigations.

To make our discussion concrete, we take the non-abelian gauge group to be either $SU(N)$ or $PSU(N)$, but our results apply equally well to other choices of non-abelian gauge group.
To make the distinction between $U(1)$ and non-abelian factors, and between $SU(N)$ and $PSU(N)$, as clear as possible, we use $F_2^i$ for $U(1)$, $G_2^s$ for $PSU(N_s)$, and $H_2^t$ for $SU(N_t)$ factors.
The axion--gauge coupling is
\beq 
S \supset \sum_{i,j=1}^k \frac{i K_{ij}}{8\pi^2} \int \theta F_2^i \wedge F_2^j  + \sum_{s=k+1}^\ell \frac{i K_{ss}}{8\pi^2} \int \theta \operatorname{Tr} \left( G_2^s \wedge G_2^s \right) + \sum_{t=\ell+1}^n \frac{i K_{tt}}{8\pi^2} \int \theta \operatorname{Tr} \left( H_2^t \wedge H_2^t \right) \, ,
\label{eq:multi-instanton_axion theory_abj_2}
\eeq 
where $K_{ss}$ and $K_{tt}$ denote diagonal entries (no implicit sum on repeated indices inside each trace term).
Let us first discuss quantization conditions on the coupling constants.
For early works on this subject, see \cite{Choi:2023pdp, Cordova:2023her}.

These are determined by requiring invariance of the action under the gauge transformation $\theta \to \theta + 2\pi$.
We obtain $K_{ij} \in \mathbb{Z}$ for $i,j = 1, \ldots, k$, $K_{ss} \in N_s \mathbb{Z}$ for $PSU(N_s)$ with $s = k+1, \ldots, \ell$, and $K_{tt} \in \mathbb{Z}$ for $SU(N_t)$ with $t = \ell+1, \ldots, n$.

One quickly realizes that in the absence of mixed terms between (1) $F_2^i$ of a $U(1)$ factor and $w_2^s$ of a $PSU(N_s)$ factor, and (2) two different $w_2^{s}$'s, the anomalous transformation of the action under a shift $\theta \to \theta + 2\pi/N$ factorizes into a purely $U(1)$ part (an $k \times k$ symmetric matrix) and a diagonal non-abelian part (an $(n - k) \times (n - k)$ diagonal matrix).
The shift in the action is then given by
\begin{equation}
\label{eq:action shift_both_U1_fract}
\begin{aligned}
S \to S + \sum_{i,j=1}^k \frac{2\pi i K_{ij}}{N} \int \frac{F_2^i \wedge F_2^j}{8\pi^2} + \sum_{s=k+1}^\ell \frac{2\pi i K_{ss}}{N} \left( n_s + \frac{N_{s} - 1}{N_s} \int \frac{w_2^s \wedge w_2^s}{2} \right) \\
+ \sum_{t=\ell+1}^n \frac{2\pi i K_{tt}}{N} \int \frac{\operatorname{Tr} (H_2^t \wedge H_2^t)}{8\pi^2} \, .
\end{aligned}
\end{equation}
Let us first discuss the invertible shift symmetry. 
Practically, this is determined by treating both $U(1)$ and fractional instanton effects, in addition to regular $SU(N)$ instantons, as genuine symmetry breaking effects.
This leads to $U(1)^{(0)} \to \mathbb{Z}_{ \gcd \left( K_{ij}, K_{ss}/N_{s}, K_{tt} \right) }^{(0)}$.
The non-invertible symmetry then corresponds to $ \mathbb{Z}_{\gcd \left( K_{ss}, K_{tt} \right)}^{(0)} \setminus \mathbb{Z}_{ \gcd \left( K_{ij}, K_{ss}/N_{s}, K_{tt} \right) }^{(0)} $ (as a set, not a group). Recall that $SU(N)$ instantons explicitly break the shift symmetry $U(1)^{(0)} \to \mathbb{Z}_{ \gcd ( K_{ss}, K_{tt} ) }^{(0)}$.

The non-invertible SDO is constructed using a suitable 3d TQFT whose 't Hooft anomaly cancels the ABJ anomaly from the $U(1)$ and fractional instantons.  
The relevant ABJ anomalies correspond to the first line (excluding the $n_s$ terms) in Eq.~\eqref{eq:action shift_both_U1_fract}.
We may view these as an $\ell \times \ell$ symmetric ABJ anomaly matrix with the first $k \times k$ block identified with $K_{ij}$, followed by an $(\ell-k) \times (\ell-k)$ diagonal block identified with the diagonal $K_{ss}$ entries (while the $SU(N_t)$ anomaly does not leave any non-invertible symmetry).
As far as the ABJ anomaly structure is concerned, this is a special case of the general analysis in Section~\ref{subsec:NIS_from_multiU1}.  
With a shift $\theta \to 2\pi/\tilde{K}$, where $\tilde{K} \equiv \gcd\left(K_{ss},K_{tt}\right)$, one simply uses the 3d TQFT in Eq.~\eqref{eq:4d_auxiliary_action_multiU1}, choosing $p_{ij}$ and $M_i$ (and similarly for indices $s$) so that (i) $ \frac{p_{ij}}{M_{ij}} = \frac{K_{ij}}{\tilde{K}} $ for $i,j = 1, \cdots, k $, and (ii) $ \frac{p_{ss}}{M_{s}} = \frac{K_{ss}(N_s-1)/N_s}{\tilde{K}}$ (no sum over $s$) for $s  = k+1, \cdots, \ell $.

Note that these conditions can always be satisfied, showing that the non-invertible symmetry is indeed 
$\mathbb{Z}_{\gcd\left(K_{ss},  K_{tt}\right)}^{(0)}$ (as a set, not a group) with $\mathbb{Z}_{ \gcd \left( K_{ij}, K_{ss}/N_{s}, K_{tt} \right) }^{(0)}$ 
being the invertible part. For the sake of completeness, we write down the non-invertible SDO explicitly
\begin{equation}
\label{eq:SDO_general}
\begin{aligned}
\mathcal{D} \left( \frac{2\pi}{\tilde{K}}, \Sigma_3\right) = \int \left[\prod_{i=1}^\ell D a_1^i\right] \exp \left[i\oint_{\Sigma_3} \left(\frac{2\pi}{\tilde{K}}*J_1 + \sum_{i,j=1}^\ell \frac{p_{ij}M_iM_j}{4\pi M_{ij}} a_1^i \wedge da_1^j \right.\right. \\
+\left.\left. \sum_{i,j=1}^k \frac{p_{ij} M_i}{2\pi M_{ij}} a_1^i \wedge F_2^j + \sum_{s=k+1}^\ell p_{ss} a_1^s \wedge w_2^s \right)\right] \, .
\end{aligned}
\end{equation}
The half-space gauging story proceeds through the same steps as in Section~\ref{subsubsec:half_space_gauging_multiU1}, with the understanding that we take the $\ell \times \ell$ symmetric ABJ anomaly matrix in 4d whose first $k \times k$ block is $K_{ij}$, followed by an $(\ell-k) \times (\ell-k)$ diagonal block given by the diagonal entries $K_{ss}$.
More explicitly, we add the following action to the theory on $x \ge 0$:
\bea
\begin{aligned}
\Delta S &= \sum_{a, b=1}^\ell \frac{i P_{ab} M_a}{2\pi M_{ab}} \int_{x \geq 0} b_2^a \wedge F_2^b + S_{\rm  MBF} \label{eq:half-space-gauging_general} \, ,\\
S_{\rm  MBF} &= \sum_{a=1}^\ell \frac{i M_a}{2\pi} \int_{x \geq 0} b_2^a \wedge d c_1^a + \sum_{a,b=1}^\ell \frac{iP_{ab} M_a M_b}{4\pi M_{ab}} \int_{x \geq 0} b_2^a \wedge b_2^b \, ,
\end{aligned}
\label{eq:M-BF action_general}
\eea 
where we introduced $F_2^b = \left(F_2^i, 2\pi w_2^s\right)$, and the symmetric integer matrix $P_{ab}$ is defined by
\beq  
P_{ab} = \left[
\begin{array}{c|c} 
p_{ij} &  \\ 
\hline 
 & p_{ss} 
\end{array}
\right] \, .
\eeq 
where $p_{ij}$ is the $k \times k$ block associated with the $U(1)$ sector and $p_{ss}$ is the $(\ell-k) \times (\ell-k)$ diagonal block associated with the $PSU(N_s)$ sector.
Here, $M_{ab} \equiv \gcd(M_a, M_b)$ as before.
The first $k \times k$ block consisting of $U(1)$ factors is identical to the setup in Section~\ref{subsubsec:half_space_gauging_multiU1}, while the $(\ell-k) \times (\ell-k)$ block consisting of $PSU(N_s)$ factors is a collection of $(\ell-k)$ copies of the discussion in Section~\ref{subsection:Fractional instantons and partial gauging}. Hence we do not repeat the full procedure here.
We instead comment on the choice of $M_s$ and $p_{ss}$ needed in the present setup.
If we consider a shift by $\theta \to \theta + 2\pi/\tilde{K}$, we can use the result of Section~\ref{subsection:Fractional instantons and partial gauging} by rewriting $2\pi/\tilde{K} = 2\pi \ell_s/K_{ss}$ with $\ell_s = K_{ss}/\tilde{K}$ for each $s$. Then, using Eq.~\eqref{eq:fractional_instanton_M_and_p}, we obtain
\beq
M_{s} = \frac{N_{s}}{\gcd(N_{s}, K_{ss} / \tilde{K})} \,, \quad p_{ss} = \frac{(N_{s} - 1) K_{ss} / \tilde{K} }{ \gcd(N_{s}, K_{ss}/\tilde{K}) } \,.
\eeq

\subsection{Global structure, CFU instantons, and non-invertible symmetry}
\label{subsec:NIS from CFU}

We finally discuss non-invertible symmetry arising from a more general type of instantons, sometimes referred to as CFU instantons~\cite{Anber:2021iip, Anber:2019nze}. The simplest example is the $1/N$-valued fractional instanton appearing in $U(N) = \frac{SU(N) \times U(1)}{\mathbb{Z}_N}$ gauge theory. The basic idea is that in $U(N)$ theory, while $\mathbb{Z}_N$ instanton configurations made only of either $U(1)$ or $SU(N)$ gauge fields are not acceptable, there still exist special combinations of the two that are admissible. For now, it may be sufficient to notice that the fact that the quotient by $\mathbb{Z}_N$ shown in the global form of $U(N)$ acts diagonally on $SU(N)$ and $U(1)$ means that any $\mathbb{Z}_N$ configuration of $SU(N)$ must be identified with a $\mathbb{Z}_N$ configuration of $U(1)$. This can be expressed as
\beq
\int_{\Sigma_2} \frac{F_2}{2\pi} = \frac{1}{N} \int w_2 \bmod 1 \quad \to \quad \frac{F_2}{2\pi} = \frac{1}{N} w_1 + X_2\,, \quad X_2 \in H^2 \left(\Sigma_4, \mathbb{Z})\right)\, ,
\label{eq:CFU_F2 to w2}
\eeq 
where $F_2$ is the $U(1)$ field strength, $w_2$ is the second Stiefel--Whitney class of $PSU(N)$, and $X_2$ is a 2-form gauge field whose integral is an integer. This interlocking leads to nontrivial $\mathbb{Z}_N$ instantons consisting of both $SU(N)$ and $U(1)$ gauge fields.
It is interesting to note that this type of instanton can appear in the Standard Model itself if the global form of the gauge group is nontrivial. Non-invertible symmetries of axion theory in the presence of a nontrivial global structure of the Standard Model gauge group were previously analyzed in~\cite{Choi:2023pdp, Cordova:2023her} (see also~\cite{Reece:2023iqn}). Here, we present a version for a general $U(N)$ (as opposed to $U(3)$ or $U(2)$ in the case of axions coupled to the Standard Model) and then determine the conditions on $p_{ij}$ and $p_{ss}$, the SPT parameters of the 3d TQFT.

The action for axion–$U(N)$ theory is given by
\beq 
S \supset \frac{i K_1}{8\pi^2} \int \theta F_2 \wedge F_2 + \frac{i K_N}{8\pi^2} \int \theta \operatorname{Tr} \left(G_2 \wedge G_2\right)\, .
\eeq 
Let us first discuss the quantization conditions on $K_1$ and $K_N$ following~\cite{Choi:2023pdp, Cordova:2023her}. Invariance of the action under $\theta \to \theta + 2\pi$ requires
\beq
\frac{K_1}{2} \int \left( \frac{1}{N} w_2 + X_2 \right)^2 + K_N \left( n_N + \frac{N-1}{N} \int \frac{w_2 \wedge w_2}{2} \right) \in \mathbb{Z}\, ,
\eeq 
where $n_N \in \mathbb{Z}$ corresponds to the integer-valued $SU(N)$ instanton number and we have used Eq.~\eqref{eq:CFU_F2 to w2}. Ignoring terms proportional to $w_2 \wedge w_2$ for the moment, the $n_N$ term implies $K_N \in \mathbb{Z}$, while the cross term proportional to $w_2 \wedge X_2$ imposes $K_1 \in N \mathbb{Z}$. Next, the term proportional to $w_2 \wedge w_2$ takes the form
\beq 
\left[ \frac{K_1}{N^2} + \frac{K_N (N -1)}{N} \right] \int \frac{w_2 \wedge w_2}{2}\, ,
\eeq
which therefore gives rise to the condition $K_1 + K_N N (N - 1) \in N^2 \mathbb{Z}$. Below, we summarize the quantization results:
\beq
K_1 \in N \mathbb{Z}\,,  \quad K_N \in \mathbb{Z}\,, \quad K_1 + K_N N (N-1) \in N^2 \mathbb{Z}\, .
\label{eq:UN_quantization}
\eeq
We are now ready to discuss the non-invertible symmetry. We consider a shift $\theta \to \theta + 2\pi/z$ and determine the condition on $z \in \mathbb{Z}$ that leaves the action invariant. Under this shift, the action transforms as
\beq 
S \to S + \frac{2\pi i K_1}{2 z} \int \left( \frac{w_2}{N} + X_2 \right)^2 + \frac{2\pi i K_N}{z} \left( n_N + \frac{N-1}{N} \int \frac{w_2 \wedge w_2}{2} \right)\,.
\label{eq:delta S in U(N)}
\eeq 
As before, ignoring the terms proportional to $w_2 \wedge w_2$ first, the $n_N$ term and the cross term proportional to $w_2 \wedge X_2$ together impose that $z = \gcd\left(K_1 / N, K_N\right)$. Next, the term proportional to $w_2 \wedge w_2$ further demands that $z$ be a divisor of $\frac{K_1 + K_N N (N - 1)}{N^2}$. Altogether, we obtain
\beq
G^{(0)}_\text{I} \equiv \mathbb{Z}^{(0)}_{\gcd\left[\frac{K_1}{N}, K_N, \frac{K_1 + K_N N (N - 1)}{N^2}\right]} \subset \mathbb{Z}^{(0)}_{K_N}\,.
\eeq 
Let us explain the meaning of this expression. First, the integer-valued $SU(N)$ instanton, denoted above as $n_N$, genuinely breaks the axion shift symmetry down to $\mathbb{Z}_{K_N}^{(0)}$. The analysis further shows that if we treat the $\mathbb{Z}_N$ instanton effect as an additional breaking (to determine the \emph{invertible} symmetry), then we obtain $G^{(0)}_\text{I}$. This is the invertible shift symmetry of the theory, while any $\mathbb{Z}_{K_N}^{(0)}$ transformations not contained in this subgroup are expected to be non-invertible symmetries. To confirm this (which was not done in previous works, e.g.~\cite{Choi:2023pdp, Cordova:2023her}), we need to identify a 3d TQFT that compensates for the $\mathbb{Z}_N$ instanton effect. In our framework, this is equivalent to determining the SPT parameters $p_{ij}$ and $p_{ss}$. To this end, it is convenient to rewrite Eq.~\eqref{eq:delta S in U(N)} for $z = K_N$ as
\beq 
\frac{\delta S}{2\pi i} = \frac{K_1}{K_N} \int \frac{X_2 \wedge X_2}{2} + \frac{\left(K_1/N\right)}{K_N} \int X_2 \wedge w_2 + \frac{\left[K_1 + K_N N (N-1)\right] / N^2}{K_N} \int \frac{w_2 \wedge w_2}{2}\,,
\label{eq:UN_ABJ_fract}
\eeq 
where we have dropped the irrelevant $n_N$ contribution. The quantization conditions in Eq.~\eqref{eq:UN_quantization} guarantee that all numerators are integers. By inspecting Eq.~\eqref{eq:4d_auxiliary_action_multiU1} and its anomaly inflow action, one finds that by considering a $2 \times 2$ version with the choice $\bkA^1 = 2\pi X_2 / M$ and $\bkA^2 = 2\pi w_2 / N$, the inflow action becomes
\beq
S_{\rm inflow} = - \frac{2\pi i p_{11}}{M} \int \frac{X_2 \wedge X_2}{2} - \frac{2\pi i p_{12}}{\gcd(M, N)} \int X_2 \wedge w_2 - \frac{2\pi i p_{22}}{N} \int \frac{w_2 \wedge w_2}{2}\,.
\eeq 
It is then straightforward to see that we have sufficient freedom to choose $p_{ij}$, $M$, and $N$ to cancel the ABJ anomaly in Eq.~\eqref{eq:UN_ABJ_fract}.

\subsection{Action on 't Hooft line operators}
\label{subsec:multi-inst_general_action on t Hooft}

Let us discuss the action of the non-invertible SDO on 't~Hooft lines.
In the general case discussed above, we have 't~Hooft lines associated with $k$ different $U(1)$ factors, and $(\ell-k)$ types of 't~Hooft operators coming from the $PSU(N_s)$ factors.
We denote the former type as $T_i(\gamma, m)$ ($i = 1, \dots, k$), where the line is defined on a closed curve $\gamma$ and $m$ labels the magnetic charge.
We write the $PSU(N_s)$ 't~Hooft lines as $\hat{T}_s(\gamma, m)$ ($s = k + 1, \dots, \ell$), where $m = 0, \dots, N_s - 1$ in this case.

The $U(1)$ 't~Hooft lines transform under the 1-form magnetic symmetry as
\begin{equation}
T_i(\gamma, m) \to T_i(\gamma, m) \exp \left(i m \oint_\gamma \lambda_1^i \right) \,.
\end{equation}
The discussion of half-space gauging clearly shows that the non-invertible SDO maps the original theory on $x<0$, where $\prod_{i=1}^k U(1)_{m,i}^{(1)}$ is the 1-form magnetic symmetry, onto a theory on $x\ge 0$ in which the 1-form magnetic symmetry $\prod_{i=1}^k \mathbb{Z}_{M_i}^{(1)} \subset \prod_{i=1}^k U(1)_{m,i}^{(1)}$ is gauged.
If we denote the 2-form gauge field for the $\mathbb{Z}_{M_i}^{(1)}$ symmetry as $\tilde{b}_2^{i}$, then under the action of the non-invertible SDO the $U(1)$ 't~Hooft line must become
\beq 
T_i(\gamma, m) \to T_i(\gamma, m) \exp \left( - i m \oint_{M_2} \tilde{b}_{2}^i \right) \, , \quad \partial M_2 = \gamma\, ,
\eeq 
in order to ensure the gauge invariance. 
On the other hand, the discussion below Eq.~\eqref{eq:half-space-gauging_multiple} shows that $\tilde{b}_2^{i} = \sum_{j=1}^{k} \frac{p_{ij} M_j}{M_{ij}} b_2^j$, implying that the $U(1)$ 't~Hooft lines transform as
\beq 
T_i(\gamma, m) \to T_i(\gamma, m) \exp \left( -\sum_{j=1}^k \frac{imp_{ij}M_j}{M_{ij}} \int_{M_2} b_2^j \right) \, .
\eeq 
Finally, substituting Eq.~\eqref{eq:multi_instanton_b2-eom} into the above expression yields
\begin{equation}
T_i(\gamma, m) \quad \text{at} \quad x < 0 \quad \to \quad  T_i(\gamma, m) \exp \left(\sum_{j=1}^k \frac{2\pi imK_{ij}}{N} \int_{M_2} \frac{F_2^j}{2\pi} \right) \quad \text{at} \quad x \geq 0\, .
\end{equation}
where we used the condition $p_{ij}/M_{ij} = K_{ij}/N$.
In other words, the 't~Hooft line $T_i(\gamma, m)$ transforms under the action of the non-invertible SDO into $T_i(\gamma, m)$ attached to a topological surface operator
$\exp\left( \sum_{j=1}^k \frac{2\pi i m K_{ij}}{N} \int_{M_2} \frac{F_2^j}{2\pi} \right)$.
This surface operator can be interpreted as a product of Wilson surfaces associated with the gauge fields $A_1^j$, each carrying fractional charge $m K_{ij}/N$.

This result is consistent with the Witten effect interpretation.
In Eq.~\eqref{eq:multi instanton_theta shift} we showed that half-space gauging shifts $K_{ij} \theta$ by $-2\pi p_{ij} / M_{ij} = -2\pi K_{ij} / N$.
By the Witten effect, a charge-$m$ $U(1)_i$ monopole acquires fractional electric charge $m K_{ij}/N$ under the gauge fields $A_1^j$, and thus becomes a dyon.

Let us now move on to the $PSU(N_s)$ 't~Hooft lines.
Since we consider only the case where the $w_2^s$ field of a $PSU(N_s)$ sector does not mix with either a $U(1)$ sector or another $PSU(N_s)$ sector, the analysis reduces to a single $PSU(N_s)$ factor.
The 2-form gauge field that gauges the $\mathbb{Z}_{M_s}^{(1)}$ symmetry transforms as $\bkB^s \equiv p_{ss} b_2^s \to \bkB^s + \frac{2\pi}{M_s} d\epsilon_1^s$. 
Hence, under the action of the non-invertible SDO, the $PSU(N_s)$ 't~Hooft line transforms as
\begin{equation}
\begin{aligned}
\hat{T}_s (\gamma, m) \to \hat{T}_s (\gamma, m) \exp \left(- i m \oint_{\Sigma_2} \bkB^s \right) & = \hat{T}_s (\gamma, m) \exp \left(- i m p_{ss} \oint_{\Sigma_2} b_2^s \right) \\ 
& = \hat{T}_s (\gamma, m) \exp \left(\frac{2\pi imp_{ss}}{M_s} \oint_{\Sigma_2} w_2^s \right) \, .
\end{aligned}
\end{equation}
We see that the $PSU(N_s)$ 't~Hooft line becomes attached to a Wilson surface carrying a $\mathbb{Z}_{M_s}$ electric center charge $mp_{ss}$.
This result is as expected from the Witten effect, in a version suitable for non-abelian gauge theory.

\section{Multi-Axions and Non-Invertible Symmetries}
\label{sec:multi-axion case}

Let us now consider the case where a $U(1)$ gauge boson couples to $n$ axions:
\begin{equation}
S = \sum_{i=1}^n \frac{f_i^2}{2} \int d \theta^i \wedge * d \theta^i + \frac{1}{2g^2} \int F_2 \wedge * F_2 + \sum_{i=1}^n \frac{i K_i}{8\pi^2} \int \theta^i F_2 \wedge F_2 \, .
\end{equation}
It turns out to be useful to perform a field redefinition to simplify the symmetry structure.
We first note that the combination $\sum_{i=1}^n K_i \theta^i$ appearing in the axion–gauge coupling motivates us to choose
\begin{align}
\tilde{\theta}^1 \equiv \frac{1}{K} \sum_{j=1}^n K_{j} \theta^{j} = \sum_{j=1}^n \tilde{K}_{j} \theta^{j}\,, \quad K \equiv \gcd\left(K_i\right)\,, \quad \tilde{K}_{j} \equiv \frac{K_{j}}{K}\,,
\end{align}
as the axion in the new basis that couples to the gauge sector. The coupling constant of $\tilde{\theta}^1$ to the gauge field is $K$. We choose a new axion basis such that $\tilde{\theta}^1$ is the only axion coupling to the gauge sector, while the remaining $(n-1)$ axions are decoupled. Note that the original periodicity $\theta^{i} \sim \theta^{i} + 2\pi$ implies $\tilde{\theta}^1 \sim \tilde{\theta}^1 + 2 \pi$, by virtue of (generalized) B\'ezout's identity. Explicitly, since $\gcd \left(\tilde{K}_i \right) = 1$, there exists a set of integers $\{ c^1, \dots , c^n \}$ such that $\sum_{i=1}^n \tilde{K}_i c^i = \gcd \left(\tilde{K}_i\right) = 1$. Hence, under $\theta^i \to \theta^i + 2\pi c^i$, the new field $\tilde{\theta}^1$ shifts exactly as desired: $\tilde{\theta}^1 \to \tilde{\theta}^1 + 2\pi$.

According to the ``unimodular completion'' (or ``primitive vector extension'')---which states that any primitive integer row vector in $\mathbb{Z}^n$ (i.e., the gcd of its entries is unity) can be completed to an $SL(n,\mathbb{Z})$ matrix---we can always find an $SL(n,\mathbb{Z})$ matrix whose first row is the reduced anomaly coefficients $\left(\tilde{K}_1, \dots, \tilde{K}_n\right)$. This is exactly what we need: the $n$-axion field space is topologically $\mathbb{T}^n$, and any change of variables respecting $\theta^{i} \sim \theta^{i} + 2\pi$ for all $i$ must be an element of $SL(n,\mathbb{Z})$. Let us denote such a transformation by $M \in SL(n,\mathbb{Z})$. The desired field redefinition is
\beq 
\left( 
\begin{array}{c} 
\theta^1  \\ \vdots \\ \theta^n
\end{array}
\right) \to \left( 
\begin{array}{c} 
\tilde{\theta}^1  \\ \vdots \\ \tilde{\theta}^n
\end{array}
\right) =
\underbrace{
\left(
\begin{array}{ccc} 
\tilde{K}_1 & \cdots & \tilde{K}_n  \\ 
\vdots & \ddots & \vdots \\ 
\cdot & \cdots & \cdot
\end{array}
\right) }_{\equiv M}
\left( 
\begin{array}{c} 
\theta^1  \\ \vdots \\ \theta^n
\end{array}
\right)\,.
\eeq 
In the new basis $\tilde{\theta}$, the unique component that couples to the gauge sector is $\tilde{\theta}^1$, and the action becomes
\begin{equation}
\begin{aligned}
S &= \sum_{i,j=1}^n \frac{\tilde{\mathcal F}_{ij}}{2} \int d\tilde{\theta}^{i} \wedge *\, d\tilde{\theta}^{j} + \frac{1}{2g^2} \int F_2 \wedge * F_2 + \frac{i K}{8\pi^2} \int \tilde{\theta}^1 F_2 \wedge F_2 \, ,  \\
\tilde{\mathcal F} &\equiv \left( M^{-1} \right)^T 
\left(
\begin{array}{ccc} 
f_1^2 &  &  \\ 
 & \ddots &  \\ 
 &  & f_n^2
\end{array}
\right) \left( M^{-1} \right) \, .
\end{aligned}
\end{equation}

\subsection{Non-invertible 0-form symmetries}

Let us first discuss the non-invertible 0-form symmetry.
Using the action in the new basis, it is straightforward to determine the Noether currents associated with the $U(1)^{(0)}_i$ shift symmetries. They are given by
\beq 
\tilde{J}_i = - i \sum_{j=1}^n\tilde{\mathcal{F}}_{ij} * d \tilde{\theta}^j\, .
\eeq 
The linear independence of all $n$ currents $\tilde{J}_i$ is guaranteed by the fact that $\det  \tilde{\mathcal{F}} \neq 0$.
From the above discussion, it follows that the invertible 0-form symmetry is broken as $\left[U(1)^{(0)}\right]^{n} \to \mathbb{Z}_K^{(0)} \times \left[U(1)^{(0)}\right]^{n-1} $, where the $\mathbb{Z}_K^{(0)}$ factor arises from the anomaly term associated with the $\tilde{\theta}^1$ component.

As we have discussed, the rational subgroup of the seemingly broken $U(1)^{(0)}_1$ symmetry is in fact realized as a non-invertible symmetry generated by the SDO:
\begin{equation}
\sdoD \left( \frac{2\pi}{N}, \Sigma_3 \right) =  \int \left[Da_{1}\right] \exp \left[ i  \oint_{\Sigma_3} \left( \frac{2\pi}{N} * \tilde{J}_1 + \frac{p M}{4\pi} a_1 \wedge d a_1 + \frac{p}{2\pi} a_1 \wedge F_2 \right) \right]\, ,
\end{equation}
where $K / N = p / M $ with $\gcd(M, p)=1$.

\subsection{Non-invertible 1-form symmetries}

Next, let us discuss the non-invertible 1-form symmetry. Since, in our new basis $\tilde{\theta}$, the axion–gauge interaction reduces to a single-axion term, we can simply adopt the steps described in Section~\ref{subsec:1-form_NIS}. Therefore, the $n$-axion theory features a non-invertible 1-form symmetry in exactly the same way as in the single-axion case.
However, there is one feature we need to address before stating the result. In Section~\ref{subsec:1-form_NIS}, the discussion was restricted to $K=1$, whereas here we consider the more general case $K \neq 1$. In that case, from the equation of motion for $A_1$, the 1-form electric symmetry $U(1)_e^{(1)}$ is reduced to an invertible $\mathbb{Z}_K^{(1)}$ symmetry together with the remaining non-invertible 1-form symmetry.
To discuss the non-invertible part, consider a 1-form symmetry transformation with parameter $\alpha = 2\pi/N$ such that $\alpha \neq 2\pi \ell/K$ for any $\ell = 1, \dots, K-1$, for some $N \in \mathbb{Z}$. If we deform the worldsheet of the non-invertible SDO, $\Sigma_2 \to \Sigma_2'$, we obtain
\beq 
\exp \left( \frac{2\pi i}{N} \int_{\Sigma_3} d * J_2 \right) = \exp \left( \frac{i K}{2\pi N} \int_{\Sigma_3} d \theta \wedge F_2 \right) \, ,
\eeq 
where $J_2 = \frac{i}{g^2} F_2$ is the (non-conserved) Noether current associated with the $U(1)^{(1)}_e$ symmetry. We need a 2d TQFT whose 't~Hooft anomaly compensates this variation. In fact, we can use the same 2d TQFT introduced in Section~\ref{subsec:1-form_NIS}, here with the replacement $N \to M$ to match our current notation:
\beq
S = \frac{i M}{2\pi} \oint_{\Sigma_2} \phi d c_1 + \frac{i p }{2\pi } \oint_{\Sigma_2} c_1 \wedge d \theta + \frac{i}{2\pi} \oint_{\Sigma_2} \phi F_2\, .
\eeq 
It suffices to choose $K/N = p/M,$ to cancel the above variation, thereby restoring the topological property of the SDO (see the inflow action Eq.~\eqref{eq:inflow for 2d TQFT}).

For our purposes with $n$ axion species, we take $\tilde{\theta}^1 = \sum_{i=1}^n \tilde{K}_i \theta^i$. The resulting SDO for the non-invertible 1-form symmetry is
\begin{equation}
\sdoD^{(1)}  \left( \frac{2\pi}{N}, \Sigma_2 \right) = \int \left[D\phi Dc_1\right] \exp \left[ i\oint_{\Sigma_2} \left(\frac{2\pi}{N} * J_2 + \frac{M}{2\pi} \phi d c_1 + \frac{p}{2\pi } c_1 \wedge d \tilde{\theta}^{1} + \frac{1}{2\pi} \phi F_2 \right)\right] \,.
\end{equation} 
In what follows, we briefly discuss how the action of the non-invertible SDO $\mathcal{D}^{(1)}\left(\frac{2\pi}{N}, \Sigma_{2}\right)$ on Wilson and ’t~Hooft line operators, as well as on axion string worldsheets, is modified.
With multiple axions $\theta^{i}$, we have $n$ species of axion string worldsheets, which we label as $S_{i}(A,w)$, where $A$ is the supporting surface and $w$ is the winding number. Note that in general, for each $i$ the supporting surface $A$ and the winding number $w$ may be different.

\begin{itemize}
    \item[(i)] For the Wilson line $W(\gamma,q)$ of electric charge $q$ supported on a closed curve $\gamma$, the non-invertible SDO acts invertibly:
    \begin{align}
     \mathcal{D}^{(1)}\left( \frac{2\pi}{N}, \Sigma_{2} \right) \quad:\quad W(\gamma,q) \to W(\gamma,q) \exp \left( \frac{2 \pi i q}{N} \right) \,.
    \end{align}

    \item[(ii)] For the 't Hooft line $ T (\gamma,m) $ of magnetic charge $m$ supported on a curve $ \gamma $, the non-invertible SDO act non-invertibly:
    \begin{align}
     \mathcal{D}^{(1)} \left( \frac{2\pi}{N},\Sigma_{2} \right) \quad:\quad  T( \gamma, m) \to  T( \gamma,m) \exp \left( \sum_{i=1}^n \frac{2\pi i mK_{i}}{N}  \int_{M_{1}} \frac{d\theta^{i}}{2\pi} \right) \,, 
    \end{align}
    where $M_{1}$ is a path connecting a point on the 't Hooft line to the intersection point of the 't Hooft line and the non-invertible SDO.

    \item[(iii)] For the action on axion string worldsheets, let us first work in the original basis $\theta$. This is convenient because it aligns with the sequence of $n$ Peccei--Quinn phase transitions during which each species of axion string worldsheets forms. As described above, the non-invertible 1-form symmetry is associated with gauging of the special linear combination proportional to $\tilde{\theta}^1 = \sum_{i=1}^n \tilde{K}_i \theta^i$; from this we can extract the action on each $\theta^i$-strings (as opposed to $\tilde{\theta}^i$-strings). Explicitly, for the axion string worldsheet $ S_{i}(A,w) $ of winding number $w$ supported on a surface $A$, the non-invertible SDO acts non-invertibly:  
    \begin{align}
    \mathcal{D}^{(1)} \left( \frac{2\pi}{N},\Sigma_{2} \right) \quad:\quad  S_{i}(A,w) \to  S_{i} (A,w) \exp \left( \frac{2\pi i w K_{i}}{N}  \int_{M_{2}} \frac{F}{2\pi} \right) \,, 
    \end{align}
    where $M_{2}$ is a surface connecting a line on the $i$-th axion string worldsheet to the intersection line between the $i$-th axion string worldsheet and the non-invertible SDO. Importantly, there are nontrivial selection rules to discuss in this case. Namely, while the special direction proportional to $\tilde{\theta}^1 = \sum_{i=1}^n \tilde{K}_i \theta_i$ undergoes a non-invertible transformation, the other $(n-1)$ independent combinations, in the sense of $SL(n,\mathbb{Z})$, transform trivially.

\end{itemize}

\section{Conclusion}
\label{sec:conclusion}

In this work, we have developed a systematic and accessible framework for analyzing non-invertible 0-form and 1-form symmetries in 4d quantum field theories, with particular emphasis on those featuring multiple instantons and multiple axions.
These types of theories appear in diverse contexts within particle physics models, and understanding non-invertible symmetries and their physical implications is therefore highly motivated.
In addition, studying quantum field theories with multiple instantons---including fractional instantons arising from nontrivial global structures---and periodic scalars is an important goal on the formal theory side.
With this in mind, our key motivation has been to bridge the formal understanding of generalized global symmetries with the concrete needs of particle phenomenology, and potentially also of condensed matter physics.
For \emph{non-invertible 0-form symmetries}, we have clarified how Adler--Bell--Jackiw anomalies with a general matrix structure $K_{ij}$, generated by multi-instanton effects, can be compensated by suitable 3d topological quantum field theories. Our approach is based on the idea of \emph{partial gauging} of 3d Chern--Simons theories.
Namely, instead of employing 3d minimal abelian topological quantum field theories with rather abstract properties, as has been extensively done in the literature, we start from a more conventional and familiar 3d Chern--Simons theory with a larger symmetry and a well-defined Lagrangian, and then gauge only an appropriate subgroup of its 1-form symmetry by coupling to the bulk 4d fields.
Through numerous examples and explicit computations, we show that this method yields an anomaly inflow action with a general anomaly coefficient and can thus be used to construct non-invertible symmetry operators in very general settings.
In particular, we demonstrate how this procedure naturally generalizes the well-known single-instanton case to systems with multiple $U(1)$ and non-abelian fractional instantons, leading to non-invertible symmetry operators whose action on local and line operators can be worked out explicitly.
This includes a detailed analysis of the transformations of various species of 't~Hooft lines, which are most conveniently studied via the half-space gauging construction of non-invertible symmetry operators. To this end, we have generalized the half-space gauging procedure to the case with multiple instantons---and hence multiple 1-form magnetic symmetries---allowing us to interpret the transformation of 't~Hooft lines in terms of the Witten effect in the presence of multiple gauge sectors.
We have also extended the analysis of \emph{non-invertible 1-form symmetries}, as well as non-invertible 0-form symmetries, to theories with multiple axions---a setting of clear phenomenological relevance given the ubiquity of axion-like particles in string compactifications and extra-dimensional models. For the non-invertible 0-form symmetry, we showed that an $SL(n,\mathbb{Z})$ change of basis reduces the symmetry analysis effectively to a single-axion case, with $(n-1)$ axions decoupled. This allows us to determine the invertible and non-invertible parts of the $n$ axion shift symmetries. For the non-invertible 1-form symmetry, we showed that the presence of $n$ axions does not complicate the story, as it merely modifies the anomaly of the 1-form electric symmetry, and the existing construction applies straightforwardly. Nevertheless, we find that the non-invertible 1-form symmetry acts nontrivially on multiple species of axion strings: among the $n$ axion strings, one special ``composite'' linear combination undergoes a non-invertible transformation, while the other $(n-1)$ combinations are blind to this non-invertible 1-form symmetry.

Looking forward, we hope that our work clarifies the concept of non-invertible symmetry and its properties, thereby promoting its broader use in both formal theory and particle physics applications. On the theoretical side, it provides a tractable framework to engineer and classify non-invertible symmetries in realistic quantum field theories, including those with nontrivial global structure or mixed gauge sectors. On the phenomenological side, our results offer a systematic approach to studying various beyond-the-Standard-Model scenarios, most notably axion models. Non-invertible symmetries introduce new selection rules, constrain the effective operator spectrum, and affect the stability and dynamics of topological defects such as axion domain walls. These insights may prove valuable in addressing long-standing challenges such as the axion quality problem and in guiding future experimental searches for axions and axion-like particles.

%Overall, we have taken steps toward a more practical and unified understanding of non-invertible symmetries in particle physics. By providing explicit, Lagrangian-based constructions that capture anomaly inflow and the topological nature of symmetry defects, we hope this work can serve both as a technical toolkit and as a conceptual bridge between formal developments and the pressing questions of beyond-Standard-Model phenomenology.

\section*{Acknowledgments}

We are grateful to T.~Daniel~Brennan, Heeyeon~Kim, and Maria~Ramos for their useful discussions.
The work of SH, HK, SML, and DS is supported by the National Research Foundation of Korea (NRF) Grant RS-2023-00211732, by the Samsung Science and Technology Foundation under Project Number SSTF-BA2302-05, and by the POSCO Science Fellowship of POSCO TJ Park Foundation. The work of SH, HK, and DS is further supported by
the National Research Foundation of Korea (NRF) Grant
RS-2024-00405629.

\appendix

%%%%%%%%%%%%%%%% 

\section{Review of Topological Spin and 't Hooft Anomaly}
\label{app:spin and anomaly}

\subsection{Spin and anomaly in 3d Chern--Simons theory}
\label{subapp:spin and anomaly_cs}

In this appendix, we review the spin of line operators and how the 't Hooft anomaly parameter $p$ appears in the spin formula. We will see that these can be used to study the symmetries and anomalies of 3d TQFTs, providing a method complementary to anomaly inflow.

\begin{figure}
    \centering
\begin{tikzpicture}

% 1st
\braid[
style strands={1}{black,thick},
style strands={2}{red,thick},
number of strands=2
]
(a) at (0,0) s_1 s_1;
\draw[red, ->] ([yshift=-0.35cm]a-rev-2-s) to ([xshift=-0.035cm,yshift=-0.45cm]a-rev-2-s);
\draw[red, ->] ([yshift=-1.2cm]a-rev-1-s) to ([yshift=-1.3cm]a-rev-1-s);
\draw[red, <-] ([yshift=+0.35cm]a-rev-2-e) to ([xshift=-0.035cm,yshift=+0.45cm]a-rev-2-e);
\node at ([yshift=+0.3cm]a-rev-1-s) {$\gamma$};
\node at ([yshift=+0.35cm]a-rev-2-s) {$\textcolor{red}{\gamma^\prime}$};
\node at ([yshift=-0.3cm]a-rev-1-e) {$V(\gamma)$};
\node at ([yshift=-0.3cm]a-rev-2-e) {$\textcolor{red}{V(\gamma^\prime)}$};

% Arrow
\draw[->, thick] ([xshift=1cm,yshift=-1.25cm]a-2-s) -- ([xshift=1.5cm,yshift=-1.25cm]a-2-s);

% 2nd
\braid[
style strands={1}{white,thick},
style strands={2}{red,thick},
number of strands=2
]
(b) at ([xshift=2.5cm]a-2-s) s_1 s_1;
\draw[black, thick] ([xshift=+0.5cm]b-rev-1-s) -- ([xshift=+0.5cm, yshift=-1.6cm]b-rev-1-s);
\draw[black, thick] ([xshift=+0.5cm, yshift=+0.6cm]b-rev-1-e) -- ([xshift=+0.5cm]b-rev-1-e);
\draw[red, ->] ([yshift=-0.35cm]b-rev-2-s) to ([xshift=-0.035cm,yshift=-0.45cm]b-rev-2-s);
\draw[red, ->] ([yshift=-1.2cm]b-rev-1-s) to ([yshift=-1.3cm]b-rev-1-s);
\draw[red, <-] ([yshift=+0.35cm]b-rev-2-e) to ([xshift=-0.035cm,yshift=+0.45cm]b-rev-2-e);

% Arrow
\draw[->, thick] ([xshift=1cm,yshift=-1.25cm]b-2-s) -- ([xshift=1.5cm,yshift=-1.25cm]b-2-s);

% 3rd
\braid[
style strands={1}{white,thick},
style strands={2}{white,thick},
number of strands=2
]
(c) at ([xshift=2.5cm]b-2-s) s_1 s_1;
\draw[red, thick] (c-rev-2-s) -- ([yshift=-0.4cm]c-rev-2-s);
\draw[red, thick] plot [smooth, tension=1] coordinates {([yshift=-0.4cm]c-rev-2-s) ([xshift=-0.15cm, yshift=-1.1cm]c-rev-2-s) ([xshift=-0.55cm, yshift=-0.75cm]c-rev-2-s) ([xshift=+0.15cm, yshift=-1.25cm]c-rev-1-s) ([xshift=-0.55cm, yshift=+0.75cm]c-rev-2-e) ([xshift=-0.15cm, yshift=+1.1cm]c-rev-2-e) ([yshift=+0.4cm]c-rev-2-e)};
\draw[red, thick] (c-rev-2-e) -- ([yshift=+0.4cm]c-rev-2-e);
\draw[white, line width=0.3cm] ([xshift=+0.8cm, yshift=-0.7cm]c-rev-1-s) -- ([xshift=+0.47cm, yshift=-0.9cm]c-rev-1-s);
\draw[white, line width=0.3cm] ([xshift=+0.4cm, yshift=-0.6cm]c-rev-1-s) -- ([xshift=+0.52cm, yshift=-0.8cm]c-rev-1-s);
\draw[black, thick] ([xshift=+0.5cm]c-rev-1-s) -- ([xshift=+0.5cm, yshift=-1.6cm]c-rev-1-s);
\draw[black, thick] ([xshift=+0.5cm, yshift=+0.6cm]c-rev-1-e) -- ([xshift=+0.5cm]c-rev-1-e);
\draw[red, ->] ([yshift=-0.35cm]c-rev-2-s) to ([yshift=-0.45cm]c-rev-2-s);
\draw[red, ->] ([xshift=-0.3cm, yshift=-1cm]c-rev-2-s) to ([xshift=-0.35cm, yshift=-0.92cm]c-rev-2-s);
\draw[red, ->] ([xshift=+0.15cm, yshift=-1.2cm]c-rev-1-s) to ([xshift=+0.15cm, yshift=-1.3cm]c-rev-1-s);
\draw[red, <-] ([xshift=-0.3cm, yshift=+1cm]c-rev-2-e) to ([xshift=-0.35cm, yshift=+0.92cm]c-rev-2-e);
\draw[red, <-] ([yshift=+0.35cm]c-rev-2-e) to ([yshift=+0.45cm]c-rev-2-e);

% Arrow
\draw[->, thick] ([xshift=1cm,yshift=-1.25cm]c-2-s) -- ([xshift=1.5cm,yshift=-1.25cm]c-2-s);

% 4th
\braid[
style strands={1}{white,thick},
style strands={2}{red,thick},
number of strands=2
]
(d) at ([xshift=2.5cm]c-2-s) 1 1;
\draw [red, thick] plot [smooth cycle, tension=1] coordinates {([xshift=+0.15cm, yshift=-1.25cm]d-rev-1-s) ([xshift=+0.5cm, yshift=-1.4cm]d-rev-1-s) ([xshift=+0.85cm, yshift=-1.25cm]d-rev-1-s) ([xshift=+0.5cm, yshift=-1.1cm]d-rev-1-s)};
\draw[white, line width=0.3cm] ([xshift=+0.54cm, yshift=-1.0cm]d-rev-1-s) -- ([xshift=+0.49cm, yshift=-1.2cm]d-rev-1-s);
\draw[white, line width=0.3cm] ([xshift=+0.46cm, yshift=-1.0cm]d-rev-1-s) -- ([xshift=+0.51cm, yshift=-1.2cm]d-rev-1-s);
\draw[black, thick] ([xshift=+0.5cm]d-rev-1-s) -- ([xshift=+0.5cm, yshift=-1.25cm]d-rev-1-s);
\draw[black, thick] ([xshift=+0.5cm]d-rev-1-e) -- ([xshift=+0.5cm, yshift=+0.95cm]d-rev-1-e);
\draw[red, ->] ([xshift=+0.45cm, yshift=-1.4cm]d-rev-1-s) to ([xshift=+0.55cm, yshift=-1.4cm]d-rev-1-s);
\draw[red, ->] ([yshift=-1.2cm]d-rev-2-s) to ([yshift=-1.3cm]d-rev-2-s);
\node at ([yshift=-1.55cm]d-rev-1-s) {$\textcolor{red}{\oint a_1}$};
\node at ([xshift=+0.5cm, yshift=-0.3cm]d-rev-1-e) {$V(\gamma)$};

\end{tikzpicture}
    \caption{Braiding of abelian anyons in 3d. In the discussion below, we consider the braiding of two $\left[V^p\right]^s$, i.e., the $\mathbb{Z}_N^{(1)}$ symmetry-generating lines of $\tqftA^{pN,1}$.} 
    \label{fig:anyon_self_braiding}
\end{figure}
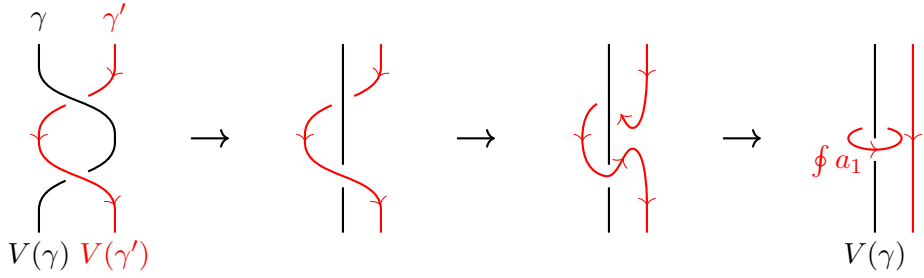

Consider a 3d TQFT with a $\mathbb{Z}_N^{(1)}$ symmetry generated by the line $V(\gamma)$. Then $V(\gamma)$ satisfies $V^N(\gamma) = 1$. If $W(\gamma')$ is a charged line with charge $q(W)$, then we have
\beq
V(\gamma) W(\gamma') = W(\gamma') e^{\frac{2\pi i q(W)}{N}}\, ,
\eeq
where we have assumed that $\gamma$ and $\gamma'$ have unit linking number.
Alternatively, one can take $V^s(\gamma)$ with an integer $s$ as the generating line, in which case
\beq
V^s(\gamma) W(\gamma') = W(\gamma') e^{\frac{2\pi i s q(W)}{N}}\, .
\eeq
The spin of $V^s$, denoted $h[V^s]$, is known to take the form \cite{Hsin:2018vcg}
\beq
h[V^s] = \frac{p s^2}{2N} \bmod 1\, , \quad p \in \{0,1,\dots,2N-1\}\, .
\eeq
As we will see momentarily, the integer parameter $p \in \mathbb{Z}_{2N}$ is nothing but the 't~Hooft anomaly parameter. A derivation of this spin formula can be found in Section~2 of \cite{Hsin:2018vcg}. Here, we derive it in a different way by explicitly computing braiding phases while assuming the factorization of the 3d $U(1)_{pN}$ CS theory Eq.~\eqref{eq:UpN_ANp_ApN_map}, $U(1)_{pN} \leftrightarrow \tqftA^{pN,1} = \tqftA^{N,p} \otimes \tqftA^{p,N}$ (if $\gcd(N,p) = 1$).
While this is not a first-principles derivation (since the assumed factorization was obtained in \cite{Hsin:2018vcg} using the spin formula), it nevertheless provides useful insights.
The 3d TQFT $\tqftA^{pN,1}$ can be described by the action (we assume a spin structure can be defined)
\beq
S = \frac{i pN}{4\pi} \int_{\Sigma_3} a_1 \wedge d a_1 \, .
\eeq 
As we discussed, a $\mathbb{Z}_{pN}^{(1)}$ symmetry is generated by $V(\gamma) = e^{i \oint_\gamma a_1}$, and the subgroup $\mathbb{Z}_N^{(1)}$ relevant for $\tqftA^{N,p}$ is generated by $V^p(\gamma) = e^{i p \oint_\gamma a_1}$. Alternatively, we can take $\left[V^p\right]^s(\gamma)$ as the generating line of $\mathbb{Z}_N^{(1)}$. We want to study the braiding of two of these $\mathbb{Z}_N^{(1)}$ symmetry lines, as shown in Figure~\ref{fig:anyon_self_braiding}.

%%%%%%%%%%%%%
%%%%%%%%%%%%%

If we define the braiding phase $\varphi = e^{i \theta \left(\text{anyon}_1, \text{anyon}_2\right)}$ by
\beq 
\vert \text{anyon}_1, \text{anyon}_2 \rangle = \varphi \vert \text{anyon}_2, \text{anyon}_1 \rangle\, ,
\eeq 
the initial and final configurations in the first diagram in Figure~\ref{fig:anyon_self_braiding} are related by $\varphi^2$. As usual, the spin is defined as $\theta = 2\pi h[V^s]$. With this definition, bosons with integer $h$ have a $+1$ braiding phase, while fermions with half-integer $h$ have a $-1$ braiding phase. Using the topological property of the line operator (an abelian anyon in our case), the first diagram can be smoothly deformed into the last configuration. If we can compute the phase of the last diagram, then we can extract the spin of the line using the above definition. The phase appearing in the last diagram can be computed as follows. First, we insert the line $\left[V^p\right]^s$ into the path integral:
\beq
\int \left[D a_1\right] e^{i ps \oint_\gamma a_1} e^{-\frac{i pN}{4\pi} \int_{\Sigma_3} a_1 \wedge d a_1} \, .
\eeq 
Then, the equation of motion for $a_1$ with the insertion of $\left[V^p\right]^s$ becomes
\beq
\frac{pN}{2\pi} d a_1 = ps \delta_\gamma\, ,
\label{eq:eom_a1_spin}
\eeq 
where $\delta_\gamma$ is a ``Poincar\'e dual delta 2-form'' to the worldline $\gamma$,\footnote{More precisely, for a submanifold $\Sigma_p \subset \Sigma_d$, $\delta_{\Sigma_p}$ is the Thom class of the normal bundle of $\Sigma_p$~\cite{Harvey:1998bx,Harvey:2005it,Brennan:2023mmt}. Equivalently, $\delta_{\Sigma_p}$ is a distributional $(d-p)$-form on $\Sigma_d$ supported on $\Sigma_p$, defined by
\begin{equation}
\int_{\Sigma_d} \omega_p \wedge \delta_{\Sigma_p} = \oint_{\Sigma_p} \omega_p \quad \text{for any smooth $p$-form $\omega_p$}\, .
\end{equation}
} which appears when we rewrite the line operator as
\beq 
e^{i ps \oint_\gamma a_1} = e^{i ps \int_{\Sigma_3} a_1 \wedge \delta_\gamma}\, .
\eeq 
Integrating Eq.~\eqref{eq:eom_a1_spin} over a 2-manifold 
$\Sigma_2$ with boundary $\partial \Sigma_2 = \gamma'$, and using the definitions 
of the intersection and linking numbers,\footnote{For a brief discussion of intersection and linking numbers, as well as the derivation of correlation functions of extended defect operators in BF theory, see Section~3 of \cite{Brennan:2023mmt}.} we obtain
\beq 
\left[ V^p \right]^s(\gamma) V (\gamma') = e^{\frac{2\pi i s}{pN}} V (\gamma')\, .
\eeq 
From this, we obtain the self-braiding of $\left[V^p\right]^s$:
\beq  
\left[ V^p \right]^s (\gamma) \left[ V^p \right]^s (\gamma') = e^{2\pi i \left( 2 h \left[ \left[ V^p \right]^s \right] \right)} \left[ V^p \right]^s (\gamma) \left[ V^p \right]^s (\gamma') = e^{\frac{2\pi i p s^2}{N }} \left[ V^p \right]^s (\gamma) \left[ V^p \right]^s (\gamma') \, ,
\eeq 
and hence the topological spin of the $\mathbb{Z}_N^{(1)}$ symmetry generating line is
\beq
h \left[ \left[ V^p \right]^s \right] = \frac{p s^2}{2N} \, .
\label{eq:topological spin}
\eeq 
Notice that, from the map $U(1)_{pN} \leftrightarrow \tqftA^{pN,1} = \tqftA^{N,p} \otimes \tqftA^{p,N}$ (if $\gcd(N,p)=1$),  we already know that the integer parameter satisfies $p \sim p+2N$ (or $p \sim p+N$ on spin manifolds) and coincides with the 't~Hooft anomaly parameter Eq.~\eqref{eq:U1_pN_inflow_partial_gauging_2}, which we reproduce below for convenience:
\beq 
S_{\rm inflow} = - \frac{2 \pi i p}{N} \int \frac{w_2 \wedge w_2}{2} \, .
\eeq 
In the remainder of this paper, we use the combination of the spin formula and the inflow action to study the symmetry and anomaly of 3d TQFTs.

\subsection{Spin and anomaly in 3d TQFT with a general 1-form symmetry}
\label{subapp:spin and anomaly_general}

Before we close this section, let us revisit the quantization conditions in Eq.~\eqref{eq:quant_period_pij_v2} within a 3d TQFT whose symmetry group is $\prod_{i=1}^n \mathbb{Z}_{N_i}^{(1)}$, following~\cite{Hsin:2018vcg}.
If we denote the line that generates the $\mathbb{Z}_{N_i}^{(1)}$ symmetry as $V_i(\gamma)$, then they satisfy the mutual braiding
\beq
V_i^{s_i} (\gamma) V_j^{s_j} (\gamma^\prime) = V_j^{s_j} (\gamma^\prime) \exp \left( - \frac{2\pi i s_i s_j m_{ij}}{N_i} \right)\, ,
\eeq
where it is understood that $\gamma$ and $\gamma'$ have nontrivial linking, and $m_{ij} \in \mathbb{Z}_{N_i}$. We note that this equation can be viewed in two different ways: (1) $\mathbb{Z}_{N_i}^{(1)}$ SDO acting on $V_j^{s_j}$, and (2) $\mathbb{Z}_{N_j}^{(1)}$ SDO acting on $V_i^{s_i}$. This yields the consistency condition
\beq
\frac{m_{ij}}{N_i} = \frac{m_{ji}}{N_j} \bmod 1 \quad\to\quad m_{ij} N_j = m_{ji} N_i \bmod N_i N_j\, .
\label{eq:NXN CS_consistency condition}
\eeq 
Defining $N_{ij} \equiv \gcd(N_i, N_j)$ and writing $N_i = N_{ij}\hat{N}_i$ and $N_j = N_{ij}\hat{N}_j$ with $\gcd(\hat{N}_i, \hat{N}_j) = 1$, the consistency condition becomes $m_{ij}\hat{N}_j = m_{ji}\hat{N}_i$.
This, in turn, implies that (i) $m_{ij}$ is an integer multiple of $\hat{N}_i$ and (ii) $m_{ji}$ is an integer multiple of $\hat{N}_j$. Equivalently,
\beq  
m_{ij} = \hat{N}_i P_{ij} = \frac{N_i P_{ij}}{N_{ij}} \,, \quad m_{ji} = \hat{N}_j P_{ji} = \frac{N_j P_{ji}}{N_{ij}} \,,
\eeq  
where $P_{ij} = P_{ji} \in \mathbb{Z}$. The equality $P_{ij} = P_{ji}$ follows by substituting the above expressions for $m_{ij}$ and $m_{ji}$ into the consistency condition~\eqref{eq:NXN CS_consistency condition}. Since $m_{ij} \in \mathbb{Z}_{N_i}$ and $m_{ji} \in \mathbb{Z}_{N_j}$, it follows that $P_{ij} \sim P_{ij} + N_{ij}$. The spins of the symmetry lines are
\beq 
h \left( \prod_{i=1}^n a_i^{s_i} \right) = \sum_{i,j=1}^n \frac{p_{ij} s_i s_j}{2 N_{ij}} \bmod 1\, , \quad p_{ij} = P_{ij} \quad \text{or} \quad P_{ij} + N_{ij}\, .
\eeq 
This shows that the $\prod_{i=1}^n \mathbb{Z}_{N_i}^{(1)}$ symmetry is characterized by the 't~Hooft anomaly matrix $p_{ij}$.
From the spin formula, it is clear that $p_{ij}$ satisfies
\beq 
p_{ii} \sim p_{ii} + 2 N_i\, , \quad p_{ij} \sim p_{ij} + N_{ij} \quad \text{for} \quad i \neq j\, .
\eeq 
The third quantization condition is obtained by imposing $V_i^{N_i} = 1$, which is necessary if the 3d TQFT is non-spin. Otherwise, the theory is defined as a spin theory and one imposes instead $V_i^{N_i} = \psi$, where $\psi$ is a transparent spin-1/2 line generating a $\mathbb{Z}_2^{(1)}$ symmetry.
Imposing $V_i^{N_i} = 1$ requires $p_{ii} N_i \in 2 \mathbb{Z}$.

\section{Boundary Line Operators and Their braiding}
\label{app:boundary lines and braiding}

In this appendix, we derive the correlation functions of boundary line operators mentioned in Eq.~\eqref{eq:braiding of boundary lines}. As discussed in Section~\ref{subsubsec:half_space_gauging_multiU1}, the gauge-invariant operators in $x \ge 0$ take the form $\tilde{W}_i = \exp\left( i \oint_{\Sigma_1} c_1^i \right) \exp\left(\sum_{j=1}^n \frac{i p_{ij} M_j}{M_{ij}} \int_{\Sigma_2} b_2^j \right)$, where $\partial \Sigma_2 = \Sigma_1$, and are reduced to gauge-invariant line operators $W_i(\Sigma_1) = \exp\left( i \oint_{\Sigma_1} c_1^i \right)$ on the boundary at $x=0$. We wish to compute the boundary correlation function $\langle W_i (\Sigma_1) W_j (\Sigma_1') \rangle$. As a first step, we rewrite the line operators as
\bea
W_i (\Sigma_1) &=& \exp \left( i \oint_{\Sigma_1} c_1^i \right) = \exp \left( i \int_{\Sigma_4} c_1^i \wedge \delta_{\Sigma_1}  \right)\, , \\
W_j (\Sigma_1') &=& \exp \left( i \oint_{\Sigma_1'} c_1^j \right) = \exp \left( i \int_{\Sigma_4} c_1^j \wedge \delta_{\Sigma_1'}  \right)\, .
\eea
Here, $\delta_{\Sigma_1}$ is the Poincar\'e dual delta 3-form, a distributional 3-form on $\Sigma_4$ supported on the boundary curve $\Sigma_1$, defined by
\beq 
\int_{\Sigma_4} w_1 \wedge \delta_{\Sigma_1} = \oint_{\Sigma_1} w_1 \quad \text{for any smooth 1-form $w_1$}\, .
\eeq 
To compute the correlation function $\langle W_i(\Sigma_1) W_j(\Sigma_1') \rangle$, we view the insertions of the lines on $\partial \Sigma_4 = \Sigma_3$ as sources for the 4d theory. Next, we choose $S_2, S_2' \subset \Sigma_4$ with $\partial S_2 = \Sigma_1$ and $\partial S_2' = \Sigma_1'$. Here, $S_2$ and $S_2'$ are Seifert surfaces.\footnote{A Seifert surface is an orientable surface whose boundary is a knot or link.} Denoting the Poincar\'e-dual delta 3-forms by $\delta_{S_2}$ and $\delta_{S_2'}$, we have\footnote{We take the following convention: for an oriented $p$-chain $C \subset X$ (here $X = \Sigma_4$), where $X$ is an $n$-manifold, its Poincar\'e-dual current $\delta_C$ is the $(n-p)$-current defined by
\beq 
\int_{X} \eta \wedge \delta_C = \int_C \eta \quad \text{for any smooth $p$-form $\eta$.}
\eeq
(rather than writing $\delta_C \wedge \eta$ in the integrand). With this convention, one has the sign identity
\beq 
d \delta_C = (-1)^p \, \delta_{\partial C},
\eeq
provided that $C$ lies in the interior of $X$ (so that no additional ambient-boundary terms appear). This can be proven by the graded Leibniz rule for differential forms---namely, for a smooth $(p-1)$-form $\alpha$,
\beq
d\left(\alpha \wedge \delta_C\right) = d\alpha \wedge \delta_C + (-1)^{p-1} \alpha \wedge d\delta_C\, ,
\eeq
and by Stokes' theorem:
\beq
-(-1)^{p-1} \int_{X} \alpha \wedge d\delta_C = \int_{X} d\alpha \wedge \delta_C = \int_C d\alpha = \int_{\partial C} \alpha = \int_{X} \alpha \wedge \delta_{\partial C}\, .
\eeq
Since this holds for all $\alpha$, one arrives at $d \delta_C = (-1)^p \delta_{\partial C}$.
} 
\beq
d \delta_{S_2} = \delta_{\Sigma_1}\, , \quad d \delta_{S_2'} = \delta_{\Sigma_1'}\, .
\eeq
The above relations show that $\delta_{S_2}$ and $\delta_{S_2'}$ are not smooth 2-forms, but rather \emph{singular 2-currents}.

To make this point clear, and also because we will need them below, let us review a few mathematical facts. First, currents can be thought of as generalized forms---the de~Rham--theoretic analogue of distributions---just as the Dirac delta $\delta(x)$ is a generalized function (a distribution) rather than an ordinary function. Here, $\delta_{\Sigma_1}$ and $\delta_{\Sigma_1'}$ act as boundary charges $q$ that turn on the current in the form $dJ = q$. More precisely (and technically), given an $n$-manifold $X$, a $k$-current $T$ is a continuous linear functional on compactly supported smooth $(n-k)$-forms:
\beq 
\langle T, \phi \rangle = \int_X T \wedge \phi \in \mathbb{R} \quad (\text{or $\mathbb{C}$})\, ,
\eeq 
for any smooth test $(n-k)$-form $\phi \in \Omega^{n-k}(X)$. Here, $\langle \cdot , \cdot \rangle$ denotes the pairing between a current and a test form. Then, every smooth $k$-form $\eta$ defines a $k$-current $T_\eta$ by
\beq
\langle T_\eta, \phi \rangle = \int_X \eta \wedge \phi\, .
\eeq 
In our discussion, it is important to keep track of the distinction between ``smooth forms'' and ``singular currents.'' Consider an oriented $p$-submanifold $Y \subset X$ and the associated Poincar\'e-dual (delta) current $\delta_Y$. Then, 
\beq 
\langle \delta_Y, \phi \rangle = \int_X \delta_Y \wedge \phi = \int_Y i^* \phi\, ,
\eeq 
where $\delta_Y$ is understood to be an $(n-p)$-current. Here, $i^*$ is the restriction to a submanifold $Y$. If $\partial Y$ is nontrivial, then one has Stokes' theorem $d\delta_Y=\delta_{\partial Y}$ as discussed above. Using the definition given above together with the properties of the wedge product and Stokes' theorem, one can show that
\bea
&\text{(i)}& \quad \langle d T, \phi \rangle = (-1)^{k+1} \langle T, d \phi \rangle\, , \\
&\text{(ii)}& \quad \langle T \wedge \alpha, \phi \rangle = \langle T, \alpha \wedge \phi \rangle\, .
\eea 
Normally, one wedges a current only with smooth forms, unless transversality makes a current--current product well defined. There is one more important fact to mention. Let us consider the case where $\partial X$ is nontrivial.
 Then, 
\begin{itemize}
    \item[(i)] Let $i:\partial X\hookrightarrow X$ be the inclusion, and $\eta$ be \emph{smooth up to the boundary}. For $S\subset\partial X$,
    \beq 
    \langle T_\eta, \delta_S \rangle = \int_X \eta \wedge \delta_S = \int_S i^* \eta\, .
    \eeq 
    \item[(ii)] Let $\omega$ be a \emph{singular current} near $\partial X$, in the sense that $d\omega = \delta_{\Sigma}$ with $\Sigma \subset \partial X$. For $S \subset \partial X$, we instead use the interior trace
    \beq 
\int_X \omega \wedge \delta_S = \int_S \operatorname{Tr}_\text{int} \omega\, .
    \eeq 
    The interior trace is defined as follows. Choose a collar $\iota:\partial X \times [0,\epsilon) \hookrightarrow X$ with coordinate $t \geq 0$. For a $k$-current $\omega$ that is locally integrable in the open collar, the interior trace $\operatorname{Tr}_\text{int}\omega$ defines a $k$-current on $\partial X$ acting on smooth test forms $\phi$ on $\partial X$. To this end, take a nonnegative bump function $\rho_\delta$ of mass $1$ (i.e.~the area under the bump is $1$), supported in $0<t<\delta$, and the projection map $\pi:\partial X \times [0,\epsilon) \to \partial X$. We then ``slice'' $\omega$ with a thin wall $dt$ just inside the boundary and let the wall approach $t=0^+$ by taking $\delta \to 0$. Technically, we have
    \beq
    \left\langle \operatorname{Tr}_\text{int}\omega , \phi \right\rangle := 
    \lim_{\delta \downarrow 0} \left\langle \omega, \pi^* \phi \wedge \rho_\delta(t)dt \right\rangle\, .
    \label{eq:interior trace}
    \eeq
    One can then show that $\operatorname{Tr}_\text{int}\omega = \delta_{\Sigma}$ if $i^*\omega = 0$. To see this, note that as $\delta \to 0$, the bump function $\rho_\delta(t)dt$ concentrates at $t=0^+$, just like the delta function. It is therefore natural to introduce a \emph{cutoff} (or step) function $\theta_\delta(t)$ defined by
    \beq 
    \theta_\delta (0) = 1\, , \quad \theta_\delta (t) = 0 \quad \text{for} \quad t \geq \delta\, , \quad d \theta_\delta (t) = - \rho_\delta (t) dt\, .
    \eeq 
    Then it is straightforward to show that
    \beq
    \left\langle \omega , \pi^* \phi \wedge \rho_\delta dt \rangle = - \langle \omega , \pi^* \phi \wedge d \theta_\delta \right\rangle = \left\langle d \omega , \left(\pi^* \phi\right) \theta_\delta \right\rangle - \left\langle \omega , d \left(\pi^* \phi\right) \theta_\delta \right\rangle.
    \eeq 
    If $i^*\omega=0$ (Dirichlet boundary condition in the sense of currents), the second term vanishes as $\delta \to 0$. This is because $\theta_\delta$ is supported in the interior $0<t<\delta$ and collapses to the boundary, while $\omega$ restricts to zero on the boundary. Therefore, we get
    \beq 
    \left\langle \operatorname{Tr}_\text{int}\omega , \phi \right\rangle = \left\langle d \omega , \left(\pi^* \phi\right) \theta_\delta \right\rangle = \left\langle \delta_{\Sigma} , \phi\, \theta_\delta (0) \right\rangle \to \left\langle \delta_\Sigma , \phi \right\rangle \quad \text{as} \quad \delta \to 0\, .
    \eeq 
    \item[(iii)] If $\omega$ is actually smooth at the boundary, then ${\rm Tr}_{\rm int} \omega = i^* \omega$. 
\end{itemize}
With all these preparations, we proceed as follows to derive the correlation function. (i) We integrate out the 1-form Lagrange multiplier field $c_1^i$ in the presence of boundary-line insertions. (ii) We solve the equation of motion for $c_1^i$ to construct a $b_2^i$-configuration. (iii) We find that the entire contribution to the topological phase comes from the SPT term proportional to $b_2^i \wedge b_2^j$, with the result given in Eq.~\eqref{eq:braiding of boundary lines}.

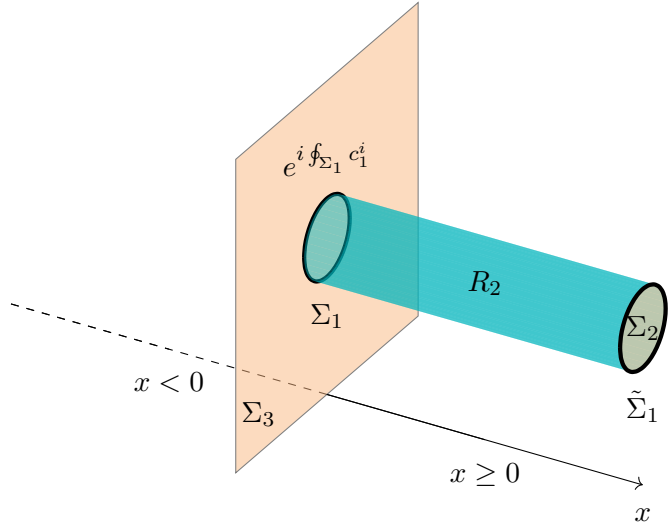
\begin{figure}
    \centering
\tdplotsetmaincoords{60}{120}
\begin{tikzpicture}[tdplot_main_coords, scale=1.2]

% --- coordinate axes ---
\draw[->] (0,0,-2) -- (0,4,-2) node[below=5pt] {$x$};
\draw[] (0,0,-2) -- (0,2,-2) node[below=5pt] {$x\ge0$};
\draw[dashed] (0,-4,-2) -- (0,0,-2) node[] {};
\draw[dashed] (0,-4,-2) -- (0,-2,-2) node[below=5pt] {$x<0$};

% --- cylinder ---
\def\r{0.5} % radius
\def\ymin{0} % bottom x
\def\ymax{4} % top x

% --- \Sigma_3 ---
\fill[Apricot,opacity=0.5,draw=black]
(-2,0,-2) -- (-2,0,2) -- (2,0,2) -- (2,0,-2) -- cycle;
\node at (1.5,0,-1.5) {$\Sigma_3$};

% --- \Sigma_1 ---
\draw[draw=black,ultra thick]
plot[domain=0:360,smooth,variable=\t]
({\r*cos(\t)},{\ymin},{\r*sin(\t)});
\node at (0,0,-1) {$\Sigma_1$};
\node at (0,0,1) {$e^{i\oint_{\Sigma_1} c_1^i}$};

% --- R_2 ---
\foreach \t in {0,10,...,350}{
\path[fill=Aquamarine,opacity=0.5,draw=none]
({\r*cos(\t)},{\ymin},{\r*sin(\t)}) --
({\r*cos(\t)},{\ymax},{\r*sin(\t)}) --
({\r*cos(\t+10)},{\ymax},{\r*sin(\t+10)}) --
({\r*cos(\t+10)},{\ymin},{\r*sin(\t+10)}) -- cycle;
}
\node at (0,\ymax*0.5,0) {$R_2$};

% --- \Sigma_2 ---
\draw[fill=Apricot,opacity=0.5,draw=none]
plot[domain=0:360,smooth,variable=\t]
({\r*cos(\t)},{\ymax},{\r*sin(\t)});
\node at (0,\ymax,0) {$\Sigma_2$};

% --- \tilde{Sigma}_1 ---
\draw[draw=black,ultra thick]
plot[domain=0:360,smooth,variable=\t]
({\r*cos(\t)},{\ymax},{\r*sin(\t)});
\node at (0,\ymax,-1) {$\tilde{\Sigma}_1$};

\end{tikzpicture}
    \caption{Illustration of the pushed-off loop and ribbon. For each boundary loop $\Sigma_1\subset\Sigma_3$, take a pushed-off copy $\tilde{\Sigma}_1$. Then choose a surface $\Sigma_2$ with $\partial \Sigma_2=\tilde{\Sigma}_1$ (note that $\Sigma_2\not\subset\Sigma_3$). Finally, choose a ribbon $R_2$ with $\partial R_2=\tilde{\Sigma}_1-\Sigma_1$.}
    \label{fig:pushed_off_ribbon}
\end{figure}

Integrating out $c_1^i$ with fixed $i$ yields
\beq
\frac{\delta S}{\delta c_1^i} = 0 \quad\to\quad
d b_2^i = \frac{2\pi}{M_i}\delta_{\Sigma_1}\, .
\label{eq:eom c1i}
\eeq
One obtains an analogous equation for $b_2^j$ by integrating out $c_1^j$, with $\Sigma_1$ replaced by $\Sigma_1'$.
Let us first discuss a particular solution to this equation, $b_2^i(p)$. For each boundary loop $\Sigma_1 \subset \Sigma_3$, we pick a ``pushed-off'' copy $\tilde{\Sigma}_1$ inside the collar $\Sigma_3 \times (0,\epsilon)$ with the boundary sitting at $\Sigma_3 \times \{0\}$. Then, we choose a $\Sigma_2 \subset \Sigma_4$ with $\partial \Sigma_2 = \tilde{\Sigma}_1$. Later, it will be important that $\Sigma_2$ is not contained in $\Sigma_3$. There is one more ingredient we need to introduce: we choose a ``ribbon'' $R_2 \subset \Sigma_3 \times (0,\epsilon)$ such that $\partial R_2 = \tilde{\Sigma}_1 - \Sigma_1$. See Figure~\ref{fig:pushed_off_ribbon} for an illustration. Then, the particular solution is given by 
\beq 
b_2^i(p) = \frac{2\pi}{M_i}\left(\delta_{\Sigma_2} - \delta_{R_2}\right)\, ,
\label{eq:particular_b2}
\eeq 
where $\delta_{\Sigma_2}$ and $\delta_{R_2}$ are the Poincar\'e-dual 2-currents of $\Sigma_2$ and $R_2$, respectively.

%%%%%%%%%%%%%
%%%%%%%%%%%%%

In fact, it is straightforward to check that the above expression solves the equation:
\beq 
d b_2^i (p) = \frac{2\pi}{M_i} \left( d \delta_{\Sigma_2} - d \delta_{R_2} \right) = \frac{2\pi}{M_i} \left[ \delta_{\tilde{\Sigma}_1} - \left( \delta_{\tilde{\Sigma}_1} - \delta_{\Sigma_1} \right) \right] = \frac{2\pi}{M_i} \delta_{\Sigma_1}\, .
\eeq 
The particular solution for $b_2^j(p)$ can be constructed in the same way, in terms of $\Sigma_2'$ and $R_2'$. 
To construct the full solution space, we note that all solutions of Eq.~\eqref{eq:eom c1i} are obtained from $b_2^i(p)$ by adding (i) an exact 2-current $d\alpha_1$, with $\alpha_1$ satisfying $i^*\alpha_1=0$ (to ensure the boundary condition), and possibly (ii) a closed 2-form $\beta_2$ with $d\beta_2=0$, again with $i^*\beta_2=0$.\footnote{In a contractible collar, or if we work locally near the boundary as in the derivation of the boundary correlation function $\langle W_i(\Sigma_1) W_j(\Sigma_1') \rangle$, the relevant relative group $H^2(\Sigma_4,\Sigma_3)$ vanishes, and such $\beta_2$ can be taken to be zero.} Therefore, the general solution takes the form 
\beq
b_2^i (\text{general}) = b_2^i (p) + d \alpha_1^i + \beta_2^i \quad \text{where} \quad i^* \alpha_1^i = 0\, , \quad d \beta_2^i = 0\, , \quad i^* \beta_2^i = 0\, .
\eeq
There is an alternative way to understand the $d\alpha_1$ term. Namely, if we change the choice of $\Sigma_2$ and $R_2$ as $\Sigma_2 \to \Sigma_2 + \partial V_3$ and $R_2 \to R_2 + \partial W_3$, the associated Poincar\'e-dual 2-currents change according to $\delta_{\Sigma_2} \to \delta_{\Sigma_2} + d\delta_{V_3}$ and $\delta_{R_2} \to \delta_{R_2} + d\delta_{W_3}$. These variations lead to a shift of $b_2^i$ by exact forms.

Note that the particular solution given in Eq.~\eqref{eq:particular_b2} on its own is not yet consistent with the topological boundary condition $b_2^i \vert_{\partial \Sigma_4} = 0$. To check this, first note that each $\Sigma_2$ lives away from $\Sigma_3$, and each $R_2$ meets $\Sigma_3$ along its edge in such a way that $\partial R_2 = \tilde{\Sigma}_1 - \Sigma_1$. The former means $i^* \delta_{\Sigma_2} = 0$, while the latter fact implies that $i^* \delta_{R_2} = - \delta_{\Sigma_1}$. %\sh{Clarify ``outward-normal-first'' to fix the convention with the minus sign.} 
This in turn results in
\beq 
i^* b_2^i (p) = \frac{2\pi}{M_i} \delta_{\Sigma_1}\, ,
\eeq 
hence it does not satisfy the required boundary condition. This issue can be fixed by properly choosing the $d\alpha_1^i$ term. We choose a 3-chain $W_3 \subset \Sigma_3 \times (0,\epsilon)$ such that $\partial W_3 = \Sigma_2 - S_2 - R_2$. The associated Poincar\'e-dual 2-current satisfies $d \delta_{W_3} = \delta_{\partial W_3} = \delta_{\Sigma_2} - \delta_{S_2} - \delta_{R_2}$. Then the following solution
\beq 
b_2^i = b_2^i (p) - \frac{2\pi}{M_i} d \delta_{W_3} = \frac{2\pi}{M_i} \left( \delta_{\Sigma_2} - \delta_{R_2} - d \delta_{W_3} \right)
\label{eq:correct_ b2}
\eeq 
satisfies the desired boundary condition. Using $i^* d \delta_{W_3} = d\left(i^* \delta_{W_3}\right) = d \delta_{S_2} = \delta_{\Sigma_1}$, one can show that 
\beq 
i^* b_2^i = \frac{2\pi}{M_i} \left[ 0 - (-\delta_{\Sigma_1} ) - \delta_{\Sigma_1} \right] = 0\, .
\eeq
Notice that $b_2^i$ still solves Eq.~\eqref{eq:eom c1i} since $d^2 \delta_{W_3} = 0$. For notational simplicity, let us define $\omega_2 = \delta_{\Sigma_2} - \delta_{R_2} - d \delta_{W_3}$, and similarly for $\omega_2'$. Let us summarize the properties of $\omega_2$ for later use:
\beq
\omega_2 : = \delta_{\Sigma_2} - \delta_{R_2} - d \delta_{W_3} \quad : \quad i^* \omega_2 = 0\, , \quad d \omega_2 = \delta_{\Sigma_1}\, , \quad \omega_2 = \delta_{S_2}\, . 
\eeq 
This shows that $\omega_2$ and $\omega_2'$ are \emph{singular 2-currents}.

Finally, to compute the correlation function of boundary line operators, one has to compute the path integral:
\beq
\begin{aligned}
\int \left[\prod_{i=1}^n d c_1^i d b_2^i \right] \exp{\left[ i\int_{\Sigma_4} \left( \sum_{i=1}^n \frac{M_i}{2\pi}  b_2^i \wedge d c_1^i + \sum_{i,j=1}^n \frac{p_{ij} M_i M_j}{4\pi M_{ij}} b_2^i \wedge b_2^j \right) \right] } \\
  \times \exp{\left[i \int_{\Sigma_4} \left(c_1^i \wedge \delta_{\Sigma_1} + c_1^j \wedge \delta_{\Sigma_1'}\right)\right]} \, .
\end{aligned}
\eeq 
Above, we have constructed the configurations $b_2^i$ and $b_2^j$, for \emph{fixed} $i$ and $j$, by integrating out $c_1^i$ and $c_1^j$ (including the boundary-line insertions). On shell, the entire correlation function is then determined by the term proportional to $b_2^i \wedge b_2^j$. Substituting the general solution Eq.~\eqref{eq:correct_ b2}, we get 
\beq 
S = \int_{\Sigma_4} \frac{2\pi i p_{ij} }{M_{ij}} \left( \omega_2 \wedge \omega_2' + \frac{1}{2} \omega_2 \wedge \omega_2 + \frac{1}{2} \omega_2' \wedge \omega_2' \right)\, .
\eeq 
The mixed term proportional to $\omega_2 \wedge \omega_2'$ gives rise to the desired result Eq.~\eqref{eq:braiding of boundary lines}, while the last two correspond to self-linking/framing phases. We first compute the mixed term: 
\beq 
\int_{\Sigma_4} \omega_2 \wedge \omega_2' = \int_{\Sigma_4} \omega_2 \wedge \delta_{S_2'} = \int_{S_2'} {\rm Tr}_{\rm int} \omega_2 = \int_{S_2'} \delta_{\Sigma_1} = \text{Link} \left( \Sigma_1, \Sigma_1' \right)\, ,
\eeq 
where in the second equality we used the interior trace rather than $i^* \omega_2$ since $\omega_2$ is a singular $2$-current, and the third equality follows from $\mathrm{Tr}_{\mathrm{int}} \omega_2 = \delta_{\Sigma_1}$ as explained below Eq.~\eqref{eq:interior trace}. Finally, we used the definition of the linking number---counting intersection points with signs determined by orientations (see \cite{Brennan:2023mmt} for a review).

The self-terms are related to the notion of framing and the topological spin of the line. For a closed oriented curve $\Sigma_1 \subset \Sigma_3$, a framing $f$ is a nowhere-zero normal vector field along $\Sigma_1$ (equivalently, a choice of a thin ribbon around $\Sigma_1$). With a framing specified, we can make a ``push-off'' $\Sigma_1^f$ by moving $\Sigma_1$ a tiny distance along $f$. Then the self-linking number with respect to the framing $f$ is
\beq 
\text{SL}_f (\Sigma_1) := \text{Link} \left( \Sigma_1, \Sigma_1^f \right) \in \mathbb{Z}\, .
\eeq
If $\Sigma_1$ bounds a surface $S_2 \subset \Sigma_3$, the ``Seifert framing'' is the one whose push-off points along the normal of $S_2$, and, relative to this choice, the self-linking number is zero: $\text{SL}_\text{Seifert} (\Sigma_1) = 0$. Following the same steps, the self-term becomes
\beq
\int_{\Sigma_4} \omega_2 \wedge \omega_2 = \int_{S_2} \delta_{\Sigma_1} = \text{SL}_\text{Seifert} (\Sigma_1 ) = 0\, .
\eeq 
Combining all, we finally get
\beq 
\left\langle W_i (\Sigma_1) W_j (\Sigma_1^\prime)  \right\rangle = \exp \left[ \frac{2\pi i p_{ij}}{M_{ij}} \text{Link} \left(\Sigma_1, \Sigma_1^\prime\right) \right] \times \exp \left[ \frac{2\pi i p_{ii}}{2 M_i} \text{SL}_f (\Sigma_1) + \frac{2\pi i p_{jj}}{2 M_j} \text{SL}_f (\Sigma_1') \right]\, ,
\eeq 
with the understanding that, for the choice $f$ equal to the Seifert framing, the framing phase can be dropped. Also, the connection between the framing phase and the topological spin can be understood by realizing that $\Sigma_1^f$ is a copy of the original line $\Sigma_1$, and the choice of framing $f$ (or ribbon) encodes the way the two strands braid. Indeed, considering a single-species case, one finds that the framing phase is identical to the topological spin Eq.~\eqref{eq:topological spin} of the line.

\section{Review of Fractional Instantons}
\label{app:fractional_instantons}

In this appendix, we review the basics of fractional instantons. To make our discussion concrete, we take a simple gauge theory with gauge group $PSU(N) = SU(N) / \mathbb{Z}_N$, which possesses fractional instantons.
$PSU(N)$ theory can be described by first considering an $SU(N)$ gauge theory with matter fields in representations with $N$-ality $N$, e.g., the adjoint representation.\footnote{Practically, $N$-ality counts the number of boxes in the Young tableau associated with the representation, taken $\bmod N$.}
Such a theory has a $\mathbb{Z}_N^{(1)}$ electric symmetry, as can be seen from the fact that the entire set of fields, including the gauge fields, is invariant under the 0-form $\mathbb{Z}_N$ center transformation.
The latter fact implies that $\mathbb{Z}_N^{(1)}$-charged Wilson lines cannot be screened by local charges and are hence topologically protected.
If we gauge the $\mathbb{Z}_N^{(1)}$ center, the resulting theory is the $PSU(N)$ gauge theory.
While gauging the $\mathbb{Z}_N^{(1)}$ center eliminates the 1-form electric symmetry, there is an emergent (dual) 1-form magnetic symmetry, which is again $\mathbb{Z}_N^{(1)}$.
One way to see this is to note that $\pi_1\left(PSU(N)\right) = \mathbb{Z}_N$.
As a result, the path integral of the $PSU(N)$ gauge theory contains a sum over a dynamical $\mathbb{Z}_N$ 2-form gauge field $B_2$, which satisfies a quantization condition consistent with $\mathbb{Z}_N^{(1)}$:
\beq
\oint_{\Sigma_2} \frac{B_2}{2\pi} = \frac{1}{N} \mathbb{Z} \, .
\eeq
This also implies the existence of $1/N$-valued fractional instantons in the $PSU(N)$ gauge theory.
To see this more explicitly, we first promote $SU(N)$ to $U(N) = \left[SU(N) \times U(1)\right] / \mathbb{Z}_N$ and later reduce it to $PSU(N)$.
The reduction can be achieved in two steps.
First, the local degree of freedom associated with the extra $U(1)$ factor is projected out by introducing a Lagrange multiplier term in the $U(N)$ theory:
\beq
S = \frac{1}{g^2} \int \text{Tr} \left(\hat{F}_2 \wedge * \hat{F}_2 \right) + \frac{i}{2\pi} \int \tilde{F}_2 \wedge \text{Tr} \left(\hat{F}_2\right) + \frac{i \theta}{8\pi^2} \int \text{Tr} \left(\hat{F}_2 \wedge \hat{F}_2\right) \, ,
\label{eq:PSUN_action_1}
\eeq
where $\hat{F}_2 = d\hat{A}_1 + \hat{A}_1 \wedge \hat{A}_1$ is the field strength of the dynamical $U(N)$ 1-form gauge field $\hat{A}_1$, and $\tilde{F}_2$ is a 2-form Lagrange multiplier field that enforces $\text{Tr}\left(\hat{F}_2\right) = 0$ via its equation of motion, thereby projecting out the $U(1)$ degree of freedom.
In order to fully reduce to the $PSU(N)$ gauge theory, we further impose the $\mathbb{Z}_N^{(1)}$ symmetry as follows.
Denoting the dynamical $SU(N)$ 1-form gauge field by $A_1$ and its field strength by
$F_2 = dA_1 + A_1 \wedge A_1$, we can write $\hat{A}_1$ as
\beq 
\hat{A}_1 = A_1 + \frac{1}{N} B_1 \mathbf{1} \, , \quad \hat{F}_2 = F_2 + \frac{1}{N} d B_1 \mathbf{1} \, ,
\eeq 
where $B_1$ is a dynamical $U(1)$ 1-form gauge field, and $\boldsymbol{1}$ is the $N \times N$ identity matrix.
Then, the $\mathbb{Z}_N^{(1)}$ symmetry acts as
\bea
A_1 \to A_1 \, , \quad B_1 \to B_1 - N \lambda_1  \quad \Rightarrow \quad \hat{A}_1 \to \hat{A}_1 - \lambda_1 \mathbf{1} \, , \quad \hat{F}_2 \to \hat{F}_2 - d \lambda_1 \mathbf{1} \, ,
\eea 
where $\lambda_1$ is the $\mathbb{Z}_N^{(1)}$ transformation parameter.
We couple the action to a 2-form BGF $B_2$ for the $\mathbb{Z}_N^{(1)}$ symmetry as
\beq
\begin{aligned}
S = \frac{1}{g^2} \int \text{Tr} \left[ \left( \hat{F}_2 - B_2 \mathbf{1} \right) \wedge * \left( \hat{F}_2 - B_2 \mathbf{1} \right) \right] + \frac{i}{2\pi} \int \tilde{F}_2 \wedge \text{Tr} \left( \hat{F}_2 - B_2 \mathbf{1} \right) \\
+ \frac{i \theta}{8\pi^2} \int \text{Tr} \left[ \left( \hat{F}_2 - B_2 \mathbf{1} \right) \wedge \left( \hat{F}_2 - B_2 \mathbf{1} \right) \right] \, .
\label{eq:PSUN_action_2_B2}
\end{aligned}
\eeq
This action is invariant under the local (gauged) $\mathbb{Z}_N^{(1)}$ transformation if we assign the transformation $B_2 \to B_2 - d\lambda_1$ (together with $\hat{A}_1 \to \hat{A}_1 - \lambda_1 \mathbf{1}$, hence $\hat{F}_2 \to \hat{F}_2 - d\lambda_1 \mathbf{1}$).
To understand the instanton spectrum, we first note that $\text{Tr}\left(\hat{F}_2\right)=dA_1$.
Using this, it is straightforward to see that the Lagrange multiplier term becomes
\beq
\frac{i}{2\pi} \int \tilde{F}_2 \wedge \left( dB_1 - N B_2 \right)\, ,
\eeq
which can be recognized as a 4d $\mathbb{Z}_N$ BF theory.\footnote{A typical form of the 4d $\mathbb{Z}_N$ BF action is $\frac{iN}{2\pi} \int B_2 \wedge \tilde{F}_2$, where $\tilde{F}_2 = d\tilde{B}_1$ is the $U(1)$ field strength. We can dualize $\tilde{B}_1$:
\beq
S = \frac{iN}{2\pi} \int B_2 \wedge \tilde{F}_2 - \frac{i}{2\pi} \int dB_1 \wedge \tilde{F}_2 \, ,
\eeq
with $B_1$ being the dual gauge field. It is introduced as a 1-form Lagrange multiplier field to impose the Bianchi identity $d\tilde{F}_2 = 0$. Thus, the 4d $\mathbb{Z}_N$ BF action appearing in the main text is this dualized version.
}
It is possible to interpret Eq.~\eqref{eq:PSUN_action_2_B2} as an $SU(N)$ gauge theory coupled to this 4d $\mathbb{Z}_N$ BF theory \cite{Kapustin:2014gua}.
A brief review of the 4d $\mathbb{Z}_N$ BF theory can be found in Appendix B of \cite{Brennan:2023kpw}, and more details can be found in \cite{Banks:2010zn, Kapustin:2014gua, Brennan:2023mmt}.
For instance, the equation of motion for $\tilde{F}_2$ sets $B_2 = dB_1/N$, showing that $B_2$ is a dynamical $\mathbb{Z}_N$ 2-form gauge field.
We are ready to discuss the instanton spectrum. The $\theta$-angle term can be written as
\beq
\begin{aligned}
S_\theta &= \frac{i \theta}{8\pi^2} \int \left[\text{Tr} \left( \hat{F}_2 \wedge \hat{F}_2 \right) - N B_2 \wedge B_2\right]  \\
& = \frac{i \theta}{8\pi^2} \int \left[\text{Tr} \left( \hat{F}_2 \wedge \hat{F}_2 \right) - \text{Tr} \left( \hat{F}_2 \right) \wedge \text{Tr} \left( \hat{F}_2 \right) + N (N-1) B_2 \wedge B_2 \right] \\
& = i \theta \left(n + \frac{N-1}{N} \int \frac{w_2 \wedge w_2}{2}\right) \, .
\label{eq:PSUN_fractional_inst}
    \end{aligned}
\eeq
In the second line, we subtracted and added the term $\text{Tr}\left(\hat{F}_2\right)\wedge \text{Tr}\left(\hat{F}_2\right)$, which allows us to express the first integral as the standard $SU(N)$ instanton number $n \in \mathbb{Z}$.
The fractional instantons are captured in the second term.
In the last line, we re-expressed the fractional-instanton part in terms of $w_2 \equiv N \frac{B_2}{2\pi}$, known as the second Stiefel--Whitney class, whose integral takes values in $\mathbb{Z}_N$:
\beq
\oint_{\Sigma_2} w_2 = 0, 1, \cdots, N-1 \, .
\eeq
In a theory with a spin structure (hence fermions can be introduced), $\int \frac{w_2 \wedge w_2}{2} \in \mathbb{Z}$.\footnote{More precisely, the generalized Pontryagin square operation can be defined as in~\cite{Kapustin:2013qsa,Hsin:2020nts}:
\beq
\begin{aligned}
\mathcal{P} \quad:\quad H^2\left(\Sigma_4, \mathbb{Z}_N\right) &\to
\begin{cases}
H^4\left(\Sigma_4, \mathbb{Z}_N\right) & \qquad\quad\ \text{for odd } N\,, \\
H^4\left(\Sigma_4, \mathbb{Z}_{2N}\right) & \qquad\quad\ \text{for even } N\,,
\end{cases} \\
w_2 &\mapsto
\begin{cases}
w_2 \cup w_2 & \text{for odd } N\,, \\
\tilde{w}_2 \cup \tilde{w}_2 - \tilde{w}_2 \cup_1 \delta\tilde{w}_2 & \text{for even } N\,,
\end{cases}
\end{aligned}
\eeq
where $\cup$ and $\delta$ are the discrete analogues of the wedge product $\wedge$ and the exterior derivative $d$ acting on differential forms.
Here, $\tilde{w}_2$ is an integer lift of $w_2$, and $\cup_i : C^p \times C^q \to C^{p+q-i}$ is the higher cup product.
When $N$ is odd, $2$ is invertible in $\mathbb{Z}_N$. When $N$ is even, $\int \mathcal{P}(w_2)$ is even (assuming a spin manifold). Thus, $\int \mathcal{P}(w_2)$ is always divisible by 2. In this paper, we use the continuum notation $\int w_2 \wedge w_2 \in 2\mathbb{Z}$ instead of $\int \mathcal{P}(w_2)$ for convenience.
}
Using this, one can clearly see that the second term in the last line of Eq.~\eqref{eq:PSUN_fractional_inst} indeed corresponds to $1/N$-valued fractional instantons.

%%%%%%%%%%%%%%%%

\bibliographystyle{jhep}
\bibliography{4dNIS_draft}
\end{document}